\documentclass[a4paper,clock]{article}
\usepackage{graphicx}
\usepackage[dvips]{epsfig}
\usepackage{ amssymb}
\usepackage{bm}
\usepackage{dcolumn}

\catcode`\@=11
\def\marginnote#1{}
\newcount\hour
\newcount\minute
\newtoks\amorpm
\hour=\time\divide\hour by60
\minute=\time{\multiply\hour by60 \global\advance\minute
by-\hour}\edef\standardtime{{\ifnum\hour<12
\global\amorpm={am}%
        \else\global\amorpm={pm}\advance\hour by-12 \fi
        \ifnum\hour=0 \hour=12 \fi
        \number\hour:\ifnum\minute<10
0\fi\number\minute\the\amorpm}}
\edef\militarytime{\number\hour:\ifnum\minute<10
0\fi\number\minute}

\def\draftlabel#1{{\@bsphack\if@filesw {\let\thepage\relax
   \xdef\@gtempa{\write\@auxout{\string
      \newlabel{#1}{{\@currentlabel}{\thepage}}}}}\@gtempa
   \if@nobreak \ifvmode\nobreak\fi\fi\fi\@esphack}
        \gdef\@eqnlabel{#1}}
\def\@eqnlabel{}
\def\@vacuum{}
\def\draftmarginnote#1{\marginpar{\raggedright\scriptsize\tt#1}}
\def\draft{\oddsidemargin -.5truein
        \def\@oddfoot{\sl preliminary draft \hfil
        \rm\thepage\hfil\sl\today\quad\militarytime}
        \let\@evenfoot\@oddfoot \overfullrule 3pt
        \let\label=\draftlabel
        \let\marginnote=\draftmarginnote

\def\@eqnnum{(\theequation)\rlap{\kern\marginparsep\tt\@eqnlabel}%
\global\let\@eqnlabel\@vacuum}  }

\def\numberbysection{\@addtoreset{equation}{section}
        \def\theequation{\thesection.\arabic{equation}}}

\def\underline#1{\relax\ifmmode\@@underline#1\else
 $\@@underline{\hbox{#1}}$\relax\fi}

\catcode`@=12
\relax

\numberbysection

\advance \topmargin by -\headheight
\advance \topmargin by -\headsep

\evensidemargin \oddsidemargin
\def\br{\begin{eqnarray}}
\def\er{\end{eqnarray}}
\def\be{\begin{equation}}
\def\ee{\end{equation}}

\def\({\left(}
\def\){\right)}

\newcommand{\bi}[1]{\bibitem{#1}}

\relax

\def\a{\alpha}

\def\b{\beta}

\def\d{\delta}
\def\D{\Delta}

\def\g{\gamma}
\def\G{\Gamma}

\def\l{\lambda}

\def\om{\omega}

\def\pa{\partial}

\def\s{\sigma}

\def\tp0{\Theta_{+}^{(0)}}
\def\tm0{\Theta_{-}^{(0)}}

\def\f#1#2#3 {f^{#1#2}_{#3}}

\def\win1{{\sf w_{1+\infty}}}

\def\Win1{{\sf W_{1+\infty}}}

\def\rlx{\relax\leavevmode}
\def\inbar{\vrule height1.5ex width.4pt depth0pt}
\def\IZ{\rlx\hbox{\sf Z\kern-.4em Z}}
\def\IR{\rlx\hbox{\rm I\kern-.18em R}}
\def\IC{\rlx\hbox{\,$\inbar\kern-.3em{\rm C}$}}
\def\IN{\rlx\hbox{\rm I\kern-.18em N}}
\def\IO{\rlx\hbox{\,$\inbar\kern-.3em{\rm O}$}}
\def\IP{\rlx\hbox{\rm I\kern-.18em P}}
\def\IQ{\rlx\hbox{\,$\inbar\kern-.3em{\rm Q}$}}
\def\IF{\rlx\hbox{\rm I\kern-.18em F}}
\def\IG{\rlx\hbox{\,$\inbar\kern-.3em{\rm G}$}}
\def\IH{\rlx\hbox{\rm I\kern-.18em H}}
\def\II{\rlx\hbox{\rm I\kern-.18em I}}
\def\IK{\rlx\hbox{\rm I\kern-.18em K}}
\def\IL{\rlx\hbox{\rm I\kern-.18em L}}
\def\one{\hbox{{1}\kern-.25em\hbox{l}}}
\def\0#1{\relax\ifmmode\mathaccent"7017{#1}%
B        \else\accent23#1\relax\fi}

\def\PRL#1#2#3{{\sl Phys. Rev. Lett.} {\bf#1} (#2) #3}
\def\NPB#1#2#3{{\sl Nucl. Phys.} {\bf B#1} (#2) #3}

\def\PRD#1#2#3{{\sl Phys. Rev.} {\bf D#1} (#2) #3}

\def\PRA#1#2#3{{\sl Phys. Rev.} {\bf A#1} (#2) #3}
\def\PRB#1#2#3{{\sl Phys. Rev.} {\bf B#1} (#2) #3}

\def\PLB#1#2#3{{\sl Phys. Lett.} {\bf #1B} (#2) #3}
\def\JMP#1#2#3{{\sl J. Math. Phys.} {\bf #1} (#2) #3}
\def\PTP#1#2#3{{\sl Prog. Theor. Phys.} {\bf #1} (#2) #3}

\def\AoP#1#2#3{{\sl Ann. of Phys.} {\bf #1} (#2) #3}

\def\RMP#1#2#3{{\sl Rev. Mod. Phys.} {\bf #1} (#2) #3}

\def\JPA#1#2#3{{\sl J. Physics} {\bf A#1} (#2) #3}

\def\JHEP#1#2#3{{\sl JHEP} {\bf #1} (#2) #3}

\def\NJP#1#2#3{{\sl New J. Phys.} {\bf #1} (#2) #3}

\def\JPCM#1#2#3{{\sl J. Phys.: Condens. Matter} {\bf #1} (#2) #3}

\def\PZ1#1#2#3{{\sl Phys. Z} {\bf #1} (#2) #3}

\def\RPP#1#2#3{{\sl Rep. Prog. Phys.} {\bf #1} (#2) #3}
\def\NC#1#2#3{{\sl Nat Commun} {\bf #1} (#2) #3}
\def\PRX#1#2#3{{\sl Physical Review X} {\bf #1} (#2) #3}
\def\AP#1#2#3{{\sl Advances in Physics} {\bf #1} (#2) #3}
\def\ARCMP#1#2#3{{\sl Annu. Rev. Condens. Matter Phys.} {\bf #1} (#2) #3}
\def\SC#1#2#3{{\sl Science} {\bf #1} (#2) #3}
\def\SR#1#2#3{{\sl Sci Rep} {\bf #1} (#2) #3}
\def\PRR#1#2#3{{\sl Physical Review Research} {\bf #1} (#2) #3}
\def\JPCS#1#2#3{{\sl J. Phys.: Conf. Ser.} {\bf A#1} (#2) #3}
\begin{document}
\begin{titlepage}
\vspace*{-1cm}

\noindent

\vskip 2cm

\vspace{.2in}
\begin{center}
{\large\bf Biorthogonal Majorana zero modes, ELC waves and soliton-fermion duality in non-Hermitian $sl(2)$ affine Toda coupled to fermions}
\end{center}

\vspace{0.1in}

\begin{center}

Harold Blas 

Instituto de F\'{\i}sica\\
Universidade Federal de Mato Grosso\\
Av. Fernando Correa, $N^{0}$ \, 2367\\
Bairro Boa Esperan\c ca, Cep 78060-900, Cuiab\'a - MT - Brazil \\

harold.achic@ufmt.br

\end{center}

\vspace{0.4in}

 
We study a non-Hermitian (NH) $sl(2)$ affine Toda model coupled to fermions through soliton theory techniques and the realizations of the pseudo-chiral and pseudo-Hermitian symmetries. The interplay of non-Hermiticity, integrability, nonlinearity, and topology significantly influence the formation and behavior of a continuum of bound state modes (CBM) and extended waves in the localized continuum (ELC). The non-Hermitian soliton-fermion duality, the complex scalar field topological charges and winding numbers in the spectral topology are uncovered. 
 The biorthogonal Majorana zero modes, dual to the NH Toda solitons with topological charges $\frac{2}{\pi} \arg{(z=\pm i)}=\pm 1$, appear at the complex-energy point gap and are pinned at zero energy. The zero eigenvalue $\l (z = \pm i)=0$, besides being a zero mode, plays the role of exceptional points (EPs), and each EP separates a real eigenvalue ${\cal A}$-symmetric and ${\cal A}$-symmetry broken regimes for an antilinear symmetry ${\cal A}\in \{ {\cal P}{\cal T}, \g_5{\cal P}{\cal T}\}$. Our findings improve the understanding of exotic quantum states, but also paves the way for future research in harnessing non-Hermitian phenomena for topological quantum computation, as well as the exploration of integrability and NH solitons in the theory of topological phases of matter.  
\end{titlepage}

\section{Introduction}

A common assumption underlying the physics of isolated systems is the Hermiticity property of the Hamiltonian. However, non-Hermitian (NH) Hamiltonians have attracted considerable interest in recent years since they are quite ubiquitous in nature and  remarkable theoretical and experimental results have been uncovered (see e.g. \cite{ashida1, bergholtz, kawabata}). They have been studied in three different approaches, a generic NH Hamiltonians with complex
eigenvalues \cite{ivanov, rivero}, (anti-)pseudo-Hermitian Hamiltonians  \cite{ali1, ali2, ali3}, and ${\cal P}{\cal T}$ symmetric Hamiltonians with real eigenvalues \cite{bender0, bender, bender1}. For related quasi/pseudo Hermitian systems see the classical paper \cite{scholtz}. Theoretical aspects and some experimental validations of nonlinear wave dynamics in ${\cal P}{\cal T}$ symmetric systems have been reviewed in \cite{konotop}. 

In physical systems with Hermitian and NH Hamiltonians a considerable amount of work on topological phases has emerged as a rapidly growing research area relevant to a variety of fields. An important concept in topological physics is the bulk-boundary correspondence in topological phases, such that the topological invariant related to the
bulk states counts the number of gapless boundary states. So, topological phenomena can be related to boundary phenomena. For a review on non-Hermitian lattice systems in condensed matter physics, focusing on the boundary phenomena, see e.g. \cite{okuma}. 
Originally found in condensed matter physics, robust edge states are currently found in practically all wave dynamics originating from a variety of physical models. For example, remarkable similarities have recently been found between the shallow water equations in their linearized form and topological phases of matter \cite{delplace}. A duality relationship between the bound state waves and the topological solitons has been uncovered for non-linear shallow water waves \cite{arxiv0}. 

The study of  topological edge modes has initially been related to Hermitian linear systems concepts, such as orthogonal eigenvectors and real energy spectra. Actually, many physical systems exhibiting these phenomena naturally present a non-linear behavior. This motivated recently  a growing interest in the confluence of nonlinearity and topology with applications to a wide range of phenomena (see e.g. \cite{jezequel} and references therein). From the quantum field theory perspective, it has been shown the presence of edge and bulk solitons related to the problem of fractionally charged solitons and electron fractionalization in $(1+1)$-dimensional systems \cite{jackiw, goldstone, mackenzie}. In various ways the concepts involved in the $(1+1)$-dimensional systems can be extended to higher dimensions.  

In this context, the exploration of the interplay of NH solitons and fermionic systems emerges as an intriguing subject in theoretical and mathematical physics. In this paper, we delve into the interplay between non-Hermiticity, nonlinearity, integrability and topology in the framework of a non-Hermitian $sl(2)$ Toda soliton coupled to Dirac fermions; so, extending the Hermitian soliton-fermion duality reported in \cite{jhep22, npb1, npb2}. The exploration of non-Hermitian systems has emerged as a fascinating and rapidly growing field, yielding intriguing insights into the fundamental nature of quantum mechanics. Non-Hermitian matrices exhibit unconventional properties  compared
with Hermitian ones: eigenstates are, in general,
nonorthogonal, instead it has been introduced the concept of biorthogonality and a complex-energy spectrum can
possess exceptional points (EPs) giving rise to spectral topology. Non-Hermiticity drastically changes the
properties of topological boundary states. Non-Hermiticity amplifies the edge states, which enables novel topologically protected states. Recently a modified biorthogonal bulk-boundary correspondence has been established that works in the presence of non-Hermiticity \cite{kunst}.

A non-Hermitian $sl(2)$ Toda model coupled to matter exhibits an extraordinary confluence of non-Hermiticity, nonlinearity, integrability and topological solitons. Central to our investigations are the concepts of pseudo-chiral and pseudo-Hermitian symmetries, which shed lights to disentangle the intricate interplay between the mentioned properties. Remarkably, a relationship between the complex energy, number wave and mass parameters in the NH soliton sector of the model can be defined on the whole complex plane through a uniformization parameter $z$, which allows us to discover that while Hermitian bound state/soliton states lie on the unit circle $|z|=1$, the non-Hermitian generalizations of that correspondence can be established on the continuous region $|z| \neq 1$ of the complex plane. In this scenario we show the emergence of biorthogonal Majorana zero modes as dual to non-Hermitian topological solitons with topological charges $\pm 1$, which manifest at the complex-energy point gap and pinned at zero energy. These zero modes, we find, hold the key to unlocking the topologically protected bulk-boundary correspondence from the integrable field theory perspective. The zero eigenvalue $\l (z = \pm i)=0$, besides being a zero mode, plays the role of an exceptional point (EP), since it separates a ${\cal P}{\cal T}$-symmetric regime with real eigenvalues
from the ${\cal P}{\cal T}$-broken regime with complex eigenvalues by varying one or more of the parameters of the model. 
 
An essential distinction between Hermitian and non-Hermitian systems is the degrees of freedom that we
have access to; nonunitary operations forbidden in Hermitian systems can be performed in non-Hermitian
systems. In other words, a change in the spectrum from real to complex increases the number of parameters that
describe the system. Since topology crucially depends  on the underlying manifold, non-Hermiticity is expected
to alter the topological classification of the physical systems such as insulators and
superconductors. A Hermitian $sl(2)$ Toda-type model coupled to matter has been found in a variety of models, such as 
the continuum limit of $s-$wave superconductor \cite{udupa}, effective theory describing Majorana bound states \cite{ardone}, a model for high $T_c$ superconductivity \cite{ferraz}, low-energy effective Lagrangian in QCD$_2$ \cite{prd1}, etc; so, our study not only becomes  of academic interest, but physically motivated. As pointed out above, pseudo-Hermiticity is a constraint
unique to non-Hermitian systems and may provide  novel
topological features.  In view of the rapid theoretical and experimental
advances in non-Hermitian physics, there has been a great
interest and an urgent need for comprehensive topological
classifications that provide reference points for experiments
and predict novel non-Hermitian topological phases.

Our exploration has led us to unveil remarkable phenomena: the existence of continuum spectra of bound state modes (CBM) and extended waves embedded within this localized continuum (ELC). These phenomena are profoundly shaped by the interplay of four key characteristics: non-Hermiticity, integrability, nonlinearity, and topology. The continuum of biorthogonal bound states emerge coupled to a continuous family of topological complex solitons, such that their topological charges depend on the energy eigenvalues. This type of  continuum of bound states in a non-Hermitian Hamiltonian have recently been uncovered also in \cite{wang2}. A Hermitian $sl(2)$ Toda system coupled to fermion, which has recently been presented in \cite{jhep22}, the in-gap and the bound states in the continuum (BIC) have quantized energies, whereas the free states form a continuum.  On the other hand, the NH extended wave states embedded into the NH localized continuum can be regarded as some kind of inverted BIC states due to the non-Hermicity of the model. Recently, these novel types of states in a NH system have been dubbed as extended states in a localized continuum (ELC) \cite{wwang, amir}.  

So, the NH models exhibit behaviors that challenge conventional understanding. Within a NH $sl(2)$ Toda theory coupled to matter integrable field theory we clarify the concept of the non-Hermitian soliton-fermion duality, revealing a profound connection between these entities which survive non-Hermiticity effects, and shed lights on the appearance of a second topological charge related to the   imaginary component of the complex scalar field. This charge arises purely as the result of the interplay between non-Hermiticity, topology, and nonlinear effects. We believe that beyond the pure theoretical inquiry, our discoveries might have important implications for the realm of practical quantum technologies. The harnessing of non-Hermitian phenomena in the context of topological quantum computation promise new frontiers in the development of quantum computing paradigms. Furthermore, our work establishes a bridge between quantum field theory, condensed matter physics and integrable systems, inviting further exploration at the crossroads of these three fields.
 
The paper is organized as follows. In the next section we present a integrable NH Toda model coupled to fermions and the relevant eigenvalue equations for the right and left eigenvectors. In the  section \ref{sec:spc11} the pseudo-chirality symmetry is discussed, as well as the concept of biorthogonality of the left and right eigenvectors.  The section \ref{sec:psh1} discusses the anti-pseudo-Hermiticity symmetry. Some theorems and properties of pseudo-Hermitian Hamiltonians are discussed in the presence of ${\cal P}$,  ${\cal T}$ and ${\cal P}{\cal T}$ transformations. In section \ref{sec:topo} we uncover the topological complex solitons and the biorthogonal spinor bound states through the tau function formalism. In section \ref{sec:real} the complex energy manifold is defined, identifying  the regions of real and complex eigenvalues, as well as the EPs of eigenvector degeneracies. The ${\cal P}{\cal T}$ and $\g_5{\cal P}{\cal T}$ symmetric Hamiltonians and their real eigenvalue regimes are identified. Moreover, the spectral topology and the associated winding number are discussed.  In section \ref{sec:class} a classification of these solutions and their symmetries are discussed. The continuum of bound states/solitons and the extended waves embedded into this localized continuum are presented. The  biorthogonality relationships between the left and right eigenvectors  are computed. In section \ref{sec:Majo1} the biorthogonal Majorana zero mode-topological soliton duality is studied by providing a mapping between the spinor and the scalar fields. The ${\cal P}{\cal T}$ symmetric region of this duality is identified  in parameter space. Section \ref{sec:discussion} presents the conclusions and discussions. In the appendix \ref{app:eqs11} we present the static equations of motion in components.

\section{The model}
\label{sec:model}
 
There is a variety of ways to introduce non-Hermiticity in the integrable models. Earlier versions of non-Hermitian affine Toda theories consider 
 a purely imaginary coupling constant which provide soliton solutions with real classical masses \cite{hollowood}. Some recent works introduce the 
non-Hermiticity by making the field of the model a complex field, as in the complex KdV,
mKdV and sine-Gordon equations which posses PT-symmetric soliton solutions with real energies \cite{fring}. Likewise, non-Hermiticity has been induced by many ways in field theory models, partially motivated by the search for 
another viewpoint to understand the topological properties of non-Hermitian systems. In particular, non-Hermiticity in the Dirac-type equations has been
 introduced through Lorentz symmetry violation and the generalization of some parameters from real to complex, e.g. the complex mass parameter \cite{ge, oh}. A generic one-dimensional Dirac model with both real and imaginary masses has been considered in optics \cite{moca} and the authors argued that those results can be tested experimentally.

Here we consider two  fermion fields chirally coupled to a complex scalar field defined by the following non-Hermitian Lagrangian\cite{matter}\footnote{Notation: $\pa^2=\pa^2_t -\pa^2_x $. $\g_0 = \(\begin{array}{cc} 0  & i\\
-i & 0\end{array}\)$, $\g_1 = \(\begin{array}{cc} 0  & -i\\
-i & 0\end{array}\)$, $\g_5 = \g_0 \g_1 = \(\begin{array}{cc} 1  & 0\\
0 & -1\end{array}\)$,  and  $\bar{\psi} = \widetilde{\psi}^{T} \g_0$;  $\bar{\chi} = \widetilde{\chi}^{T} \g_0$.} 
\br
\mathcal{L} &=&\;   \frac{1}{2}\partial_\mu \Phi \partial^\mu \Phi + i \bar{\psi} \gamma^\mu\partial_\mu\psi 
  - M(\bar{\psi}\psi \cos(\beta \Phi)+i \bar{\psi}\gamma_5\psi \sin(\beta \Phi)), \label{lag1}
\er  
with $\Phi$ being a complex field and $\widetilde{\psi}$ and $\psi$ two Dirac spinors. Below we will consider a complex mass parameter $M\in \IC$ and a real $\beta$ as the coupling strength between the fermions and the complex scalar. We define the model (\ref{lag1})  as a {\sl non-Hermitian $sl(2)$ affine Toda model coupled to Dirac fields} (NHTD). The general model (\ref{lag1}), with two Dirac spinors $\psi$ and $\widetilde{\psi}$ and a complex field $\Phi \in \IC$, belongs to a wide class of integrable soliton theories, the so-called  $\hat{sl}(2)$ affine Toda systems coupled to matter fields \cite{matter}. In the Hermitian case, i.e.  $\Phi \in \IR$ and $\widetilde{\psi} = \psi^{\star}$ with $M \in \IR$, some properties such as the soliton solutions, the soliton-fermion duality, as well as its symplectic structures were discussed in \cite{npb1, npb2, aop1}. We will consider the non-Hermitian model (\ref{lag1}) and show that the dynamics of the  system exhibits a continuum of  kink type states and its associated biorthogonal spinor bound states (CBM). In addition, it will be shown the appearance of extended wave states (ELC) embedded into the continuum of bound states. In this context we will realize the various symmetry properties of the model such as the pseudo-chirality and its associated biorthogonal bound states, such that the generalized non-Hermitian particle-hole (NHPH) symmetry holds. 

The non-Hermiticity in the model (\ref{lag1}) can be traced to the presence of at least one of the next conditions.  First, $\Phi \neq \Phi^{\star}$, where $\star$ stands for complex conjugation. Second, $\widetilde{\psi} \neq \psi^{\star}$. Third, in clear contrast to the Hermitian case one has the complex mass parameter $M$. The presence of any of these conditions would make the Lagrangian non-Hermitian. As we will see below,  the interplay between these conditions gives rise to interesting new results related to non-Hermitian systems. Among the new results of the NHTD model (\ref{lag1}), beyond the known results in the literature for the Hermitian case, one has the emergence of new type of topological solitons associated to the imaginary component of $\Phi$ and their related spinor bound states, as well as the appearance of extended wave states (ELC) in the continuum of bound states supported by the complex scalar and spinor fields. So, one can argue that the novel solitons and extended waves associated to the imaginary sector $Im(\Phi)$ will be generated due to the interplay between the non-Hermiticity, topological properties and nonlinearity of the model, as compared to the Hermitian counterpart of the model \cite{jhep22} which possesses kink solutions of the sine-Gordon type and BIC states associated to real $\Phi$.

Regarding the physical motivation to introduce the complex parameter $M$ one can justify
in hindsight by some physical and mathematical concepts. As we will show below,  the points $w \equiv \frac{k}{M} = \pm 1$ in the $w$ complex plane will become the branch point singularities or exceptional points (EPs) of the multivalued complex energy function with the dispersion relation $\l(w) =\pm i M \sqrt{w^2-1}$, in the enlarged complex parameter space manifold with $k$ a complex wave number. The zero modes are defined by $\l =0$, and so,  in this case one has $k = \pm M$. Taking the complex number notation $z \equiv z_1+i z_2$, one notices that for wave solutions with the $x-$dependence of type $e^{k x}$ the real part of the wave number $k_1$ would be associated to the localized bound states and the imaginary part  $k_2$ to the extended wave states. The introduction of the complex $M$ will give rise to new phenomena, such as the appearance of the ELC states. In fact, the complex part $M_2$ will play an important role in this type of states, as we will show in section {\ref{sec:class2}} the EPs separate the ${\cal P} {\cal T}$-symmetric phase with real eigenvalues and  the spontaneously broken phase by varying the parameter $\frac{k_2}{M_2}$, when the remaining parameters are held fixed.

We will see below that some solutions for the spinors in the ${\cal P} {\cal T}$-symmetry broken regime, exhibit the dissipation/amplification factors of type $e^{\l_2 t}$. This behavior calls for an alternative interpretation of the physical origin of the non-Hermiticity in the realm of a topological non-Hermitian system without resorting to the gain/loss concept which makes sense in optics, for example following the recent alternative approach introduced by means of the skin effect of reflected waves \cite{franca}. We provide a physical interpretation of the non-Hermiticity as the effect of the kink on the spinor bound states through the appearance of phase and scale factors in the spinor bilinear amplitude as it evolves from $x= -\infty$ to $x= +\infty$ in the presence of the kink. For the special cases of the non-Hermitian model (\ref{lag1}) with anti-linear ${\cal P} {\cal T}$ and $\g_5{\cal P} {\cal T}$ symmetric Hamiltonians, respectively,  we will uncover the parameter subspaces related to the scalar and spinor solutions with real eigenvalues $\l \in \IR$ which come in pairs $\{\pm \l_1, \l_1 \in \IR\}$, such that the pseudo-chiral and pseudo-Hermitian symmetries hold. 

Related to the points discussed above one can define a topological charge associated to the scalar field as ${\cal Q}_{top} \equiv \frac{\b}{2\pi} (\Phi(x=+ \infty)-\Phi(x=-\infty))$. For certain field configurations the both real and imaginary components of $\Phi$ will develop non-vanishing boundary values, and so, this could imply a non-trivial complex charge ${\cal Q}$. This charge will be regarded as a complex topological charge. In fact, by introducing the so-called uniformization complex variable $z$ through $\l = - \frac{M}{2} (z + z^{-1})$, we will show below a profound connection between this charge and the topological properties of the complex manifold in parameter space expressed by the remarkable relationship ${\cal Q}_{top} = \pm \frac{2}{\pi i } \log{z}$.  So, one can argue that the appearance of this complex topological charge is due to the interplay between the non-Hermitian, topological and nonlinear properties of the model (\ref{lag1}). 

The Hermitian version of the model (\ref{lag1}) has been studied in the literature. In this case one has a real scalar field, the fermions satisfy the reality  condition $\widetilde{\psi} = \psi^{\star}$ and $M$ is a real mass parameter. In that case the model has been defined  by a real Lagrangian and associated Hermitian hamiltonian \cite{npb1, aop1, npb2, prd1, jhep22}. The Hermitian model has been shown to describe the low-energy effective Lagrangian of QCD$_2$ with one flavor and $N$ colors \cite{prd1} and the BCS coupling in spinless fermions in a two dimensional model of high T superconductivity in which the solitons play the role of the Cooper pairs \cite{ferraz}. Moreover, the model has earlier been studied as a model for fermion confinement in a chiral invariant theory \cite{chang} and the mechanism of fermion mass generation without spontaneously chiral symmetry breaking in two-dimensions \cite{witten}. Recently, through analytical and numerical methods the Hermitian version of the model (\ref{lag1}) plus a scalar potential has been shown to exhibit Majorana zero mode-soliton duality and in-gap and
BIC bound states, such that the Noether, topological and a novel
nonlocal charge densities satisfy a formula of the Atiyah-Patodi-Singer-type \cite{jhep22}.
 
The equations of motion  become
\br
\label{eq11}
\partial^2\Phi - M\beta\widetilde{\psi}\psi\sin(\beta \Phi) +i M \beta \widetilde{\psi}\gamma_5\psi \cos(\beta \Phi)  =0\\
i\gamma_\mu\partial^\mu\psi - M(\psi \cos(\beta \Phi)+i\gamma_5\psi \sin(\beta \Phi))=0\label{eqs11}\\
i\gamma_\mu\partial^\mu\widetilde{\psi}-M(\widetilde{\psi} \cos(\beta \Phi) - i\gamma_5\widetilde{\psi}\sin(\beta \Phi))=0\label{eqs22}
\er
Let us define
\br
\label{psi1}
\psi &\equiv& e^{-i\lambda t} \xi ,\,\,\,\,\,\,\,\,\,\,\,\,\,\,\,\,\widetilde{\psi} \equiv e^{i \lambda t} \eta\\
\xi &=&  
\(\begin{array}{c}
 \xi_{A} \\ 
\xi_{B}
\end{array}\), \,\,\,\,\,\,\,\, 
\eta = \(\begin{array}{c}
\eta_{A} \\ 
\eta_{B}
\end{array}\),
\er
with $\l$ being a complex parameter.
The equations (\ref{eqs11})-(\ref{eqs22}) can be written in matrix form as
\br
\label{H1}
H \xi &=& \l \, \xi,\\
\label{H11}
\widetilde{H} \eta  &=& \l\, \eta \er
where
\br \label{NH}
H = \(\begin{array}{cc} i \frac{d}{dx} & i M e^{-i \b \Phi}\\
 -i M e^{i \b \Phi} & -i \frac{d}{dx}  \end{array} \),\,\,\,\,\,\,\,  \widetilde{H} = \(\begin{array}{cc} -i \frac{d}{dx} & -i M e^{i \b \Phi}\\
 i M e^{-i \b \Phi} & i \frac{d}{dx}  \end{array} \).
\er 
Notice that the equations above define a pair of eigenvalue equations for the non-Hermitian operators  $H$ and $\widetilde{H}$ with right eigenvectors $\xi$ and $\eta$, respectively.  Recently, the Hermitian version of the model (\ref{H1}) has been considered as the continuum limit of a one-dimensional system of spin-orbit coupled Dirac Hamiltonian and a $s-$wave superconducting pairing, such that the $s-$wave pairing is represented by $-i M e^{i \b \Phi}\, (\Phi \in \IR)$  \cite{udupa}. Moreover, the system (\ref{H1}) has been considered in \cite{ardone} as an effective theory describing Majorana bound states such that the function $i M e^{-i \b \Phi}$ is related to the superconducting order parameter.
 
One can show that the NH operators $H$ and $\widetilde{H}$ are related by
\br
\label{hh1}
\G H \G =\widetilde{H},\,\,\,\,\,\, \G = i \g_1.
\er
Due to the last relationship (\ref{hh1}),  given an eigenvector  $\eta$ of $\widetilde{H}$ with eingenvalue $\l$ one can get an eigenvector $\G \eta$ of $H$ with the same eigenvalue $\l$. In fact, from (\ref{H11})  and taking into account (\ref{hh1}) one can get  
\br
\G H \G \eta = \l  \eta\,\,\, \rightarrow \,\,\,   H (\G \eta)  =  \l  (\G \eta). 
\label{xieta1}
\er
So, comparing (\ref{xieta1}) with (\ref{H1}) one can write
\br
\label{AB1}
\xi = \G \eta\,\,\,\,\, \Rightarrow    \,\,\,\,\, \xi_A = \eta_B,\,\,\,\,\,\,\xi_B = \eta_A.
\er 

From (\ref{eq11}) and taking into account the Ansatz (\ref{psi1})  one can get a static version of the complex scalar field equation of motion
\br
\label{sec1}
 - \pa^2_x \Phi  +  M \b [ e^{i \b \Phi}\,  \eta_{B} \xi_A + e^{-i \b \Phi}\,  \eta_{A} \xi_B]  =0.
\er
 
We will examine the static solutions satisfying the relationship
\br
\label{equi1}
\pa_x \Phi &=& -\frac{\b}{2} [ \eta_{A} \xi_A +  \eta_{B} \xi_B],\\
&=&-\frac{\b}{2} [ \eta_{3} \xi_3 -  \eta_{4} \xi_4 + \eta_{1} \xi_1 -  \eta_{2} \xi_2 + i (\eta_{3} \xi_4 +  \eta_{4} \xi_3 + \eta_{1} \xi_2 +  \eta_{2} \xi_1)], \label{equi11}
\er
where in the last equation we have written the real and imaginary parts of the  bilinear current terms using the component of $\xi$ and $\eta$ defined in (\ref{xi11}) and (\ref{eta11}).

Remarkably, the first order equation (\ref{equi1})  together with the first order equations of motion for the Dirac spinors (\ref{H1})-(\ref{H11}) imply the second order equation for $\Phi$ in (\ref{sec1}). Therefore, a given set of solutions of the system of static first order differential equations (\ref{H1})-(\ref{H11})  and  (\ref{equi1}) will also be solutions of the second order equation for $\Phi$. The equation (\ref{equi1}) suggests a certain kind of equivalence relationship between the $x-$derivative of the scalar and the fermion bilinear in the NH model. In fact, this is a NH analog of the equivalence between the $U(1)$ Noether and topological soliton currents considered in the literature in the Hermitian case of the model \cite{jhep22}. In both condensed matter (e.g. Bogoliubov-de Gennes (BdG) and gap equations ) and high energy physics (e.g. Gross-Neveu (GN) and Nambu-Jona-Lasinio models) there have been studied analogous relationships to (\ref{equi1}) by self-consistency requirements with (\ref{H1})-(\ref{H11}) which amounts to write e.g. $i M e^{i \b \Phi} \sim [ \xi_{A}^\star \xi_B +  \eta_{A}^\star \eta_B]$ \cite{nitta}. Actually, for a real scalar and one Dirac fermion these relationships resemble to the bosonization rules which map some fermion bilinears to a topological current and exponentials of the scalar \cite{stone}.  

One can notice that the imaginary component of the bilinear current in (\ref{equi11}) vanishes provided that
\br
\label{etaxi}
\eta = \pm \xi^{\star}\,\, \Rightarrow \,\, \Big\{\begin{array}{c} Case\, I: \, \eta_{3}=\xi_3,\,\,\, \eta_{4}=-\xi_4,\,\, \eta_{1} = \xi_1,\,\,  \eta_{2} =- \xi_2,\\
Case\, II:\,
\eta_{3}=-\xi_3,\,\,\, \eta_{4}=\xi_4,\,\, \eta_{1} = -\xi_1,\,\,  \eta_{2} = \xi_2.\end{array}\Big\}
\er
Then, for the cases (\ref{etaxi}) the scalar field must be real, i.e.  $\Re{(\Phi)} = \Phi_1\, (\Im{(\Phi)}=0)$. So, this implies $\widetilde{H} = H^{\star}$ (for real $M$)  and one recovers a Hermitian model with 
\br
H^{\dagger} = H,\,\,\,\,\, \widetilde{H}^{\dagger} = \widetilde{H},
\er
and real eigenvalues $\l\in \IR$. In the both cases (\ref{etaxi}) the imaginary part of the  second order equation (\ref{sec1})  for the scalar field identically vanishes for $\Phi_2 = 0$ and $M_2=0$. The bilinear terms of the spinor components are presented in (\ref{eq10}). 

However, for $M_2 \neq 0$, $\eta \neq \pm \xi^{\star}$ and $\Phi_2 =0$ the eq. (\ref{equi11}) provides the general constraint $(\eta_{3} \xi_4 +  \eta_{4} \xi_3 + \eta_{1} \xi_2 +  \eta_{2} \xi_1)=0$ and  the second order eq.  (\ref{eq10}) will provide an additional constraint (\ref{const12}).

In the case of NH Hamiltonians it is convenient to define the so-called left eigenvectors, since in general they are not the same as the relevant  Hermitian conjugated  right eingenvectors. So, the left eigenvectors  $\bar{\xi}_j^T$ and $\bar{\eta}_j^T$ are defined as
\br
\label{lefteigen1}
\bar{\xi}_j^T H = \l_j  \,\bar{\xi}^T_j,\,\,\,\,\,\bar{\eta}_j^T \widetilde{H} = \l_j  \,\bar{\eta}^T_j
\er
Next, we will consider certain solutions satisfying 
\br
\label{equiv22}
\pa_x \bar{\Phi} = -\frac{\b}{2} [ \bar{\eta}_{A} \bar{\xi}_A +  \bar{\eta}_{B} \bar{\xi}_B]  ,
\er
where we have introduced a complex scalar field $\bar{\Phi}$, and $\bar{\eta}^T \equiv (\bar{\eta}_{A}\, ,\, \bar{\eta}_B)\,$  and $\bar{\xi}^T \equiv (\bar{\xi}_{A}\, ,\,\bar{\xi}_B).$ In general one has that the left and right eigenvectors are different, i.e.
\br
\label{lrdiff}
\bar{\eta} \neq \eta ,\,\,\, \bar{\xi} \neq \xi,
\er
and, so, the scalar field  $\bar{\Phi}$ can be considered to be in general different from $\Phi$ in (\ref{equi1}). However, we will see below that the left and right eigenvectors satisfy some symmetry relationships, which imply a relationship between the scalars  $\Phi$ and $\bar{\Phi}$ provided that the equation (\ref{equiv22}) holds. So, the second order differential equation for the scalar $\bar{\Phi}$ becomes 
\br
\label{sec11}
  \pa^2_x \bar{\Phi}  +  M \b [ e^{i \b \bar{\Phi}}\,  \bar{\eta}_{A} \bar{\xi}_B + e^{-i \b \bar{\Phi}}\,  \bar{\eta}_{B} \bar{\xi}_A]  =0.
\er
Notice that  the equations (\ref{sec1}) and (\ref{sec11}) have different origins since the first one has been derived from the Lagrangian, whereas the second one is a consequence of the equivalence relationship (\ref{equiv22}) which has been assumed to hold between a topological current associated to $\bar{\Phi}$ and the left-eigenvector current density, respectively.  However, we will see below that for  $\l$ and $-\l$ eigenvalues the relevant right and left eigenvectors will be related by symmetry arguments, and so,  those equations will be related by the so-called pseudo-chirality symmetry. So, we are establishing some equivalences between topological currents and biorthogonal fermion currents represented by the correspondence relationships (\ref{equi1}) and (\ref{equiv22}), respectively.  

Let us discuss the degrees of freedom associated to the non-Hermiticity of the model. So, one can write the identities
\br
\label{commu}
[H\,,\, H^{\dagger}] &=& \(\begin{array}{cc} S & R \\ 
 R^{\star}  & - S \end{array} \) \,\,\,\,\, \mbox{and} \,\,\,\,\,\, \,\,\,\,\, [\widetilde{H}\, ,\widetilde{H}^{\dagger}] = \(\begin{array}{cc} -S & R^{\star} \\ 
 R  & S \end{array} \), \er
with
\br
\label{commu1}
S &=& 2|M|^2 \sinh{(2 \b \Phi_2)},\,\,\,\,\,\,\,\,\,\, \Phi \equiv \Phi_1 + i \Phi_2, \\
R&=& -2 \b e^{-i \b \Phi_1} [M_1 \cosh{(\b \Phi_2)} + i M_2 \sinh{(\b \Phi_2)}  ] \, \pa_x \Phi_2. \label{commu2}
\er
So, the Hamiltonians are nonnormal, i.e. $[H, H^{\dagger}] \neq 0$ ($[\widetilde{H}\, ,\widetilde{H}^{\dagger}] \neq 0$), then one has to distinguish right eigenstates  from the corresponding left eigenstates as stated in (\ref{lrdiff}). Moreover, the appearance of the exceptional points (EPs) in the eigenvalue spectra requires, aside from non-Hermiticity, the Hamiltonians to be nonnormal \cite{jan1}. 

It is clear that for vanishing $\Phi_2 =0$ one has $R=S=0$; and so, the identities  (\ref{commu}) imply
\br
\label{comm1}
[H\, , \, H^{\dagger}] =0,\,\,\,\,\, \mbox{and}\,\,\,\,\,\,[\widetilde{H}\, ,\, \widetilde{H}^{\dagger}]=0.
\er 
So, in this case the Hamiltonians become normal. Therefore, $H$ ($\widetilde{H}$) and $H^{\dagger}$ ($\widetilde{H}^{\dagger}$) have the same
eigenvectors and the eigenvalues of $H^{\dagger}$ ($\widetilde{H}^{\dagger}$) are the complex conjugates of the eigenvalues of
$H$ ($\widetilde{H}$). If in addition one imposes $M \in \IR$ in (\ref{NH}) one has that  $H^{\dagger} = H$ ($\widetilde{H}^{\dagger}=\widetilde{H})$ and the relevant eigenvalues will be real numbers. Moreover, by setting $\Phi_2 =0$ into the imaginary part of the equations (\ref{equi1})-(\ref{equi11}) one can get  $\xi_A \xi_B - \xi_A^{\star} \xi_B^{\star} =0$. Therefore, in order to have a Hermitian Hamiltonian one must have
\br
\label{hermicond}
\Phi_2 =0,\,\,\,\,\,\,\,\,\,\, \mbox{Im}{(\xi_A \xi_B)} =0,\,\,\,\,\,\,\,\,\, M\in \IR.
\er   
Below we will consider a physical origin of the non-Hermiticity by studying the spinor field configurations associated to the bilinear $(\xi_A \xi_B)$ ($\mbox{Im}{(\xi_A \xi_B)} \neq 0$) in the presence of the field $\Phi_2\neq 0$. 

\section{Pseudo-chirality and biorthogonality of the eigenvectors}
\label{sec:spc11}

For a generic complex scalar field $\Phi$ the operators $H$ and $\widetilde{H}$ are non-Hermitian with a spectra of complex eingenvalues $\l$. Recently, there have been various attempts to realize
particle-hole symmetry in non-Hermitian (NH) systems.  However, due to the fact that these effective particles
and holes have complex energies in a NH system, they do not
need to fulfill the Fermi-Dirac statistics. Non-Hermitian Hamiltonian matrices exhibit more types of symmetries than the Hermitian counterparts \cite{okuma}. So, in order to examine  certain extensions of the Noether's theorem, it has been proposed distinct ways to define the inner product in NH systems. An inner product, without the complex conjugation, has been proposed in \cite{rivero} which focuses on asymmetric non-Hermitian Hamiltonians. So, let us write the  next inner product 
\br
\label{nhinn}
(\xi_i|\xi_j) \equiv \xi_i^T \xi_j. 
\er
With this inner product and using an alternative form of the Ehrenfest's theorem for the time evolution of $(\psi|{\cal G}|\psi)$ one can show that the operator ${\cal G}$ is a constant of motion provided that $H$ satisfies \cite{rivero}
\br
\label{spc1}
 {\cal G} H  {\cal G}^{-1} = - H^{T},
\er
where $H^T$ stands for matrix transpose. For the NH Hamiltonian $H$ in (\ref{NH})  one has that 
\br
\label{G1}
{\cal G} = -i \g_0.
\er 
In \cite{rivero} the symmetry (\ref{spc1}) has been defined as
pseudo-chirality, and it guaranties, as we will see below, a
symmetric spectrum about the origin of the complex energy
plane, i.e. for each $\l$ there exists $-\l$,  similar to the consequence of  particle-hole symmetry in the hermitian counterpart of the model (\ref{lag1}) studied recently in \cite{jhep22}. 

Below we will follow the biorthogonal approach to describe the orthogonality properties of eigenvectors and eigenstates \cite{bergholtz, rivero}. So, let us define the right eigenvectors $\xi_j$ and the relevant left eigenvectors  $\bar{\xi}_j^T$  as
\br
\label{eig1}
H \xi_i= \l_i  \, \xi_i, \,\,\,\,\,\,\,\,\,\,\bar{\xi}_i^T H = \l_i  \,\bar{\xi}^T_i. 
\er 
The eigenvectors $\xi_i$ and $\bar{\xi}_i$  are different in general but share the same eigenvalue $\l_i$. Moreover, they  satisfy the biorthogonality relationship \cite{rivero}
\br
\label{bior}
(\bar{\xi}_j|\xi_i) \equiv  \bar{\xi}_j^T \xi_i = \d_{ji},
\er
which holds away from a zero mode ($\l_j =0$). In fact, considering the both eigenvalue equations in (\ref{eig1}) for $\xi_i$ and $\bar{\xi}_j^T$, respectively,   one can write the identity
\br
\bar{\xi}_j^T H \xi_i = \l_i  \,\bar{\xi}_j^T \xi_i,\,\,\,\,\,\bar{\xi}_j^T H \xi_i = \l_j  \,\bar{\xi}^T_j \xi_i \,\,\,\,\, \Rightarrow\,\,\,\,\, 0 = ( \l_i -  \l_j) \, \bar{\xi}_j^T \xi_i.
\er 
Then, from this equation one can see that the biorthogonality relationship (\ref{bior}) is satisfied for $\l_i  \neq \l_j$. The normalization $(\bar{\xi}_i|\xi_i) =1\, (\l_i \neq 0)$ will be considered in the sections below in the context of the soliton type solutions of the model.   

Next, for the Hamiltonian $H$ in (\ref{NH}) which possesses the property (\ref{spc1}), one can show the following property  
\br
\label{HT}
H^T [{\cal G} \xi_{\l_j}]  =  - {\cal G} H \xi_{\l_j}  = - \l_j [{\cal G} \xi_{\l_j}].
\er 
Since the second eq. in  (\ref{eig1}) can be written as $H^T \bar{\xi}_{-\l_j} = -\l_j \bar{\xi}_{-\l_j}$, then from (\ref{HT}) one can write 
\br
\label{lrxi1}
\bar{\xi}_{-\l_j} = {\cal G} \xi_{\l_j}.
\er
This relation suggests that for each right eigenstate $\xi_{\l_j}$ with eigenvalue $\l_j$, one can find a left eigenstate $\bar{\xi}_{-\l_j}$ for the relevant eigenvalue $-\l_j$, such that in components one has the next relationships
\br
\label{ABl}
\bar{\xi}_{-\l_j A} =\xi_{\l_j B},\,\, \bar{\xi}_{-\l_j B} = - \xi_{\l_j A}.
\er
In terms of the Dirac field $\psi_{\l_j}$ defined in (\ref{psi1}), and  defining $ \bar{\psi}_{-\l_j} \equiv  e^{-i \l_j t} \bar{\xi}_{-\l_j}$, one can write 
\br
\label{lpm}
\bar{\psi}_{-\l_j} =  {\cal G}  \psi_{\l_j}.
\er
So, the operator ${\cal G}$ maps a right Dirac eigenstate with eigenvalue $\l_j$ to one of the left eigenstates with corresponding eigenvalue $-\l_j$. This relationship exhibits the so-called NH particle-hole symmetry of the system and then one expects that the `particles' and `holes' will have  identical spectra, since the spectra of $H$ is symmetric about the origin of the complex eigenvalue plane. So, for certain zero modes one can formulate the NH analog of the Majorana fermions for a system of this kind since there exists a biorthogonal `particle-hole' symmetry. 

A Majorana fermion in a Hermitian model is defined by considering the particle and hole with the same energy as
a single particle. Here we generalize the Majorana zero modes to NH model which exhibits topological solitons, and show a general but quite different properties from their Hermitian counterparts. Then, following analogous reasoning, from the relationship (\ref{lpm}) and for zero-mode solutions ($\l_j=0$), one can write
\br
\label{NHmajorana}
\bar{\psi}_{0} =  {\cal G}  \psi_{0}.
\er
The biorthogonal states obeying the condition (\ref{NHmajorana}) will be regarded as NH Majorana zero modes. These states are well known to lead to interesting phenomena in condensed matter and we will search for them as  bound states for the biorthogonal states.  

Notice that the biorthogonal relationship (\ref{bior}) is different from the NH inner product (\ref{nhinn}). For example, the product (\ref{bior}) vanishes at the zero mode $\l_j=0$, since taking into account (\ref{NHmajorana}), one has $\bar{\xi}_{0A} =\xi_{0B},\,\, \bar{\xi}_{0B} = - \xi_{0A}$, which implies
\br
\nonumber
(\bar{\xi}_0|\xi_0) & \equiv & \bar{\xi}_0^T \xi_0 \\
&=&0.\label{xi00}
\er 
On the other hand, the product (\ref{nhinn}) does not vanish at the zero mode $\l_j=0$, i.e. it becomes
\br
\label{nhinn0}
(\xi_0|\xi_0) &\equiv& \xi_0^T \xi_0\\
&=& (\xi_{0A})^2  + (\xi_{0B})^2, \label{nhinn00}
\er
which is a complex number in general. This is in contradistinction to the Hermitian case in which an inner product has been defined to provide a real value for the Hermitian Majorana zero modes \cite{jhep22}. 

Similar constructions as above can be performed for the second Dirac field $\widetilde{\psi}$. Since the pseudochirality property (\ref{spc1}) also holds for the second operator $\widetilde{H}$, i.e.
\br
\label{spc1c}
 {\cal G} \widetilde{H} {\cal G}^{-1} = - \widetilde{H}^{T},
\er
then one can follow throughout all the discussions above for the right eigenvectors $\eta$ and the relevant left eigenvectors $\bar{\eta}^T$
\br
\label{eig22}
\widetilde{H} \eta_i= \l_i  \, \eta_i, \,\,\,\,\,\,\,\,\,\,\bar{\eta}_i^T \widetilde{H} = \l_i  \,\bar{\eta}^T_i. 
\er
So one can show
\br
\label{etaG1}
\bar{\eta}_{-\l_j} = {\cal G} \eta_{\l_j}.
\er
So, for each right eigenstate $\eta_{\l_j}$, one can find a left eigenstate $\bar{\eta}_{-\l_j}$, which in components become
\br
\label{lretas1}
\bar{\eta}_{-\l_j A} =\eta_{\l_j B},\,\, \bar{\eta}_{-\l_j B} = - \eta_{\l_j A}.
\er
For the Dirac field $\widetilde{\psi}$ as defined in (\ref{psi1}) ($\widetilde{\psi} = e^{i \l t} \eta$) and  taking into account  (\ref{etaG1}) one can written  
\br
\label{lpm1}
\overline{\widetilde{\psi}}_{-\l_j} =  {\cal G}  \widetilde{\psi}_{\l_j}.
\er
So, one can argue that this relationship also exhibits the NH particle-hole symmetry for the second Dirac field $\widetilde{\psi}$ and then allows us to discuss the Majorana zero modes.  Following analogous reasoning from the relationship (\ref{lpm1}) and for zero-mode solutions ($\l_j=0$) one can write
\br
\label{NHmajorana1}
\overline{\widetilde{\psi}}_{0} =  {\cal G}  \widetilde{\psi}_{0}.
\er
The biorthogonal states obeying the condition (\ref{NHmajorana1}) will be regarded as the second NH Majorana zero mode of the system. The analog of the product (\ref{bior}) becomes
\br
\label{bior2}
(\bar{\eta}_i|\eta_j) = \bar{\eta}_i^T \eta_j = \d_{ij}.
\er
This product vanishes at the zero mode point $\l_j=0$, since taking into account (\ref{NHmajorana1}), one has $\bar{\eta}_{0A} =\eta_{0B},\,\, \bar{\eta}_{0B} = - \eta_{0A}$, which implies
\br
\nonumber
(\bar{\eta}_0|\eta_0) & \equiv & \bar{\eta}_0^T \eta_0 \\
&=&0.  \label{eta00}
\er  
At the zero mode one has $\l=0$, at this point the bi-orthogonal conditions (\ref{bior}) and (\ref{bior2})  break down, and the set of states $\{\bar{\xi}_0$, $\xi_0\}$ and $\{\bar{\eta}_0$, $\eta_0\}$  coalesce, i.e. they become orthogonal, and the Hamiltonian becomes non-diagonalizable, see e.g. \cite{sato, das}. The results (\ref{xi00}) and (\ref{eta00}) will be verified below for spinor bound state eigenvectors. Notice that at the point $\l =0$ we will have an energy degeneracy for the both types of eigenvector spaces, left and right. In fact, the set of right and left eigenvectors $\{\xi_0\,,\, \eta_0\}$ and  $\{\bar{\xi}_0\,,\, \bar{\eta}_0\}$, respectively, correspond to the vanishing eigenvalue  $\l=0$. 

At this point, let us examine the relationships between the currents equivalence equations  (\ref{equi1}) and (\ref{equiv22}) in view of the pseudo-chirality symmetry discussed above. So, taking into account the relationships (\ref{ABl}) and (\ref{lretas1}) one can write 
\br
\label{lrcurr11}
\bar{\eta}_{-\l,A} \,\bar{\xi}_{-\l,A} +  \bar{\eta}_{-\l,B} \, \bar{\xi}_{-\l,B} =  \eta_{\l,A} \, \xi_{\l,A} +  \eta_{\l,B} \, \xi_{\l,B}.
\er  
Then, due to the identity (\ref{lrcurr11}) one can argue that the both scalars $\Phi$ and $\bar{\Phi}$ satisfy the identity
\br
\label{PPhi}
\bar{\Phi} = \Phi, 
\er
provided that they satisfy the equations (\ref{equi1}) and (\ref{equiv22}) for the right-eigenvectors with $\l$ eigenvalue and the left-eigenvectors with $-\l$ eigenvalue, respectively. Likewise, the second order differential equations (\ref{sec1}) and (\ref{sec11}) can be mapped to each other using the  pseudo-chiral symmetry relationships (\ref{ABl}) and (\ref{lretas1})  of the right and left eigenvectors for eigenvalues $\l$ and $-\l$, respectively, and the relationship (\ref{PPhi}) for the scalars which holds for field configurations satisfying the currents equivalence equations (\ref{equi1}) and (\ref{equiv22}). 

\section{Anti-pseudo-Hermiticity,  ${\cal P}$,  ${\cal T}$ and ${\cal P}{\cal T}$  transformations} 
\label{sec:psh1}

Breaking the Hermiticity property leads to a larger number of symmetries since the non-Hermitian Hamiltonian  (\ref{NH}) exhibits the properties  $H \neq H^{\dagger} $ and $H^{\star} \neq H^{T}$. In that scenario the time-reversal (TR) and particle-hole symmetries constitute two inequivalent representations, i.e. depending on whether
the symmetry considers conjugation or transposition of $H$, respectively (see e.g. \cite{wang1}). Notice that the symmetry property considering the transposition of $H$ has already been discussed as the pseudo-chiral  symmetry in the last section.  Next, we consider the other inequivalent representation by taking the conjugation of $H$. 

\subsection{Anti-pseudo-Hermiticity}

Let us consider a TR operator ${\cal T }$ such that ${\cal T } z {\cal T}^{-1} = z^{\star}$ and ${\cal T }^2 = \pm 1$ and so, ${\cal T }$ is an anti-linear operator. Next, apply this operator on the both hand sides of the pseudo-chirality relationship  (\ref{spc1}) for a generic complex scalar fied  $\Phi$ and complex parameter $M$. So, one can write 
\br
\label{psc2}
\s_{-} H \s_{-}^{-1}&=& - H^{\dagger},\\
\s_{-} & \equiv &  {\cal G }  {\cal T}, \label{psc2i}
\er
where $ {\cal G }$ is given in (\ref{G1}). Notice that $\s_{-}$ is an antilinear operator. Following the notations and conventions in the current literature one has that the equations (\ref{psc2})-(\ref{psc2i}) define $H$ to be an anti-pseudo-(anti)-Hermitian Hamiltonian\footnote{We denote $\s_{\pm}$ for pseudo-, and pseudo-(anti)-Hermitian cases, respectively, such that $\s_{\pm} H \s_{\pm}^{-1} = \pm H^\dagger.$ Further generalizations are possible, $\s_{\pm}$ operators can be either linear or anti-linear. In our case $\s_{-}$ is an anti-linear operator. So, $H$ can be dubbed as an anti-pseudo-(anti)-Hermitian Hamiltonian} \cite{das, ali2}. A similar property can be discussed for the companion Hamiltonian $\widetilde{H}$, since it also satisfies the pseudo-chirality relationship (\ref{spc1c}); so, $\widetilde{H}$ is also an anti-pseudo-(anti)-Hermitian Hamiltonian
\br
\label{psc2c}
\s_{-} \widetilde{H} \s_{-}^{-1}&=& - \widetilde{H}^{\dagger}.
\er
Next, let us define the right and left eigenvector equations for the operator $H^{\star}$
\br
H^{\star} \chi_j= \l_j \, \chi_j,\,\,\,\,\hat{\chi}^{T}_i H^{\star} = \l_i \, \hat{\chi}^{T}_i.\label{eig11}
\er 
Taking the transpose of the left eigenvalue equation in (\ref{eig11}) and subsequently considering the pseudo-anti-Hermiticity relationship (\ref{psc2}) one can write 
\br
\label{hdag1}
H^{\dagger} \hat{\chi}_j &= & \l_j \, \hat{\chi}_j, \\
\label{hdag11}
 - \s_{-} H \s_{-}^{-1}  \hat{\chi}_j &=& \l_j\, \hat{\chi}_j, \\
\label{hdag111}
H^\star ( {\cal G }^{-1}  \hat{\chi}_j  ) &=&  - \l_j\,  ( {\cal G }^{-1}  \hat{\chi}_j ),
\er
where (\ref{hdag11}) follows upon using (\ref{psc2}) into (\ref{hdag1}). In order to get (\ref{hdag111}) we have used $  {\cal T} H  {\cal T}^{-1} = H^\star$ and multiplied on the left by ${\cal G }^{-1}$ on the b.h.s. of  (\ref{hdag11}). So, comparing the last relationship (\ref{hdag111})  and the right eigenvalue equation in (\ref{eig11}) one gets
\br
\label{chim1}
\chi_{\widetilde{j}} =  {\cal G }^{-1}  \hat{\chi}_{j},
\er
where $\chi_{\widetilde{j}} $ denotes a right eigenvector with eigenvalue  $(-\l_{j})$. 

Notice that one can also relate the above left eigenvector $\hat{\chi}^{T}$ to the left eigenvector of type $\bar{\xi}^T$ in (\ref{eig1}) for the same eigenvalue $\l_j$. In fact, by taking  the complex conjugate of (\ref{eig1}) one has 
\br
\bar{\xi}_j^{\dagger} H^\star&=&  \l_j^\star \bar{\xi}_j^{\dagger},
\er
which upon comparing to the relevant equation in (\ref{eig11}) provides the left eigenvector of $H^{\star}$
\br
\hat{\chi}^{T}_j = \bar{\xi}_j^{\dagger}
\er
with eigenvalue $\l_j^{\star}$.
Moreover, using the symmetry (\ref{psc2}) and the right eingenvector equation (\ref{eig1}) one can write the following left  eingenvector equations
\br
\label{left2}
(\xi^{\dagger} \s_{-}) H = -\l^{\star} (\xi^{\dagger} \s_{-}),\\
\label{left2i}
(\xi^{\dagger}  {\cal G } ) H^{\star} = -\l^{\star} (\xi^{\dagger}  {\cal G } ),
\er
where in order to get the last relationship we have used the form of $\s_{-}$ in (\ref{psc2i}).
So, by comparing (\ref{left2i}) with the left-eigenvector equation in (\ref{eig11}) one can conclude that for each left eigenvector $\hat{\chi}^{T}_j$ with eigenvalue $\l_j$ one must have another eigenvector $(\xi^{\dagger}  {\cal G } )$ with eigenvalue $ -\l^{\star}_j$, i.e.
\br
\label{chihat1}
\hat{\chi}^{T}_{-\l_j^{\star}} =  (\xi^{\dagger}_{\l_j}  {\cal G } ),
\er 
with $\xi_j$ being the right eigenvector of $H$ with eigenvalue $\l_j$ defined in (\ref{eig1}). 

Next, let us write a relationship between the right-eigenvectors of the Hamiltonians $H$ and $H^{\star}$. From (\ref{chim1}) and (\ref{chihat1}) one can write
\br
\label{chixi1}
\chi_{\l} = - (\xi_{\l^{\star}})^{\star}. 
\er
Analogous construction can be performed for the companion Hamiltonian $\widetilde{H}$. So, let us define the right and left eigenvector equations for $\widetilde{H}^{\star}$ 
\br
\widetilde{H}^{\star} \phi_j= \l_j \, \phi_j,\,\,\,\,\hat{\phi}^{T}_j \widetilde{H}^{\star} = \l_j \, \hat{\phi}^{T}_j.\label{eig11c}
\er 
So, taking the transpose of the left eigenvalue equation in (\ref{eig11c}) and subsequently considering the pseudo-anti-Hermiticity relationship (\ref{psc2c}) one can show the next relationship
\br
\label{phim1}
\phi_{\widetilde{j}} =  {\cal G }^{-1}  \hat{\phi}_{j},
\er
where $\phi_{\widetilde{j}} $ denotes a right eigenvector with eigenvalue  $(-\l_{j})$.
 
Similarly,  one can relate the above left eigenvector $\hat{\phi}^{T}$ to the left eigenvector of type $\bar{\eta}^T$ in (\ref{eig22}) for the conjugate eigenvalue $\l_j^\star$
\br
\hat{\phi}^{T}_j = \bar{\eta}_j^{\dagger}.
\er
Moreover, using the symmetry (\ref{psc2c}) and the right eingenvector equation (\ref{eig22}) one can write the following left  eingenvector equations
\br
\label{left2c}
(\eta^{\dagger} \s_{-}) \widetilde{H} = -\l^{\star} (\eta^{\dagger} \s_{-}),\\
\label{left2ic}
(\eta^{\dagger}  {\cal G } )  \widetilde{H}^{\star} = -\l^{\star} (\eta^{\dagger}  {\cal G } ).
\er
So, by comparing (\ref{left2ic}) with the left-eigenvector equation in (\ref{eig11c}) one can conclude that for each left eigenvector $\hat{\phi}^{T}_j$ with eigenvalue $\l_j$ one must have another eigenvector $(\eta^{\dagger}  {\cal G } )$ with eigenvalue $ -\l^{\star}_j$, i.e.
\br
\label{phieta1}
\hat{\phi}^{T}_{-\l_j^{\star}} =  (\eta^{\dagger}_{\l_j}  {\cal G } ),
\er 
with $\eta_j$ being the right eigenvector of $\widetilde{H}$ with eigenvalue $\l_j$ defined in (\ref{eig22}).
  
So, in contradistinction to the pseudo-chiral symmetry (\ref{spc1}) which ensures that for each energy level $\l$ there exists the opposite $-\l$, i.e. the pairs $(\l, -\l)$, the  symmetries (\ref{psc2}) and (\ref{psc2c}) ensure the
energy levels in pairs with identical imaginary part and opposite real part, i.e.  the pairs $(\l , -\l^{\star})$. 
Remarkably, in our constructions of the next section, we will provide analytic realizations of these eigenvalue and eigenvector symmetries for a continuum of spinor bound states and extended wave states in the localized continuum.  

We write the biorthogonality relationship in this case. So, from (\ref{eig11}) one can get 
\br
\hat{\chi}^{T}_i H^{\star} \chi_j= \l_j \, \hat{\chi}^{T}_i \chi_j,\,\,\,\,\hat{\chi}^{T}_i H^{\star} \chi_j= \l_i \, \hat{\chi}^{T}_i \chi_j,
\er
which implies 
\br
0 = (\l_i -\l_j) \hat{\chi}^{T}_i \chi_j \,\,\,\,  \Rightarrow \,\,\,\,  \hat{\chi}^{T}_i \chi_j  = \d_{ij}.
\er
Similar construction from (\ref{eig11c}) allows one to write
\br
0 = (\l_i -\l_j) \hat{\phi}^{T}_i \phi_j \,\,\,\,  \Rightarrow \,\,\,\,  \hat{\phi}^{T}_i \phi_j  = \d_{ij}.
\er

Next, let us write a relationship between the left-eigenvectors of the Hamiltonians $\widetilde{H}$ and $\widetilde{H}^{\star}$. From (\ref{phim1}) and (\ref{phieta1}) one can write
\br
\label{chixi}
\phi_{\l} = - (\eta_{\l^{\star}})^{\star}. 
\er 
\begin{table}[ht]
\centering
\caption{Pseudo-chiral and anti-pseudo-(anti-)Hermitian symmetries and representations}
\begin{tabular}[t]{lll}
\hline
&Pseudo-chiral & Anti-pseudo-Hermitian \\
\hline
Symmetry  &  ${\cal G } X {\cal G }^{-1}=-X^T$,\,\, $X \equiv\{ H,\widetilde{H}\} $ &$\s_{-} X \s_{-}^{-1}= -X^\dagger$,\,\,   $X \equiv\{ H,\widetilde{H}\} $\\
Right-eigenvectors & $H \xi = \l\, \xi,\,\,\,\,\widetilde{H}\eta= \l \,\eta$ & $H^\star \chi = \l\, \chi ,\,\,\,\, \widetilde{H}^\star \phi = \l \,\phi $\\
 & $\G H \G ^{-1} = \widetilde{H}$\,$\rightarrow$\,$\eta = \G \xi$,\,\,$\bar{\eta} = \G \bar{\xi}$  & $\G H^\star \G ^{-1} = \widetilde{H}^\star$\,$\rightarrow$\, $\phi = \G \chi$,\,\,$\hat{\phi} = \G \hat{\chi}$ \\
Left-eigenvectors & $\bar{\xi}^T H = \l \, \bar{\xi}^T,\,\,\,\,\bar{\eta}^T \widetilde{H}= \l \, \bar{\eta}^T$ & $\hat{\chi}^T H^{\star} = \l \, \hat{\chi}^T,\,\,\,\,\hat{\phi}^T \widetilde{H}^\star= \l\,  \hat{\phi}^T$\\
& $\G H \G ^{-1} = \widetilde{H}$\,$\rightarrow$\, \,\,$\bar{\eta} = \G \bar{\xi}$  & $\G H^\star \G ^{-1} = \widetilde{H}^\star$\,$\rightarrow$\,  \,\,$\hat{\phi} = \G \hat{\chi}$ \\
Eigenvalues &  $\{ \l,\,\,\,\,-\l\}$  & $\{\l,\,\l^{\star},\,-\l^{\star}\}$ \\
R/L-eigenvectors, $(\l , -\l)$  & $\bar{\xi}_{-\l} = {\cal G } \xi_{\l},\,\,\,\,  \bar{\eta}_{-\l} = {\cal G } \eta_{\l}$ &  $\chi_{-\l} = {\cal G }^{-1} \hat{\chi}_{\l}$,\,\,\,\, $\phi_{-\l} = {\cal G }^{-1} \hat{\phi}_{\l}$\\
R/L-eigenvectors, $(\l , \pm \l^{\star})$  & -  & $\hat{\chi}_{-\l^\star} = - {\cal G } \xi^{\star}_{\l}$, \,\,\,$\hat{\phi}_{-\l^\star} = - {\cal G } \eta^{\star}_{\l}$\,\,\,\,\,\,\,\, \\
   & - &  $\hat{\chi}_{\l^\star} = - \bar{\xi}^{\star}_{\l}$, \,\,\,\,\,\,\,\,\,\,\,\,\,\,\,$\hat{\phi}_{\l^\star} = - \bar{\eta}^{\star}_{\l}$\,\,\,\,\,\,\,\, \\
   & - &  $\chi_{\l} = - (\xi_{\l^{\star}})^{\star}$, \,\,\,\,\,\,\,\,\,\,\,$\phi_{\l} = - (\eta_{\l^{\star}})^{\star}$\,\,\,\,\,\,\,\, \\
Biorthogonality & $\bar{\xi}^T_i\xi_j = \d_{ij},\,\,\,\,\bar{\eta}^T_i\eta_j = \d_{ij}$&  $\hat{\chi}_i^T\chi_j = \d_{ij},\,\,\,\,\hat{\phi}_i^T\phi_j = \d_{ij} $\\
Majorana modes ($\l=0$) & $\bar{\xi}^T_0\xi_0 = 0,\,\,\,\,\bar{\eta}^T_0\eta_0 = 0$ & $\hat{\chi}_0^T\chi_0 = 0,\,\,\,\,\hat{\phi}_0^T\phi_0 = 0 $ \\
\end{tabular}
\end{table}
The Table 1 summarizes the main results related to the pseudo-chiral and $(\s_{-})$-anti-pseudo-Hermitian symmetries and their relevant representations, respectively, for the set of Hamiltonians $\{H, \widetilde{H}\}$ and  $\{H^\star, \widetilde{H}^\star\}$. So, we have organized, for each type of symmetry, the corresponding eigenvectors and their mutual relationships, as well as the allowed values of the spectral parameter $\l$ permitted by the relevant symmetry. The  corresponding biorthogonality relationships and the appearance of the  Majorana zero modes at $\l=0$ and the coalescence of the eigenvectors are also included.

\subsection{Some properties of pseudo-Hermitian Hamiltonians}

Let us discuss some theorems and properties of pseudo-Hermiticity and anti-pseudo-Hermiticity for NH Hamiltonians following the approach in \cite{ali1, ali2, ali3, ali4}, which are relevant to our model. A NH Hamiltonian $H$ is said to be a $\zeta$-pseudo-Hermitian Hamiltonian if one has 
\br
\label{pseudo1}
\zeta H \zeta^{-1}  = H^{\dagger},
\er
for $\zeta$ being a linear invertible Hermitian operator. Notice that for $\zeta = {\bold 1}$ one recovers a Hermitian Hamiltonian, i.e. $H^{\dagger}=H$. On the other hand, a $\s-$anti-pseudo-Hermitian Hamiltonian is defined as \cite{ali2, ali4}
\br
\label{pseudo2}
\s  H \s^{-1}  =  H^{\dagger}, 
\er
for $\s$ being an anti-linear Hermitian automorphism. These types of anti-linear operators appearing in the definition of pseudo-Hermicity have further been classified into $\s_{\pm}$,  Hermitian \cite{ali4} and anti-Hermitian ones  \cite{ali2}, respectively. The anti-linear operator $\s_{-}$ in (\ref{psc2}) belongs to the anti-Hermitian type in this classification; so, $H$  in (\ref{NH})  which satisfies  (\ref{psc2}), can be dubbed as $(\s_{-})-$anti-pseudo-(anti-)Hermitian Hamiltonian. An equivalence between a $\zeta-$pseudo-Hermiticity and a $\s-$anti-pseudo-Hermiticity has been established using spectral decomposition techniques in \cite{ali2, ali4}. 

Next, we summarize the main properties of pseudo-Hermitian operators \cite{ali1, ali3} to be useful for our discussions of the spectra of the pseudo-Hermitian Hamiltonian (\ref{NH}). Let $H : {\cal H}\rightarrow {\cal H}$ be a diagonalizable linear operator acting on a Hilbert space ${\cal H}$ and $\zeta :{\cal H}\rightarrow {\cal H}$ be a linear Hermitian automorphism. Then

1). $H$ is pseudo-Hermitian if and only if its eigenvalues are real or come in complex-conjugate
pairs.

2). if H is pseudo-Hermitian with respect to two linear Hermitian automorphisms $\zeta_1$ and $\zeta_2$,
then $\zeta_1^{-1} \zeta_2$ generates a symmetry of H, i.e. 
\br
\label{symm11}
 [\zeta_1^{-1} \zeta_2 , H] =0.
\er 
 
Moreover, in \cite{ali2} it has been shown that every diagonalizable Hamiltonian possessing an antilinear symmetry is necessarily pseudo-hermitian. This applies, in particular, to diagonalizable
PT-symmetric Hamiltonians. The proof of the equivalence of pseudo-hermiticity and the presence of antilinear
symmetries has been performed provided that the Hamiltonian is diagonalizable and has a discrete spectrum. It used the fact that such Hamiltonian $H$ is anti-pseudo-hermitian, i.e. there is an antilinear, Hermitian, invertible operator as $\s$ in (\ref{pseudo2}). The analogous properties relevant to field theory models, to our knowledge, have not yet been proved; however, it is expected that some of them must hold, with relevant modifications, for the NH Hamiltonians in the field theory context.  In fact, we will see below that the eigenvalues of the field theory NH model described by the $(\s_{-})-$anti-pseudo-Hermitian Hamiltonian $H$ in (\ref{NH}) come in complex-conjugate pairs and some of the Hamiltonian symmetries (\ref{symm11}) can be realized provided that the complex scalar $\Phi$ components satisfy some definite parities under space inversion. 

\subsection{${\cal P }$, ${\cal T }$, ${\cal PT}$ and $\g_5{\cal PT}$  symmetries}
\label{PT1}
So, in order to examine additional symmetries we consider the ${\cal P }$, ${\cal T }$ and ${\cal PT}$ symmetry transformations of the NH Hamiltonian (\ref{NH}). Let us define the next symmetry transformations for the scalar field components $\Phi_a$, the complex parameter $z$ and the space coordinate $x$
\br
\Phi &\equiv& \Phi_1 + i \Phi_2,\\
\label{scalar12p}
{\cal P } \Phi_a {\cal P } &=& \nu_a \Phi_a,\, \,\,\,\, \,\,\nu_a = \pm 1,\,\,\,a=1,2.\\
{\cal P } x {\cal P } &=& -x,\,\,\,\,\,\,\,{\cal T } z {\cal T}^{-1} = z^{\star}.
\er
Next, we will assume the scalar field components $\Phi_a\, (a=1,2)$ to possess definite parities $\nu_{a}= \pm 1$ as in  (\ref{scalar12p}). So, let us consider the following cases: {\bf Case A.} $\nu_1=+1,\,\, \nu_2 = +1$. {\bf Case B.} $\nu_1=+1,\,\, \nu_2 = -1$. {\bf Case C.} $\nu_1=-1,\,\, \nu_2 = +1$. And {\bf Case D.} $\nu_1=-1,\,\, \nu_2 = -1$. The {\bf Case E} requires a new shifted parity-operator and scalar fields with $\nu_1=-1,\,\, \nu_2 = 1$ relevant parities (see below).

One can show the next relationships  for the Hamiltonian $H$ in (\ref{NH}) upon making the relevant transformations in its matrix representation. 

{\bf Case A.} $\nu_1=+1,\,\, \nu_2 = + 1$. In this case one has
\br
{\cal P } \g_5  H \g_5 {\cal P }^{-1} = - H.
\er
This implies the anti-symmetry $\{{\cal P } \g_5  , H\}=0$.

{\bf Case B.} $\nu_1=+1,\,\, \nu_2 = - 1$. 
In this case one has for $M$ real  
\br
\label{B1}
{\cal P} H {\cal P}^{-1}  = H^{\dagger}.
\er
Making use of the anti-pseudo-Hermiticity relationship (\ref{psc2}) together with (\ref{B1}) one can get the anti-symmetry $\{\s_{-}^{-1}{\cal P} \,,\, H\}=0.$

{\bf Case C.} $\nu_1=-1,\,\, \nu_2 = +1$.  For $M$ real ($M= M_1,\, M_1 \in \IR$)  one has

\br
\label{c11}
{\cal P}{\cal T} \g_5 H [{\cal P }{\cal T } \g_5]^{-1} = H.
\er
Then, in this case the NH Hamiltonian $H$ possesses the antilinear symmetry ${\cal P }{\cal T } \g_5$, i.e. $[{\cal P }{\cal T } \g_5\,,\, H]=0$. So, the antilinear $\g_5{\cal P }{\cal T } $ symmetry (\ref{c11}) must be taken into account together with the pseudo-chirality and anti-pseudo-Hermiticity of the model in order to examine its spectral properties in this case.

{\bf Case D.} $\nu_1=-1,\,\, \nu_2 = -1$.  

In this case for  $M \in \IC$  one has the relationship 
\br
\label{alps21}
\bar{\s}_{-} H \bar{\s}_{-}^{-1}&=& - H^{\dagger},\\
\bar{\s}_{-}  &\equiv& {\cal P}{\cal T}.
\er 
Notice that $\bar{\s}_{-}$ is an antilinear operator. So, this Hamiltonian is a (anti)-pseudo-anti-Hermitian Hamiltonian as in (\ref{psc2}), however in this particular case under the $ {\cal P}{\cal T}$ operator $\bar{\s}_{-}$. Considering the both pseudo-anti-Hermiticity relationships (\ref{alps21}) and  (\ref{psc2}) one can show that 
\br
\label{alps22}
\s^{-1}_{-} \bar{\s}_{-} H (\s^{-1}_{-} \bar{\s}_{-} )^{-1} &=& H\\
({\cal P} {\cal G})  H ({\cal P} {\cal G})^{-1} &=& H,\label{alps221}
\er
where the product $\s^{-1}_{-} \bar{\s}_{-} = {\cal P} {\cal G}$ has been inserted in the eq. (\ref{alps221}). So, (\ref{alps221}) implies the symmetry  
\br
\label{symm22}
[H, {\cal P} {\cal G}] = 0.\er
Then, (\ref{symm22}) establishes a symmetry of type $\s^{-1}_{-} \bar{\s}_{-}$ for a Hamitonian which is pseudo-anti-Hermitian with respect to two different anti-linear operators $\s_{-}$ and $\bar{\s}_{-}$. So, it is an analogous symmetry to the one in (\ref{symm11}) which establishes a type of symmetry ($\zeta_1^{-1} \zeta_2$) for pseudo-Hermitian Hamiltonians, i.e. with  pseudo-Hermiticity property with respect to two different linear Hermitian operators $\zeta_{1}$ and $ \zeta_{2}$.

{\bf Case E.}  $\nu_1=-1,\,\, \nu_2 = +1$,  for $M$ purely imaginary  ($M = i M_2\, (M_2 \in \IR)$).

Let us consider a shifted space-inversion operator ${\cal P }_{s}$ such that
\br
\label{pt10}
{\cal P }_{s}(\Phi(\hat{x})) &=& \Phi(-\hat{x}),\,\,\,\,\,\,\, \hat{x} \equiv x - x_0,\,\,\, x_0 = \mbox{const.}\\
\label{pt1}
\Phi_1(-\hat{x}) &=& - \Phi_1(\hat{x}),\\
 \Phi_2(-\hat{x})  &=& \Phi_2(\hat{x}). \label{pt2}
\er
The above transformation defines an odd $\Phi_1$ and an even $\Phi_2$ functions. Therefore, one can show that the Hamiltonian $H$ in (\ref{NH}) satisfies the following symmetry
\br
\label{ptsym}
[{\cal P }_{s} {\cal T}\, ,\, H ] =0,
\er
So,  (\ref{ptsym})  defines a  ${\cal P } {\cal T}$-type symmetry of $H$ provided that the field components transform as defined in (\ref{pt1})-(\ref{pt2}). Moreover, one can define a (partial) ${\cal P}{\cal T} $ symmetric Hamiltonian in the case 
\br
\label{pt11}
\Phi_1(-\hat{x}) &=& - \Phi_1(\hat{x}),\\
 \Phi_2(-\hat{x})  &\neq & \Phi_2(\hat{x}). \label{pt22}
\er
Analogous partially ${\cal P}{\cal T}$ symmetric non-Hermitian
Hamiltonians associated to chains of $N$ coupled harmonic oscillators have been considered \cite{benderpar}, such that
an eigenfunction having the partial ${\cal P}{\cal T}$ symmetry  is associated with a real eigenvalue, and  the real spectrum is part of a larger spectrum containing
complex-conjugate pairs. 

Moreover, similar steps as above can be followed for the complex conjugate Hamiltonian $H^{\star}$ in each of the cases $A, B, C, D$ and $E$ discussed above. The results are summarizes in the Table 2, which shows that the both Hamiltonians satisfy the same symmetry in each case.  We will find below some spinors, solitons and extended wave solutions satisfying the symmetries above. 
 
\begin{table}[ht]
\centering
\caption{ ${\cal P }$, ${\cal T }$,  ${\cal PT}$ and $\g_5{\cal PT}$ symmetries of $X \in \{H, H^{\star}\}$}
\begin{tabular}[t]{llll}
\hline
Case &\,\,\,\,\,Parity:$(\nu_1,\,\nu_2)$  &  $H$ and $H^{\star}$ properties & (Anti-)symmetry for $X \in \{H, H^{\star}\}$\\
\hline
A  &  \,\,\,\,\,\,\,\,\,\,\,\,\,\,\,\,\,\,\,$\(  + \,\,\,\, +\)$  & Pseudo-chiral+Anti-pseudo-Hermitian & $\{\g_5{\cal P} \,,\, X\}=0$\,\\ 
B  &  \,\,\,\,\,\,\,\,\,\,\,\,\,\,\,\,\,\,\,$\(  + \,\,\,\, -\)$  & Anti-pseudo-Hermitian & $\{\s_{-}^{-1}{\cal P} \,,\, X\}=0$\,\,$(M\in \IR$)\\ 
C  &  \,\,\,\,\,\,\,\,\,\,\,\,\,\,\,\,\,\,\,$\(- \,\,\,\,\, +\)$  & Pseudo-chiral+Anti-pseudo-Hermitian &$[\g_5{\cal P} {\cal T }  \,,\, X]=0$\, \\
D  & \,\,\,\,\,\,\,\,\,\,\,\,\,\,\,\,\,\,\,$\( -\,\,\,\,\, -\)$ & Anti-pseudo-Hermitian &$ [{\cal P} {\cal G }\,,\,X]=0$\\
E  & \,\,\,\,\,\,\,\,\,\,\,\,\,\,\,\,\,\,\,$\( -\,\,\,\,\, +\)$ & ${\cal P}_s {\cal T }$ Symmetric &$ [{\cal P}_s {\cal T }\, ,\,X]=0$,\,
\end{tabular}
\end{table}
In the Table 2 we summarize the main results of the cases $A, B, C, D$ and $E$ discussed above. We present the interplay between the ${\cal P }$, ${\cal T }$, ${\cal PT}$  and $\g_5{\cal PT}$ symmetry transformations together with the pseudo-chiral and anti-pseudo-Hermiticity properties, which imply (anti-)symmetries of the Hamiltonians $H$ and $H^{\star}$ (last column of the Table 2) for each set of parities ($\nu_1,\,\nu_2$) of the scalar fields $\Phi_1$ and $\Phi_2$, respectively.  The case $E$  presents a shifted space-inversion  parity of the scalar fields and the anti-linear ${\cal P}_s {\cal T }$ symmetry of $H$ and $H^{\star}$. Note that in the cases $C$ and $E$ the both type of Hamiltonians exhibit anti-linear symmetries, i.e. the $\g_5{\cal P} {\cal T }$  and ${\cal P} {\cal T }$ symmetries, respectively. These properties will be used below in order to examine the symmetry unbroken phases with real eigenvalues and the broken phases of the model.  

Below, in the context of the tau-function approach, we will perform the construction of the right and left eigenvectors satisfying the pseudo-chirality and pseudo-Hermitian properties discussed above, and for a special solution we will show a NH Majorana zero mode-soliton duality description of the model, such that the static scalar field is a self-consistent solution of the equation (\ref{equi1}) and (\ref{equiv22}).  In fact, we will realize the theorem 1) above  by showing that the both  NH Hamiltonians in  (\ref{NH}) possess complex-conjugate pairs of eigenvalues. 

\section{NH topological solitons coupled to the fermions}
\label{sec:topo}

In this section we find exact analytical solutions of the system of eqs. (\ref{H1})-(\ref{H11}) and (\ref{sec1}) associated to the NH model (\ref{lag1}) and search for nontrivial topological configurations for the complex field $\Phi$ interacting with the bound states formed by the two fermion fields $\xi$ and $\eta$. The Hermitian case has been considered before by  defining a submodel for $\Phi$ being a real scalar and by imposing  the reality condition $\widetilde{\psi} \equiv  \psi^{\star}$ to the model (\ref{lag1}); so, the reduced model defined a Dirac field interacting with real Toda solitons \cite{matter, witten, aop1, npb2, prd1, npb1, jhep22}.  

\subsection{Topological solitons and spinor bound states}

The soliton solutions, as well as the bound states of the Hermitian version of this model has been considered in \cite{jhep22} through the tau function formalism. Here we perform the construction of the general complex 1-solitons and spinor bound state configurations of the non-Hermitian model (\ref{lag1}). In this context, for a suitable subspace in parameter space, it will emerge also some extended wave states in the localized continuum. 
 
Since the pair of first order differential equations (\ref{H1})-(\ref{H11}) for the spinors plus the first order equation (\ref{equi1}) for $\Phi$ imply the second order differential eq. for the complex scalar (\ref{sec1}), each set of solutions of the first order system of equations will be solutions of the second order differential equation (\ref{sec1}). So, it is sufficient to consider the system of equations (\ref{H1})-(\ref{H11}) and  (\ref{equi1}) in order to find the complex solitons and related bound state solutions.   

In the next steps we rewrite the equations of motion of the model in terms of the tau functions $\tau_{0, 1}$, $\tau_{A,  B},\widetilde{\tau}_{A, B}$  as
\br
\label{tau01}
e^{i \b \Phi} &=& \rho\, (\frac{\tau_0}{\tau_1})^2,\\
\label{tauab}
\xi_{A} &=& \frac{\tau_A}{\tau_0},\,\,\,\xi_{B} = \frac{\tau_B}{\tau_1},\,\,\,\eta_{A} = \frac{\widetilde{\tau}_A}{\tau_1},\,\,\,
\eta_{B} = \frac{\widetilde{\tau}_B}{\tau_0},
\er
with $\rho$ being a constant complex parameter. Next, substituting the field components (\ref{tau01})-(\ref{tauab}) into the system of first order equations (\ref{H1})-(\ref{H11}) and  (\ref{equi1}) one has, respectively 
 \br
\label{tau11}
i \l \tau_1 \tau_B - M \rho\, \tau_0 \tau_A + \tau_B \tau'_1 - \tau_1 \tau'_B &=& 0\,\\
\label{tau22}
i \l \tau_0 \tau_A + M \rho^{-1} \tau_1 \tau_B - \tau_A \tau'_0 + \tau_0 \tau'_A &=& 0\,\\
\label{tau33}
i \l \tau_0 \widetilde{\tau}_B + M  \rho^{-1} \tau_1 \widetilde{\tau}_A - \widetilde{\tau}_B \tau'_0 + \tau_0 \widetilde{\tau}'_B &=& 0\,\\
\label{tau44}
i \l \tau_1 \widetilde{\tau}_A - M  \rho\, \tau_0 \widetilde{\tau}_B + \widetilde{\tau}_A \tau'_1 - \tau_1 \widetilde{\tau}'_A &=& 0.
\er
and 
\br
\label{tau011}
2i (\tau'_0 \tau_1-\tau_0 \tau'_1) + M \b (\tau_A \widetilde{\tau}_A+\tau_B \widetilde{\tau}_B) =0.
\er
Next, let us assume the following tau functions
\br
\label{tau01x}
\tau_1 &=& 1 + a\, e^{2 k x},\,\,\,\,\tau_0 = 1 + b \, e^{2 k x},\\
\label{tauabx}
\tau_A &=& c \,e^{ k x},\,\,\,\tau_B = d \, e^{ k x},\\
\label{tauabxx}
\widetilde{\tau}_A &=& \widetilde{c} \, e^{ k x},\,\,\,\widetilde{\tau}_B = \widetilde{d}\,  e^{ k x},
\er 
with the complex parameters $a,b,c,d, \widetilde{c}, \widetilde{d}$ and $k$. So, the above tau functions satisfy the system of equations  (\ref{tau11})-(\ref{tau44}) and (\ref{tau011}) provided that
\br
\label{para1}
a &=&  \frac{\widetilde{c} c}{2M} \frac{\b^2 z^3}{(z^2-1)^2},\,\,\,\,\,\,\,
\widetilde{d} = \frac{\widetilde{c}\, c}{d},\,\,\,\,\,\,\,
 \rho = -i \frac{d}{c} z,\\
\label{para111}
 b  &=&\frac{\widetilde{c} c}{2M} \frac{\b^2 \, z}{(z^2-1)^2},\,\,\,\,\,
z \equiv -\frac{\l}{M} \pm \sqrt{\frac{\l^2}{M^2} - 1} \\
\label{lvr}
\l & = & -\frac{M}{2} (z + z^{-1}),\,\,\,\,\,\,\, k = \frac{i M}{2} (z-z^{-1}).
\er
From the above equations one can write  
\br
\label{lmk1}
\l^2 + k^2 = M^2.
\er
The equation (\ref{lmk1}) can be regarded as a dispersion relationship between the complex energy $\l$ and the complex wave number $k$ of the quasi-particles for a fixed value of the complex mass parameter $M$. Notice that this equation is invariant under the transformations 
\br
\label{lkm1}
\l  &\rightarrow&  -\l\\
k & \rightarrow & -k. \label{lkm2}
\er
So, the eigenvalues come in pairs of type $(\l, -\l)$. This is in accordance to the pseudo-chirality symmetry property discussed in section \ref{sec:spc11}. Then, one can argue that the eigenvectors corresponding to this pair of eigenvalues realize the  pseudo-chirality  symmetry (\ref{spc1}) and the left $\bar{\xi}_{-\l}$ and right eigenvectors $\xi_{\l}$ can be related by (\ref{lrxi1}).
 
Next, one can take the complex conjugate of the dispersion relationship (\ref{lmk1}) and get  
\br
\label{lmk1star}
(\l^\star)^2 + (k^{\star})^2 = (M^\star)^2.
\er 
Since this equation defines a new eigenvalue $\l^\star$, one has that the eigenvalues come in complex-conjugate pairs. Notice that (\ref{lmk1star}) is invariant under $\l^{\star}  \rightarrow  -\l^{\star},\,\,\,k^{\star}  \rightarrow - k^{\star}$. So, from (\ref{lmk1}) and (\ref{lmk1star}) and their relevant symmetries under inversion $\l \rightarrow - \l$ and $\l^{\star} \rightarrow - \l^{\star}$, one can argue that the solutions above describe the corresponding left and right eigenvectors with the set of eigenvalues $\{ \l, -\l, \l^{\star}, -\l^{\star}\}$. 

One can see that the complex dispersion relationship (\ref{lmk1star}) arises from the tau function representation of the eigenvalue equation  in  (\ref{eig11}) for the right eigenvector $\chi$ and the operator $H^\star$. In fact, by taking the complex conjugate of $H \xi = \l \xi$ in (\ref{eig1}) and the relevant tau function components in (\ref{tau01})-(\ref{tauab}) and subsequent substitution into the corresponding tau function  equations  analogous  to (\ref{tau11})-(\ref{tau44}) one can get (\ref{lmk1star}).

In fact, these solutions become the realizations of the pseudo-chiral and pseudo-Hermitian symmetries of the model discussed in the sections \ref{sec:spc11} and \ref{sec:psh1}, respectively. Remarkably, the relevant  spectral properties associated to the symmetries (\ref{spc1}) and (\ref{psc2}) are realized above by providing the set of eigenvectors with eingenvalues $\{ \l, -\l \}$ and $\{\l, -\l^{\star}\}$, respectively. Then, the eigenvalues come in complex conjugate pairs and their reversions. This is in accordance to the first theorem 1) which must hold for pseudo-anti-Hermitian Hamiltonians, once the equivalence between the pseudo-Hermicity and the anti-pseudo-Hermicity properties  is assumed \cite{ali2, ali4}.
 
The energy $\l$ and wave number $k$ in (\ref{lvr}) are defined in terms of the uniformization complex variable $z$. The Fig.1 shows some singular points such as the pole at the origin $z=0$ ($|\l| \rightarrow + \infty$) and the zero modes at $z=\pm i$ of the eigenvalue $\l(z)$. The points $z=\pm 1$ correspond to the edge states $k=0\rightarrow \l=\pm M$.  Note that the contour $\G$ (outer circle) encircles the zero modes $\l = 0 $ located at the points $z = \pm i$ and the pole at $z=0$ in the complex plane $z$. The winding number of $\l$ will we computed in (\ref{wind111}) as a $\G$ contour integral in the complex $z-$plane. In the Hermitian case ($M=M_1, \l = \l_1$) the bound states/solitons lie on the unit circle $|z|=1$ (inner circle), and the scattering states are allowed for $z \in \IR$ ($k = i k_2$).  

Next, we define the complex parameters as follows 
\br
\label{M12}
M & \equiv & M_1 + i\, M_2,\\
\label{l12}
\l & \equiv &  \l_1 + i \, \l_2,\\
\label{k12}
k & \equiv & k_1 + i \, k_2,
\er 
and
\br
a = |a| e^{i \a_1},\,b = |b| e^{i \a_2},\,\rho = |\rho| e^{i \d},\,\,c = |c| e^{i \zeta_1},\,\,d = |d| e^{i \zeta_2}. \er
So, from the eqs.  (\ref{para1})-(\ref{para111}) the complex variable $z$ can be written as
\br
\label{zz1}
z = \sqrt{|a|/|b|}\, e^{i \s},\,\,\,\,\s \equiv \frac{\a_1 - \a_2}{2}.
\er
Next, taking into account  (\ref{M12})-(\ref{k12}) and (\ref{para1})-(\ref{lvr}) one can write
\br
\label{zk12}
z  = \sqrt{\frac{(\l_1-k_2)^2 +(\l_2 + k_1)^2}{|M|^2}}  \, e^{i \s},\,\,\,\,\s \equiv \arctan{\Big[\frac{M_2(k_2-\l_1)+ M_1 (k_1+\l_2)}{M_2(k_1+\l_2) - M_1 (k_2-\l_1)}\Big]}.
\er
\begin{figure}
\centering
\includegraphics[width=8cm,scale=1, angle=0,height=5cm]{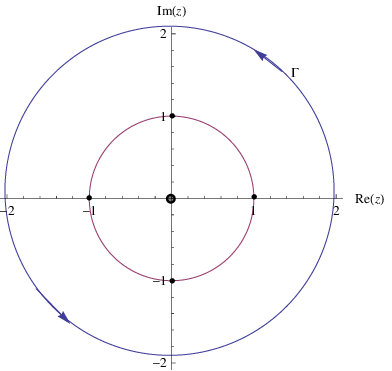}
\parbox{5in}{\caption{The complex plane defined by the uniformization variable $z$. The complex energy $\l(z)$ (\ref{lvr}) exhibits a pole at the origin $z=0$ and the zeros at $z=\pm i$ ($|z|=1, \s= \pm \pi/2$). The points $z=\pm 1$ correspond to the edge states $k=0\, (\l=\pm M)$. The Hermitian ($M \in \IR, \l \in \IR$) bound states/solitons lie on the unit circle $|z|=1$ (inner circle) and the scattering states lie on $z \in \IR$. The winding number of $\l(z)$ will we computed in (\ref{wind111}) as a $\G$ (outer circle) contour integral in the complex $z-$plane.}}
\end{figure}
Notice that from (\ref{lmk1}) one can consider a complex function $\l = \pm \sqrt{M^2-k^2}$ defined on a Riemann surface consisting of two sheets, with $\pm M$ being the branch points, which are necessary to make the square-root function single valued. However, the convenient parametrization in the complex $z$ variable defined in (\ref{lvr}) avoids the introduction of this two-sheet surface, see below more about this construction. Below, we will define a convenient complex variable $w= k/M$ in order to introduce the EPs at the  branch points $w= \pm 1$.

The scalar field component solutions from (\ref{tau01}) can be written as   
\br
\label{phi1tau}
\Phi_1 &=& \frac{2}{\b} \arctan{\Big[ i \, \frac{e^{-i \frac{\d}{2}} \tau_0^{\star} \tau_1 - e^{i \frac{\d}{2}} \tau_0 \tau_1^{\star}}{e^{i \frac{\d}{2}} \tau_0 \tau^{\star}_1 + e^{-i \frac{\d}{2}} \tau_0^{\star} \tau_1 }\Big]}\\
\Phi_2 &=&  - \frac{1}{\b}\log{\Big[ |\rho| \frac{\tau_0^{\star} \tau_0}{\tau_1^{\star} \tau_1}\Big]}. \label{phi2tau}
\er
Next, taking into account the tau functions  (\ref{tau01x})-(\ref{tauabxx}) one can write
\br
\Phi_1& = & 
\frac{2}{\b} \arctan{\Big[\frac{\sin{\frac{\d}{2}} + (|b| \sin{(\Xi_2(x))} - |a| \sin{(\Xi_1(x))} e^{2 k_1 x} - |a| |b| \sin{(\Theta ) } e^{4 k_1 x} }{\cos{\frac{\d}{2}} + (|b| \cos{(\Xi_2(x))} +  |a| \cos{(\Xi_1(x))} e^{2 k_1 x} + |a| |b| \cos{(\Theta ) } e^{4 k_1 x} } \Big]}  
\label{phi1sol}\\
\Xi_1(x) & \equiv & 2 k_2 x + \a_1 - \frac{\d}{2},\,\,\,\,\Xi_2(x) \equiv 2 k_2 x + \a_2 + \d/2,\,\,\,\,\,\,\Theta \equiv \a_1 - \a_2 - \frac{\d}{2} \nonumber
\er
and
\br
\Phi_2 & = & - \frac{1}{\b}\log{\Big[|\rho| \,\,  \frac{1+ 2 |b| \cos{(2 k_2 x + \a_2 )} e^{2 k_1 x}  + |b|^2 e^{4 k_1 x}   }{1+ 2 |a| \cos{(2 k_2 x + \a_1 )} e^{2 k_1 x}  + |a|^2 e^{4 k_1 x}   }\big]} \label{phi2sol}.
\er
For the wave number with real component different from zero, i.e. $k_1 \neq 0$, one can define the next topological invariant for each component of the complex scalar field as 
\br
\label{top1}
\Phi_1 (+\infty)-\Phi_1 (-\infty) &=& \pm \frac{2}{\b} (\a_1 - \a_2),\\
&=&  \pm \frac{4}{\b} \, \s, \\
&=&  \pm \frac{4}{\b} \, \arg{z} \label{top11}\\
\label{top2}
\Phi_2 (+\infty)-\Phi_2 (-\infty) &=& \pm \frac{2}{\b} \log{\frac{|b|}{|a|} },\\
&=&  \pm \frac{4}{\b} \log{|z|} \label{top22}
\er
where $\s$ ($\s = \arg{z}$)  and $|z|$ are the argument and modulus of the complex parameter $z$, respectively, presented in (\ref{zz1})-(\ref{zk12}). Remarkably, from the results (\ref{top11}) and (\ref{top22}) one can write the next  suggestive formula
\br
e^{i \b (\Phi(+\infty)-\Phi(-\infty) )}  &=& e^{i \b (\Phi_1 (+\infty)-\Phi_1 (-\infty)) } e^{-\b (\Phi_2 (+\infty)-\Phi_2 (-\infty) )} \\
&=& z^{\pm 4}.
\er 
This implies that 
\br
\label{topzz}
\Phi(+\infty)-\Phi(-\infty) = \pm \frac{4}{i \b} \log{z}. 
\er
Next, we will regard the fist order differential equation (\ref{equi1}) as  an equivalence relationship between the topological charge density associated to the complex  scalar and the spinor current density defined by a sum of the spinor bilinear terms. So, integrating (\ref{equi1}) in the whole real line, one can define the next topological charges 
\br
[\Phi_1 + i \Phi_2]\Big|_{x= -\infty}^{x= +\infty} &=&-\frac{\b}{2} \int^{+\infty}_{-\infty} ( \eta_{A} \xi_A +  \eta_{B} \xi_B)\, dx,\\
{\cal Q}_1 &\equiv & \frac{\b}{2\pi} [\Phi_1( +\infty) - \Phi_1(  -\infty)] \label{Q1}\\
{\cal Q}_2 &\equiv & \frac{\b}{2\pi} [\Phi_2(  +\infty) - \Phi_2(  -\infty)].\label{Q2}
\er
Then, one can write
\br
{\cal Q}_1 &\equiv & \mbox{Re}\Big[-\b^2 \int^{+\infty}_{-\infty} \xi_A \xi_B\, dx\Big]\label{Q11}\\
{\cal Q}_2 &\equiv & \mbox{Im}\Big[-\b^2 \int^{+\infty}_{-\infty} \xi_A \xi_B\, dx\Big]\label{Q22}
\er
where the relationship (\ref{xieta1}) between the $\xi$ and $\eta$ fields has been used.

In some cases  the solutions $\Phi_{1,2}$ exhibit vanishing topological charges. In fact, for $\a_2 = \a_1$ ($\arg{z} =0$) one has $\Phi_1 (+\infty)=\Phi_1 (-\infty)$, while for $|b|=|a|$ ($|z|=1$) one gets  $\Phi_2 (+\infty)=\Phi_2 (-\infty)$. These would imply ${\cal Q}_{1} =0$ or ${\cal Q}_{2} =0$ in (\ref{Q1})-(\ref{Q2}), respectively.
 
Next, let us discuss the analytical properties of the multi-valued function $\l$ associated to the complex plane defined by the variable $z$. The so-called uniformization variable $z$ has been introduced for the energy $\l$ and wave number $k$ in (\ref{lvr}). This parametrization avoids the introduction of the Riemann surface consisting of two sheets with relevant branching points as mentioned above. Notice that the relationship (\ref{lmk1}) will hold for any arbitrary complex number $z$ in the  entire complex plane excluding the origin, i.e.  $z \in \IC \backslash \{0\}$. 

From  (\ref{lvr}) one can show the next useful identity 
\br
\label{z2}
\l & = & -\frac{M}{2} (z + z^{-1})\,\,\,\mbox{and}\,\,\,\, k = \frac{i M}{2} (z-z^{-1})\,\,\, \ \Rightarrow \,\,\,z^2 = \frac{\l + i k}{\l - i k},\,\,\,\,  (M \neq 0).
\er    
Notice that the phase and modulus of the uniformization complex variable $z$ encode the topological charges of the two types of solitons as shown in (\ref{top11}) and (\ref{top22}). So, we write the complex variable $z$ in polar form such that the modulus $|z|$ and $\arg{z}$  depend on the real and complex parts of $\l$ and $k$ as follow
\br
\label{polar}
z = \(\frac{|\l + i k|}{|\l - i k|}\)^{1/2}\, e^{i \s},\,\,\,\,\,\, \s =\arctan\Big\{ \frac{|k|^2-|\l|^2 + |\l - i k| |\l + i k|}{\l\, k^{\star} + \l^{\star} k}\Big\}.
\er
Then, using the results (\ref{top11}) and (\ref{top22}) together with the polar decomposition  (\ref{polar}) into the formulas  (\ref{Q1}) and (\ref{Q2}) for the topological charges one has
\br
\label{Q1lk}
{\cal Q}_1 &=& \pm  \frac{2}{\b \pi} \, \arctan\Big\{ \frac{|k|^2-|\l|^2 + |\l - i k| \, |\l + i k|}{\l\, k^{\star} + \l^{\star} k}\Big\}, \\
{\cal Q}_2 &=& \pm  \frac{1}{\b \pi}  \log{\frac{|\l + i k|}{|\l - i k|}}. 
\label{Q2lk}
\er
Therefore, one has a family of localized continuum states with complex energy $\l$ and wave number $k$ defining two types of solitons with topological charges provided in (\ref{Q1lk})-(\ref{Q2lk}). The second topological soliton with charge ${\cal Q}_2$ owns its existence purely to the non-Hermicity and non-linearity of the model. In fact for $\l$ and $k$ real one has a trivial topological charge, i.e. from (\ref{Q2lk}) one has that ${\cal Q}_2 =0$. So, the appearance of the second soliton shows an interesting phenomena in which the interplay between non-linearity and non-Hermicity of the model gives rise to a new state provided by the composite topological soliton $\Phi$ with components (\ref{phi1sol}) and (\ref{phi2sol}) and the next spinor bound states 
\br
\label{xiA}
\xi_A &=&|c| e^{-i(k_2 x + \a_2-\zeta_1)}\,  \frac{e^{k_1 x} \( |b| e^{2 k_1 x} + e^{i(2k_2 x + \a_2)}\)}{(1+|b| e^{2 k_1 x})^2- 2 |b| e^{2 k_1 x}(1-\cos{(2 k_2 x+ \a_2)})},\\
\label{xiB}
\xi_B &=&|d| e^{-i(k_2 x + \a_1-\zeta_2)}\, \frac{e^{k_1 x}  \( |a| e^{2 k_1 x} + e^{i(2 k_2 x + \a_1)}\)}{(1+|a| e^{2 k_1 x})^2- 2 |a| e^{2 k_1 x}(1-\cos{(2 k_2 x+ \a_1)})},
\er 
where the tau functions  (\ref{tau01x})-(\ref{tauabxx}) have been inserted into (\ref{tauab}) in order to write the spinor solutions (\ref{xiA})-(\ref{xiB}). The components $\eta_A$ and $\eta_B$ can be shown to satisfy the relationships (\ref{AB1}), so from (\ref{xiA})-(\ref{xiB}) one can get $\eta_A = \xi_B$ and $\eta_B= \xi_A$. 

The Fig. 2 shows a representative plot of the above type of localized states. In this figure the spinor bound states become localized inside the corresponding topological complex solitons supported by the both scalar field components. Qualitatively inspecting the Fig. 2 one can argue that for a general set of soliton parameters the solution does not present any of the ${\cal P T}-$like symmetries discussed in the Table 2, since the scalar fields do not have definite parities.    

\begin{figure}
\centering
\includegraphics[width=6cm,scale=1, angle=0,height=3cm]{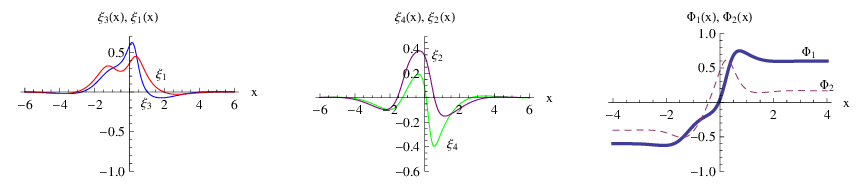}
\parbox{5in}{\caption{A representative of the continuum of bound states (CBM). Topological complex soliton $\Phi=\Phi_1+i \Phi_2$ and bound state  $\xi^T =(\xi_3+i \xi_4,\, \xi_1+i \xi_2)$  for $|a|=2,|b|=1.5, |c|=1, |d|=1.3, \zeta_1 = 1.2, \zeta_2=1.5, \a_1=1.1,\a_2=1.7, k_1=1, k_2=1,\b=1,|\rho|=1.5, M_1 = 4.04,\,M_2=1.49$.}}
\end{figure}

Moreover, the $x-$integrated current density in (\ref{equi1}) can be written as
\br
\label{bilint1}
 \int_{-\infty}^{+\infty} [ \eta_{A} \xi_A +  \eta_{B} \xi_B]\, dx &=& \frac{|c| |d|}{|b|} \Big\{\frac{(\log{\frac{|a|}{|b|}})^2+(\a_1-\a_2)^2}{(k_1^2+k_2^2) \Big[(\frac{|a|}{|b|})^2+1- 2 \frac{|a|}{|b|} \cos{(\a_1-\a_2)}\Big]}\Big\}^{1/2}\, \, e^{i \D}\\
\label{bilint2}
\D&\equiv& \zeta_1 + \zeta_2 + \d_o +\a_o + \nu,\,\,\,\,\, \,\,\,\,\,\,\,\,\,\,
\nu  = - \arctan{[\frac{k_2}{k_1}]},\\
\label{bilint22}
\d_o&=& \arctan{ \Big\{\frac{\a_1-\a_2}{\log{|a|}-\log{|b|}}} \Big\},\,\,\a_o = \arctan{\Big\{\frac{|b| \sin{\a_2}-|a| \sin{\a_1}}{|a| \cos{\a_1}-|b| \cos{\a_2}}}\Big\}.\er

So, the above integration (\ref{bilint1}) provides a complex number such that the real part will be proportional to ${\cal Q}_1$ and the imaginary part to ${\cal Q}_2$ as in (\ref{Q11})-(\ref{Q22}), respectively. The equations (\ref{Q1lk})-(\ref{Q2lk}) together with (\ref{bilint1}) will provide two relationships for the parameters. Additional relationships which  must be satisfied by the parameters come  from the biorthogonality relationship $(\bar{\xi}_{\l}|\xi_{\l})$ in (\ref{bior}) which will be computed below for any complex eigenvalue $\l$. 

Notice that for the Hermitian case one has $|a|=|b|,\,k_2 =0$ and since $\tau_0 = \tau_1^{\star}$ in (\ref{tau01x}) one can take $\a_1 = \frac{\pi}{2},\,\a_2 = \frac{3\pi}{2}$, and then $\nu =0,\,\d_o = \frac{\pi}{2},\,\a_o = \frac{\pi}{2}$. So, it must be $\zeta_1+\zeta_2 = n \pi$ in order to fix $e^{i \D} = \pm 1$, such that the charge ${\cal Q}_2$ vanishes.  

So, the right-eigenvector fermion and the complex topological soliton $\Phi$ currents, which satisfy the equivalence relationship  (\ref{equi1}), as well as  the relevant equivalence of the left-eigenvector fermion current to the scalar field $\bar{\Phi}$ topological current in (\ref{equiv22}), can be regarded as the analogs of the biorthogonal generalized bulk-boundary correspondence in non-Hermitian condensed matter systems \cite{kunst}.  

\subsubsection{A physical interpretation of the non-Hermiticity: kink effect on spinor bound states}

One can see that an essential distinction between a Hermitian Toda model coupled to fermion and
its non-Hermitian counterpart is the degrees of freedom that one can access to; i.e. nonunitary transformations forbidden in the Hermitian system can be realized in its non-Hermitian version. In fact, the change in the spectrum from real parameters $(\l_1, k_1, M_1)$ to the complex ones ($\l, k , M$) increases the number of parameters that
describe the system. Since topology mainly depends on the underlying manifold, non-Hermiticity is expected
to alter the topological classification of the solitons related to the scalar field components $\Phi_1$ and $\Phi_2$. So, in order to classify the topological sectors of the NH model, the pseudo-chiral and anti-pseudo-Hermitian symmetries will play an important role. 

Next let us discuss the physical origin of the non-Hermiticity. Some physical effects which introduce non-Hermiticity into Hermitian systems are,  gain and loss, dissipation of energy in open quantum systems, finite life-time of resonances, etc. In this context, it is usual to ascribe some physical meaning to the real and imaginary parts of a complex eigenvalue, i.e. the real
part codifies the excitation energy, whereas the imaginary part encrypts the decay or amplification
rate.  However, in our constructions above, one can see that the modulus and phase of the uniformization complex parameter $z$ in (\ref{z2}) acquire some physical meaning. In fact, as mentioned above, the phase, $\arg{z}$, and modulus, $|z|$, encode the topological charges of the scalar fields $\Phi_1$ and $\Phi_2$ in (\ref{top11}) and (\ref{top22}), respectively. Remarkably, the circle $|z|=1$ will correspond to the trivial topological charge associated to the field $\Phi_2$. As we will discuss below, the bound states in the Hermitian system (for $\Phi_2=0$)  lie on the unit circle $|z|=1$. So, one can argue that the topological soliton associated to $\Phi_2$ is purely of NH origin. An alternative interpretation into the realm of non-Hermitian physics, which involves neither gain nor loss, has recently been introduced in \cite{franca}, in which the complex
eigenvalues emerge from the amplitudes and phase differences of waves back scattered from the boundary
of insulators. 

Below we discuss the physical origin of the non-Hermiticity of the model in the framework of the tau function approach. We will show that the complex eigenvalues emerge due to the interactions of the spinor components with the non-vanishing component $\Phi_2$ of the scalar field. Firstly, let us identify the parameter values associated to the Hermiticity condition (\ref{hermicond}) by inspecting the vanishing of $\Phi_2$ for the solution in (\ref{phi2sol}) and the identity $\mbox{Im}{(\xi_A \xi_B)}=0$ for the spinor solutions in (\ref{xiA})-(\ref{xiB}). So, the solutions (\ref{phi2sol}) and (\ref{xiA})-(\ref{xiB}) satisfying the Hermiticity conditions (\ref{hermicond}) possess the parameter values  
\br 
|\rho|=1,\,\,\,|a|=|b|,\,\,\,\, \a_1 = -\a_2,\,\,\,\,\zeta_1= -\zeta_2,\,\,\,\, k_2=0,\,\,\,|z|=1. \label{condtau}
\er
Then, the wave number $k$ and the energy eigenvalue $\l$ become real numbers. In fact, using the parameter values (\ref{condtau}) and the relationships (\ref{lvr}) one can get
\br 
k_1 &=& - M_1 \sin{\s},\,\,\,\,\,\,\l = \pm M_1 \cos{\s},\,\,\,\,\,\,z = e^{i \s},\\
M_2 &=& - \frac{k_2}{|z|}\, \mbox{csc}(\s), \\
&=& 0.
\er 
So,  in order to describe the non-Hermitian sector of the model one must require the condition  
\br
\label{k2m2}
k_2 \neq 0,\,\,\,\,\,\,M_2 \neq 0,\,\,\,\,\,\,\,\, |z| \neq 1,\,\,\,\,\, \l \in \IC,
\er 
in the tau function construction developed above.

Next, let us examine the asymptotic behavior of the spinor components interacting with the scalar field and the role played by the parameters $M_2$ and $k_2$ as the spinor field interpolates from $x=-\infty$ to $x=+\infty$ in the presence of the nontrivial configuration $\Phi_2$, which arises for the non-Hermitian conditions imposed on the parameters in (\ref{k2m2}). So, considering $k_1>0$ in (\ref{xiA})-(\ref{xiB}) one can write the following asymptotic quantities 
\br
\xi_A(x \rightarrow -\infty) &\Rightarrow &c\, e^{k_1 x} ,\\
\xi_A(x \rightarrow +\infty) &\Rightarrow & \frac{c}{b} e^{-ik_2 x }\,  e^{-k_1 x},
\er
and 
\br
\xi_B(x \rightarrow -\infty) &\Rightarrow & d \, e^{k_1 x} ,\\
\xi_B(x \rightarrow +\infty) &\Rightarrow & \frac{d}{a} e^{-ik_2 x}\,  e^{-k_1 x}.
\er
One notices that the asymptotic forms of the relevant spinor bilinear $\xi_A \xi_B$ changes from  $-\infty$ to $+\infty$ as
\br
\nonumber
\xi_A \xi_B (x \rightarrow -\infty)   &\Rightarrow& \xi_A \xi_B (x \rightarrow +\infty) \\
 c d\, e^{2 k_1 x}  &\Rightarrow& \frac{cd}{a b} \,\, e^{-2 (k_1+i k_2) x }  = \frac{64 cd}{c^2 \widetilde{c}^2 \b^4} \frac{|k|^4}{|M|^2}\, e^{i(4\d_k-2\d_m)}\,\, e^{-2k x}, \label{phfac12}
\er
where
\br
\label{dkm1}
M\equiv |M| e^{i\d_m},\,\,\,k \equiv |k| e^{i\d_k},\,\,\,\, \d_m \equiv \mbox{arctan}(\frac{M_2}{M_1}),\,\,\,\d_k \equiv \mbox{arctan}( \frac{k_2}{k_1}),
\er
and the constant factor $\frac{1}{a b}$ has been computed from (\ref{para1})-(\ref{lvr}). Note that $\xi_A \xi_B$ in (\ref{phfac12}) has developed the phase factors $ e^{i(4\d_k-2\d_m)}\,e^{-2 i k_2 x}$ as it interpolates from $x=-\infty$ to $x=+\infty$ in the presence of the scalar field $\Phi$ with non-vanishing component $\Phi_2$. Remarkably, the emerging phase factors carry the complex components $k_2$ and $M_2$ which characterize the NH sector of the model. In this way, we have identified a physical origin of the non-Hermiticity property of the model as the appearance of a phase factor and the scale factor $\sim \frac{1}{ab}$ in the amplitude of the spinor bilinear in the  presence of the scalar field with non-vanishing imaginary component $\Phi_2$. An analogous property has been reported in the context of condensed matter in \cite{franca} as mentioned above.

Moreover, the bilinear $\widetilde{\psi}^T \psi$ can be written as a time independent expression as  
\br
\label{current111}
\widetilde{\psi}^T \psi  &=& - \frac{\b}{2} [(e^{i \l t}\eta_A) (e^{-i \l t} \xi_A) +  (e^{i \l t}\eta_B) (e^{-i \l t} \xi_B)] \\
&=& - \b \xi_A \xi_B\label{current222}.
\er
This quantity enters the equivalence relationship between the spinor current density and the topological charge density associated to the complex  scalar  (\ref{Q1})-(\ref{Q2}), with  topological charges defined in (\ref{Q11})-(\ref{Q22}) as the integral of the real and imaginary components of (\ref{current222}). So, the non-Hermiticity of the topological and spinor density currents equivalence in the model can be traced to the mechanism discussed above, as an alternative interpretation which involves neither gain nor loss.

\section{Real eigenvalues and  anti-linear ${\cal P}{\cal T}$ and $\g_5{\cal P}{\cal T}$ symmetries}
\label{sec:real}

Summarizing our results, so far, one has that the tau-function Ansatz (\ref{tau01})-(\ref{tauab}) together with  (\ref{tau01x})-(\ref{tauabxx}) provide the solution $\{\Phi,\xi,\eta\}$($\eta =\Gamma \xi$) which realizes the pseudo-chiral and anti-pseudo-Hermitian symmetries of the NH system defined by the Hamiltonians $H$ and $\widetilde{H}$. In fact, in order to realize the pseudo-chirality one can see in the Table 1 that once the right-eigenvectors $\{\xi_{\l},\eta_{\l}\}$ are known one can construct the left-eigenvectors as $\bar{\xi}_{\l} = {\cal G} \xi_{-\l}$ and $\bar{\eta}_{\l} = {\cal G} \eta_{-\l}$. Likewise, the anti-pseudo-Hermitian symmetry is realized by constructing the right and left eigenvectors of $H^{\star}$ and $\widetilde{H}^{\star}$, respectively, as $\chi_{\l} =-\xi^{\star}_{-\l^{\star}}$ and $\hat{\chi}_{\l} = -{\cal G} \xi^{\star}_{-\l^{\star}}$, and supplemented by $\phi = \Gamma \chi_{\l}$ and $\hat{\phi} = \Gamma \hat{\chi}_{\l}$ for the relevant eigenvectors of $\widetilde{H}^{\star}$.    

In this section and in the sections below, we will be interested in searching for scalar and spinor solutions with real eigenvalues. So, we search for${\cal A}$-symmetric Hamiltonians under anti-linear operators ${\cal A}$, i.e. $[{\cal A}, X]=0\,( X \in \{H, H^{\star}\}$). We have seen above that the model exhibits the anti-linear ${\cal P}{\cal T}$ and $\g_5{\cal P}{\cal T}$ symmetries under special circumstances; for example in the cases C and E of Table 2, respectively,  in which the scalar field components exhibit the parities $(-, +)$. So, in this case one would resort to the known anti-linear symmetry argument in order to determine the unbroken ${\cal A}$-symmetric phase of the model with real eigenvalues. As is usual in quantum and quantum field theoretical models the broken regime must be discarded, as infinite gain in time is not physically acceptable. 

The reality of the spectrum can be explained  by a standard argument assuming an antilinear symmetry of $H$, see e.g. \cite{fring1}. Assume that there exists an antilinear operator ${\cal A}$ satisfying
\br
\label{Ah1}
[H, {\cal A} ] = 0,
\er
and
\br
\label{Axi1}
{\cal A} \xi = e^{i\zeta} \xi,
\er
with $\xi$ denoting the eigenvectors of $H$ and the eigenvalues of ${\cal A}$ being pure phases. Note that ${\cal A}$ is an involution, i.e. ${\cal A}^2 = I$. From (\ref{Ah1})-(\ref{Axi1}) one can conclude  that the eigenvalues of $H$ are real since one has 
\br
\label{reald1}
e^{i\zeta} \l \xi = e^{i\zeta} H \xi = H {\cal A}  \xi = {\cal A} H \xi = {\cal A}\, \l\, \xi = \l^{\star}  {\cal A} \xi =   \l^{\star} e^{i\zeta} \xi \rightarrow \l \in \IR.
\er
When only the relation (\ref{Ah1}) holds and (\ref{Axi1}) does not hold, i.e. ${\cal A} \xi \neq e^{i\zeta} \xi $, the ${\cal A}-$symmetry is spontaneously broken and some of the eigenvalues emerge in complex conjugate pair. In order to see this, let us assume 
\br
\label{Axi12}
{\cal A} \xi_{(1)} = \xi_{(2)} ,\,\,\,\, H \xi_{(1)}  = \l_{(1)}  \xi_{(1)} ,\, H \xi_{(2)}  = \l_{(2)}  \xi_{(2)} ,\er
then one has
\br
\label{complex12}
\l_{(1)}  \xi_{(1)}  = H  \xi_{(1)}  = H  {\cal A} \xi_{(2)}    =  {\cal A} H \xi_{(2)}   =  {\cal A} \l_{(2)}  \xi_{(2)}   = \l_{(2)} ^{\star}  {\cal A}\xi_{(2)}   = \l_{(2)} ^{\star}  \xi_{(1)}  \rightarrow   \l_{(1)}  = \l_{(2)} ^{\star} .
\er
These different results allow  one to distinguish three distinct regimes: 

i) the $ {\cal A} -$symmetric regime with real eigenvalues (\ref{reald1}) when  (\ref{Ah1})-(\ref{Axi1}) hold,  

ii) the spontaneously broken $ {\cal A}$ regime with complex conjugate pairs of eigenvalues (\ref{complex12}) when  (\ref{Ah1}) and (\ref{Axi12}) hold, and 

iii) the ${\cal A}-$broken regime with complex and unrelated
eigenvalues such that also property (\ref{Ah1}) does not hold. 

Note that the discussions above can directly be reproduced for the Hamiltonian $H^{\star}$, since according to Table 2 this Hamiltonian also exhibits the  $ {\cal A}-$symmetries, and then satisfies (\ref{Ah1}) in each case. In fact, the condition (\ref{Axi1}) can be verified since from Table 1 the eigenvector $\chi$ is related to the eigenvector $\xi$ as $\chi_{\l} = - (\xi_{\l^{*}})^{\star}$. So, one has
\br
H \xi_{\l} = \l \, \xi_{\l} \rightarrow  H \xi_{\l^*} &=& \l^{\star} \, \xi_{\l^*}\\
H^{\star} (\xi_{\l^*})^{\star} &=& \l \, (\xi_{\l^*})^{\star}\\
H^{\star} \chi_{\l} &=& \l \, \chi_{\l}
\er
and
\br
{\cal A} \xi_{\l} = e^{i\zeta} \xi_{\l} \rightarrow {\cal A} \xi_{\l^*} &=& e^{i\bar{\zeta}} \xi_{\l^*}\\
 {\cal P} (\xi_{\l^*})^{\star} &=& e^{i\bar{\zeta}} \xi_{\l^*}\\
{\cal P}{\cal T} (\xi_{\l^*})^{\star} &=& e^{-i\bar{\zeta}} (\xi_{\l^*})^{\star}\\
{\cal P}{\cal T} \chi_{\l} &=& e^{-i\bar{\zeta}} \chi_{\l}.
\er
Then, the eigenvector $\chi_{\l}$ of $H^{\star}$ is also an eigenvector of ${\cal P}{\cal T}$ operator with eigenvalue $e^{-i\bar{\zeta}}$. Similar steps can be followed in the case $\g_5 {\cal P}{\cal T}$. Below, we will check the above statements by identifying the anti-linear operators ${\cal A}$ and Hamiltonians $H$ and $H^{\star}$ satisfying (\ref{Ah1}) for relevant scalar fields and the associated spinor eigenstates.   

Moreover, as we have seen above the model presents a rich variety of topological solitons and related bound states without presenting definite parities under the space inversion operator ${\cal P}$ for the scalar field components. For example, inspecting qualitatively the Fig. 2 one can see that the scalar solitons do not exhibit definite parities. So, some questions arise regarding the conditions under which the NH model (\ref{lag1}) would present real eigenvalues and the existence of points for pseudo-Hermiticity breaking. We examine these issues firstly by considering the dispersion relation (\ref{lmk1}). So, taking  $M = |M| e^{i \d_m}$ and $k= |k| e^{i \d_k}$  in (\ref{lmk1}) one can write
\br
\label{dmdk1}
\l^2 &=& e^{i(\d_m+\d_k)} \Big[ |M|^2 e^{i(\d_m-\d_k)} - |k|^2 e^{-i(\d_m-\d_k)}\Big],\er
and
\br
\l &=& \pm |M| \, e^{i(\d_m+\d_k+\phi)/2} \left\{[\frac{|k|^2}{|M|^2}-1]^2 + 2 \frac{|k|^2}{|M|^2} \sin^2{(\d_k-\d_m)}\right\}^{1/4},\label{dmdk2}
\er
with
\br
\label{phi11}
\tan{\phi} \equiv [\frac{1+\frac{|k|^2}{|M|^2}}{1-\frac{|k|^2}{|M|^2}}] \tan{(\d_m-\d_k)}
\er
In order to get a real eigenvalue $\l \in \IR$ one must write the condition 
\br
\label{realpara1}
\d_m+\d_k+\phi = 2\pi n,\,\,\,\, n\in \IZ.
\er   
So, from (\ref{phi11}) and (\ref{realpara1}) one gets the eigenvalue reality condition as 
\br
\label{EPs1}
 \tan{(\d_m)}= (\frac{k_1}{M_1})^2 \tan{(\d_k)}.
\er
Note that the real spectrum satisfying (\ref{EPs1}) is part of a larger spectrum containing
complex-conjugate pairs. Then, one can argue that for the parameter values such that $\d_m+\d_k+\phi \neq 2\pi n $, some
 of the eigenvalues start to become complex, i.e. the parameters not satisfying (\ref{EPs1}) give rise to complex eigenvalues. One notices in (\ref{dmdk2}) that the zero mode can be defined by the condition $|M|=|k|$,\,\,$\d_m -\d_k =  n \pi,\,n \in \IZ$. In particular, the zero mode satisfies (\ref{EPs1}) such that $k = \pm M$. 
 
Notice that in (\ref{dmdk2}) the composed parameters $\frac{|k|^2}{|M|^2}$ and $\d_{km}\equiv \d_{k}-\d_{m}$ play a fundamental role, since at $\frac{|k|^2}{|M|^2}=1$ and $\d_{km} = n \pi,\,n \in \IZ$, one recovers the zero mode $\l = 0$. So, it is convenient to introduce a new complex  $\omega-$plane parametrization in order to analyze  the singular points, such as the the zero modes and the EPs. So, let us consider
\br
\label{om1}
\omega &\equiv& \frac{k}{M}\\
&=& |\omega| e^{i \d_{km}},\,\,\,\,\,\, \d_{km} \equiv \d_k-\d_m,\label{om11}
\er
where the phases $\d_k$ and $\d_m$ have been defined in (\ref{dkm1}).  
Next, let us examine the dispersion relation (\ref{lmk1}) in the complex plane $\om$. So,  (\ref{lmk1})  can be written as
\br
\label{lom1}
\l = \pm |M|  e^{i(\d_m + \pi/2)} \, \sqrt{\om^2 - 1}.
\er
In fact, the eq.  (\ref{lom1}) is equivalent to the expression in (\ref{dmdk2}). It is well known that exceptional points arise on the energy Riemann surface corresponding to branch point singularities where two (or more) states become degenerate. So, let us examine the singular points of the energy eigenvalue defined in (\ref{lom1}). The energy
surface becomes non-analytic at the branch points $\om = \pm 1\, (|w|=1 , \d_{km} = n \pi,\, n \in \IZ)$ where one has the square-root singularities. Then, one can define the points $w=\pm 1$ as the exceptional points (EPs) on the complex eigenvalue surface. So, one can consider two branch cuts running along the real line in the complex plane $\om$ between the points $-\infty$ and $-1$ and the other one, between the points $1$ and $+\infty$. At the EPs the eigenvectors also coalesce and satisfy the relationship $(\bar{\xi}_{\l_{EP}}|\xi_{\l_{EP}}) =0$ which will be computed in the section \ref{sec:biorth}. In the Fig 3. we plot the eigenvalue $\l$ ($-\l$) $\mbox{vs}\,\, \d_{km}$ of the eq. (\ref{dmdk2}) with real component as red (orange) lines and imaginary component as blue (green) lines. With parameters $|\om|=\frac{|k|}{|M|}=1,\,M_1=1, \d_m = 0.15.$ The dashed lines represent the real eigenvalues $\l =\pm \l_1+ i \l_2\, (\l_2=0)$. Notice that the points of the dashed line must satisfy the real condition requirement for the relevant parameters (\ref{EPs1}). The bold dots at $\d_{km} = n \pi\,(n\in \IZ)$ represent the eigenvalues coalescence points (EPs). Notice that the same pattern repeats at a period of $4\pi$.  

\begin{figure}
\centering
\includegraphics[width=8cm,scale=1, angle=0,height=8cm]{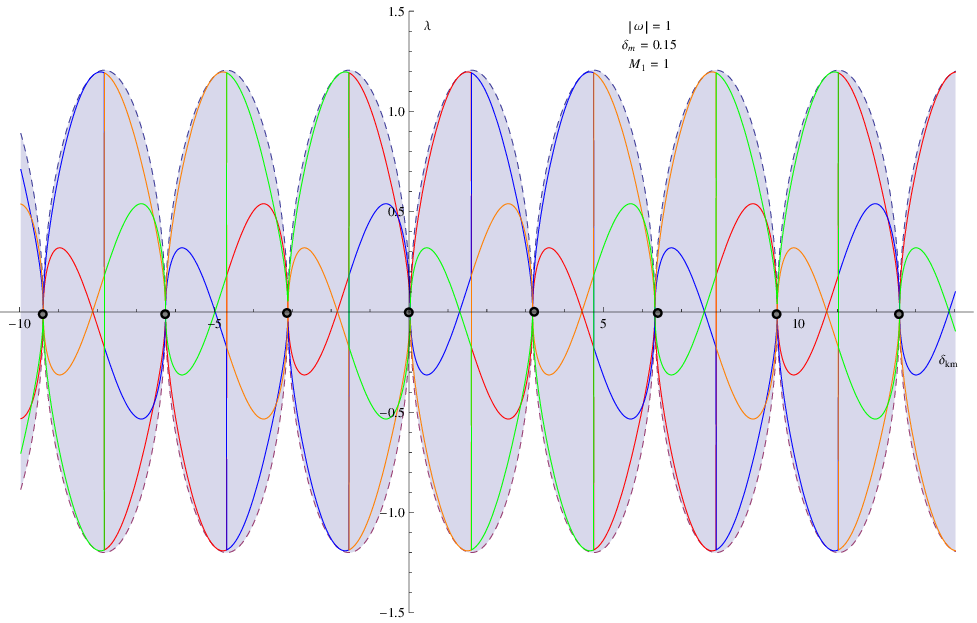} 
\parbox{5in}{\caption{The eigenvalue $\l$ ($-\l$) $\mbox{vs}\,\, \d_{km}$ with real component as red (orange) lines and imaginary component as blue (green) lines of (\ref{dmdk2}). For $|\om|=\frac{|k|}{|M|}=1,\,M_1=1, \d_m = 0.15.$ The dashed lines represent real eigenvalues $\l =\pm \l_1+ i \l_2\, (\l_2=0)$. The dots at $\d_{km} = n \pi\,(n\in \IZ)$ represent the eigenvalues coalescence points (EPs). Notice that the same pattern repeats at a period of $4\pi$.}}
\end{figure} 

Next, let us identify  the EPs points in the $z$-parametrization. From (\ref{lvr}) and (\ref{lom1}) one can see that the mapping from the $z-$plane to the $\om-$plane is defined by
\br
\label{tr22}
\om^2 = \frac{1}{2} - \frac{1}{4} (z^2 + \frac{1}{z^2}).
\er
So, the exceptional points at $ \om= \pm 1$ correspond to the points $z = \pm i$. Notice that these points also define the zero modes. Then, one can argue that the dispersion relation $\l(k, M)$ can be suitably parametrized by $w$ in order to define the EPs and the zero modes; whereas, the the $z-$parametrization is suitable to exhibit the relevant zero-modes and the topological charges of the scalar fields. Bellow, we will see that the $z-$parametrization is also useful in the computations of the winding numbers of eigenvalues and the biorthogonality relationships of the left and right eigenvectors.      

An exceptional point separates a ${\cal P}{\cal T}$-symmetric regime with real eigenvalues
from the ${\cal P}{\cal T}$-broken regime with complex eigenvalues by varying one or more of the parameters of the model. So, we analyze the ${\cal P}{\cal T}$ symmetry phase transition in the $\om$ parametrization $\om = \om_1 + i \om_2 $ of the dispersion relationship (\ref{lom1}). Let us consider the next two cases. 
 
{\bf Case A.} \,\,  $\d_m \in \{ 0, \pm \pi\}  \rightarrow  M = M_1.$

So, by setting this value into  (\ref{lom1})  one has
\br
\label{lom1A}
\l = \pm\, i \, M_1 \, \sqrt{(\frac{k_1+ i k_2}{M_1})^2 -1}.
\er
One can split it in  two sub-cases .  {\bf Case A1.} $\frac{k_1}{M_1} \rightarrow 0$. So, one has
\br
\label{lom1A1}
\l = \pm M_1 \, \sqrt{(\frac{k_2}{M_1})^2  + 1}.
\er 
It gives  a completely real eigenvalue and an unbroken ${\cal P}{\cal T}$ symmetry. The next {\bf Case A2.} $\frac{k_2}{M_1} \rightarrow 0$. It provides the expression
 \br
\label{lom1A2}
\l = \pm i\, M_1 \, \sqrt{(\frac{k_1}{M_1})^2  - 1}.
\er 
Notice that in (\ref{lom1A2}) the energy becomes real when $|\frac{k_1}{M_1}| <1$; $|\frac{k_1}{M_1}| > 1$ is
the region of broken ${\cal P}{\cal T}$ symmetry and $|\frac{k_1}{M_1}| < 1$  is the region of unbroken ${\cal P}{\cal T}$ symmetry. In the Fig. 4 we present a representative plot of the function (\ref{lom1A2}). The real parts $\l_1$ are shown as dashed lines and the imaginary parts $\l_2$ are shown as solid lines. The ${\cal P}{\cal T}$ phase transition occurs at the EPs for $|\frac{k_1}{M_1}| = 1$. At the points $\frac{k_1}{M_1}=\pm 1$ the real eigenvalues merge and become complex.

\begin{figure}
\centering
\includegraphics[width=8cm,scale=1, angle=0,height=5cm]{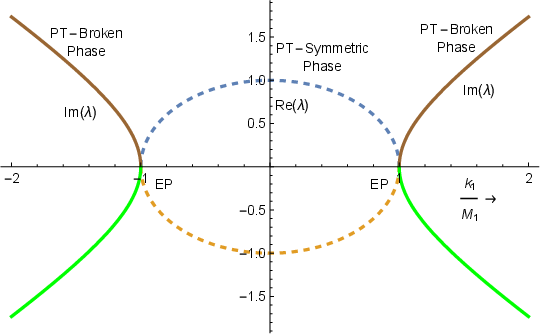} 
\parbox{5in}{\caption{{\bf Case A2.} ${\cal P}{\cal T}$ symmetry phase transition at the EPs $\frac{k_1}{M_1} = \pm 1$. It is set  $M_1 =1$ in (\ref{lom1A2}). The eigenvalue changes with $\frac{k_1}{M_1}$. The real parts $\l_1 = \mbox{Re}(\l)$ are shown as dashed lines and the imaginary parts $\l_2= \mbox{Im}(\l) $ are shown as solid lines.}}
\end{figure}

{\bf Case B.} \,\,  $\d_m = \pm \frac{\pi}{2} \rightarrow  M = i M_2.$

So, setting this value into  (\ref{lom1})  one has
\br
\label{lom11}
\l = \pm M_2 \, \sqrt{(\frac{k_2 - i k_1}{M_2})^2 -1}.
\er
Likewise, one can split it in  two sub-cases .  {\bf Case B1.} $\frac{k_2}{M_2} \rightarrow 0$. So, one has
\br
\label{lom1B1}
\l = \pm i\, M_2 \, \sqrt{(\frac{k_1}{M_2})^2 + 1}.
\er
Then, it gives  a purely imaginary eigenvalue and a broken ${\cal P}{\cal T}$ symmetry. The next case {\bf Case B2.} $\frac{k_1}{M_2} \rightarrow 0$. So, one has
\br
\label{lom1B2}
\l = \pm M_2 \, \sqrt{(\frac{k_2}{M_2})^2 -1}.
\er
So, the behavior of the eigenvalue depends on the parameter $\frac{k_2}{M_2}$. Notice that in (\ref{lom1B2}) the energy becomes real when $|\frac{k_2}{M_2}| >1$; $|\frac{k_2}{M_2}| < 1$ is
the region of broken ${\cal P}{\cal T}$ symmetry and $|\frac{k_2}{M_2}| > 1$  is the region of unbroken ${\cal P}{\cal T}$ symmetry. In the Fig. 5 we present a representative plot of the function (\ref{lom1B2}). The real parts $\l_1$ are shown as dashed lines and the imaginary parts $\l_2$ are shown as solid lines. The ${\cal P}{\cal T}$ phase transition occurs at the EPs for $|\frac{k_2}{M_2}| = 1$. At the points $\frac{k_2}{M_2}=\pm 1$ the real eigenvalues merge and become complex.  

\begin{figure}
\centering
\includegraphics[width=8cm,scale=1, angle=0,height=5cm]{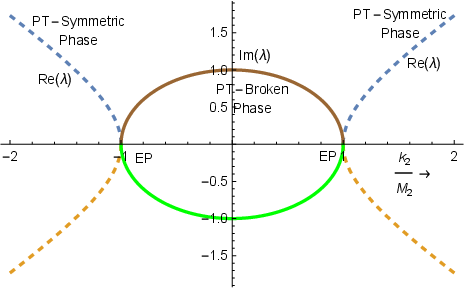} 
\parbox{5in}{\caption{{\bf Case B2.} ${\cal P}{\cal T}$ symmetry phase transition at the EPs $\frac{k_2}{M_2} = \pm 1$. It is set  $M_2 =1$ in (\ref{lom1B2}). The eigenvalue changes with $\frac{k_2}{M_2}$. The real parts $\l_1 = \mbox{Re}(\l)$ are shown as dashed lines and the imaginary parts $\l_2= \mbox{Im}(\l) $ are shown as solid lines.}}
\end{figure} 
 
It is also interesting to discuss the real component of the energy eigenvalue $\l$ in the complex plane $z$ parametrization of (\ref{lvr}). So, let us examine the family of curves in the complex planes $z$ which give rise to the real energy eigenvalues. So, in the complex $z-$plane parametrization one can write
\br
\label{lambreal}
\l_1 (\s) &=& \pm |M|\,  \sqrt{\frac{(1+(\frac{M_2}{M_1})^2) (\sin{\s})^2}{(\tan{\s})^2 - (\frac{M_2}{M_1})^2}},\,\,\,\,\,\,\,\,\,\l_2 = 0,\\
|z|^2 &=& \frac{\tan{\s} - \frac{M_2}{M_1}}{\tan{\s} + \frac{M_2}{M_1}}, \,\,\,\,\,\,\,\, \s \in \, (-\pi+|\d_m|\,,\, -|\d_m|) \cup  (|\d_m| \, , \, \pi-|\d_m|),\,\,\,\,\, z = |z| e^{i \s}.
\label{lambreal1}
\er
Then, the NH model (\ref{lag1}) supports real eigenvalues for the parameter subspace defined by the relationships (\ref{lambreal})-(\ref{lambreal1}).  The Fig. 6 shows a family of curves in the complex plane $z$ defined by (\ref{lambreal})-(\ref{lambreal1}) and  parametrized by $\frac{M_2}{M_1}$. Notice that the family of curves intersect at the singular point $z=0$ ($\l_1 \rightarrow +\infty$), and at the points of the complex plane $z$ corresponding to the zero mode $\l(z=\pm i) = 0$. Note that these last points also represent the EPs. Remarkably, one has  $|z| \in (0,+\infty)$ for a fixed value of $\frac{M_2}{M_1}$. However, some regions  are prohibited for the phase $\s$ in the complex plane, as $\s \notin [-|\d_m| , |\d_m|]$ and $ \s \notin \{[-\pi ,  -\pi +|\d_m|] \cup [\pi-|\d_m| ,  \pi ]\}$.  In the  limit $\frac{M_2}{M_1} \rightarrow 0$ one has a unit circle at the center, and then $|z|=1$, which implies $\l_1 = \pm |M| \cos{\s}\, (\s \in [-\pi, + \pi]).$ 

The Fig. 7 shows $\l_1\, \mbox{vs}\, \s$ for $\frac{M_2}{M_1}=0.05$ (left) and for $\frac{M_2}{M_1}\rightarrow 0 $ (right) defined by (\ref{lambreal}). As mentioned above, some intervals for $\s$ are missing since $\l_1$ is not defined on these regions (Fig. 7 left). Notice that the pseudo-chiral symmetry is maintained since the real eigenvalues come  in pairs  $\pm \l_1$.

So, our findings add new results to the  current
exploration whether the non-Hermitian topological models 
can exhibit real energy spectra. The studies in the literature along this line 
have used mainly the ${\cal P }{\cal T}$ symmetry in order to preserve the real
spectra when non-Hermiticity is relatively weak. Recently, it has been put forward a systematic way to construct non-Hermitian topological systems exhibiting energy spectra that are always real, regardless of weak or strong non-Hermiticity \cite{long}. The classification of the eigenspace of the anti-pseudo-Hermitian Hamiltonians $H$ and $\widetilde{H}$, such as the regions of real and complex eigenvalues and  the points of eigenvector degeneracies, have become important in order to identify the points of pseudo-Hermiticity breaking. Recently, a new treatment to deal with pseudo-Hermiticity breaking has been developed in \cite{melkani} by classifying the eigenspace of pseudo-Hermitian matrices using an intertwining operator such as $\s_{-}$ in (\ref{psc2})-(\ref{psc2i}) and showing that such symmetry breaking occurs if and only if eigenvalues of opposite kinds collide on the real axis of the complex eigenvalue plane. Below, we will provide the conditions for symmetry breaking to occur by considering  the properties of the eigenstates of the relevant anti-linear operators ${\cal A}$ ($ {\cal A} \in \{{\cal P }{\cal T}, \g_5{\cal P }{\cal T}\})$ and the pseudo-Hermitian Hamiltonian $H$ presented above when discussing the properties (\ref{Ah1})-(\ref{complex12}), following the known approach in \cite{ashida1, bender1, fring1}.   
  
\begin{figure}
\centering
\includegraphics[width=8cm,scale=1, angle=0,height=7cm]{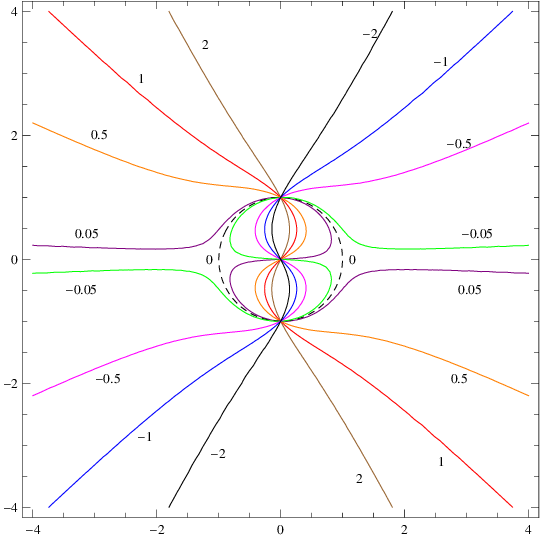}
\parbox{5in}{\caption{Family of curves in the complex plane $z$ for $M_1=1$ defined by the real energy condition (\ref{lambreal1}). For these points $\l$ becomes real $\{\l_1 \in \IR , \, \l_2=0\}$. They are parametrized by $M_2=0, \pm 0.05,\pm 0.5, \pm 1, \pm 2$. The unit circle at the center (dashed line) emerges in the limit $\frac{M_2}{M_1} \rightarrow 0$, and so, in this case $|\s| \leq \pi$. The points $z = \pm i$ represent the EPs and $z=0$ is a pole such that $\l_1 \rightarrow +\infty$.}}
\end{figure}

\begin{figure}
\centering
\includegraphics[width=8cm,scale=1, angle=0,height=4cm]{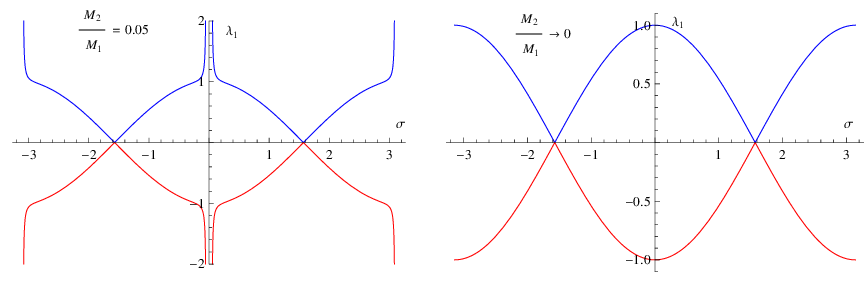}
\parbox{5in}{\caption{The plot $\l_1\, \mbox{vs}\, \s$ for $\frac{M_2}{M_1}=0.05$ (left) and for $\frac{M_2}{M_1}\rightarrow 0 $ (right) defined by (\ref{lambreal}).   $\l_1$ is not defined on the intervals $|\s| \leq \d_m$ and $\{-\pi \leq \s \leq -\pi +\d_m\} \cup \{\pi-\d_m \leq \s \leq \pi \}\,  (\d_m \approx 0.05)$ of the left figure.}}
\end{figure}

\subsection{${\cal P}{\cal T}$ and $\g_5{\cal P}{\cal T}$ symmetric solutions}
\label{subsec:pt}

We study the ${\cal P}{\cal T}$ and $\g_5{\cal P}{\cal T}$ symmetric sectors of the model. First, we find the symmetric Hamiltonians, as indicated
in Table 2, and then examine  the symmetry breaking which will be visible in the properties of the relevant spinor eigenfunctions to which they are related.

Next,  we consider  ${\cal P}{\cal T}$  symmetric Hamiltonians associated to scalar solitons with definite parities under a shifted space-inversion transformation. So, let us  assume the relationship $\d = \a_1-\a_2$ and define the parameters  
\br
\label{parcent1}
x_{0} \equiv \frac{\a_1+\a_2}{4 k_2},\,\,\,\,\, r \equiv e^{\frac{k_1}{k_2}(\a_1+\a_2)},\,\,\,\, k_2 \neq 0,
\er
where $x_0$ indicates the position of the soliton center.  So, taking into account (\ref{phi1sol})  and (\ref{parcent1}) one can write 
\br
\Phi_1(x-x_0) &=& \frac{2}{\b} \arctan{\left\{ \tan{(\frac{\d}{2})} \Big[\frac{\sinh{(2 k_1 (x-x_0))}+ \frac{e^{-\frac{1}{2}\log{r}}(|b|-|a|)}{2 \sin{(\d/2)}} \sin{(2 k_2 (x-x_0))}}{ \cosh{(2 k_1 (x-x_0))} + \frac{e^{-\frac{1}{2}\log{r}}(|b|+|a|)}{2 \cos{(\d/2)}}  \cos{(2 k_2 (x-x_0))} } \Big]\right\}} \label{phi1imp} 
\er
It is clear that this solution satisfies
\br
\label{par12i}
 \Phi_1(-(x-x_0)) = - \Phi_1(x-x_0). \er
So, it is an odd function under a shifted parity inversion symmetry ${\cal P}_s$ defined in  (\ref{pt10})-(\ref{pt2}). 

Next, we examine  the scalar component $\Phi_2$. So, the attempt to construct the $\Phi_2$ component from the solution (\ref{phi2sol}) with even parity under the symmetry above merely leads to a constant solution
\br
\label{phi2i}
\Phi_2 (x) = -\frac{1}{\b} \log{ |\rho| }.  
\er
Therefore, this complex scalar solution with the components (\ref{phi1imp}) and (\ref{phi2i}) makes the Hamiltonian to be ${\cal P}_s{\cal T}$  symmetric according to (\ref{ptsym}). However, the spinor solution is not an eigenstate of the  ${\cal P}_s{\cal T}$ operator. In fact, from (\ref{xiA})-(\ref{xiB}) one has
\br
{\cal P}_s{\cal T}  (\xi)\,  \neq \, \xi, 
\er
provided that $k_1\neq 0, k_2 \neq 0$. This case corresponds to the properties (\ref{Ah1}) and (\ref{Axi12})-(\ref{complex12}). So, the ${\cal P}_s{\cal T}$ symmetry is spontaneously broken and some of the eigenvalues will emerge in complex conjugate pairs.  

For completeness, we construct a component $\Phi_2$ with odd parity. So, for the special choice of parameters $\a_2=-\a_1,\,\,\, x_0 =0,\,|a|=|b|=1,\,|\rho| =1$ one can get 
\br
\Phi_2 (x) = -\frac{1}{\b} \log{\left\{\frac{\cosh{(k_1 x)} +\cos{(k_2 x-\a_1)} }{\cosh{(k_1 x)} +\cos{(k_2 x+\a_1)}}\right\}}
\er 
By inspecting the above expression one can see that $\Phi_2(-x) = - \Phi_2(x)$, i.e. it is an odd function with the soliton centered at the origin $x_0 =0$. However, in this special case, the first component $\Phi_1(x)$ in (\ref{par12i}) remains odd.  So, it corresponds to the {\bf case D} of the symmetry classification (see also Table 2) with scalar parities $(- , -)$ and so, the symmetry (\ref{symm22}) must hold, i.e. $[H, {\cal P} {\cal G}] = 0$. Since this symmetry is not an antilinear symmetry it belongs to the ${\cal A}-$broken regime iii) of the discussion above, i.e. since (\ref{Ah1}) also does not hold it is a system with complex and unrelated eigenvalues.  

However,  in the case above with scalar parities $(- , \mbox{undefined})$ one can regard the Hamiltonian to be {\sl partially} ${\cal P}{\cal T}$ symmetric according to the definition (\ref{pt11})-(\ref{pt22}).  It has been reported that the ${\cal P}{\cal T}$ symmetry is not the sole symmetry allowing real eigenvalues  for non-Hermitian
Hamiltonians. For example for a system of chains of $N$ coupled harmonic oscillators the entire spectrum may be completely real in some regions  of the parameter space, even though the system is only partially ${\cal P}{\cal T}$
symmetric, such that a phase transition point
exists beyond which the energy spectrum is only partially
real \cite{benderpar}. Another instance is the non-Hermitian version of the SSH model, which does not respect
${\cal P}{\cal T}$ symmetry; nevertheless, it has a real energy spectra in certain parameter subspace. Due to lack of ${\cal P}{\cal T}$  symmetry, its energy spectrum and exceptional points depend on the choice of certain 
boundary conditions \cite{amir}. Those interesting {\sl partial} ${\cal A}$-symmetric regimes deserve a further consideration and they are beyond the scope of the present paper.   

Below, we will consider the following special cases:  the case  $k_2 =0$ in this section and  the case $k_1 =0$ in section \ref{sec:class2}. In the both cases we will find unbroken ${\cal A}-$symmetric sectors with real eigenvalues. 

Next, we examine the $\g_5 {\cal P}{\cal T}$ symmetric sector of the solutions. Let us assume $k_2 =0$ and 
\br
\label{parcent122}
|a|=|b|=1,\,\, \a_1-\a_2 = \d,\,\,\, k_2=0\, \rightarrow \d_k = \arctan{(k_2/k_1)} = 0.
\er
So, taking into account  (\ref{phi1sol})  and (\ref{parcent122}) one can write
\br
\Phi_1(x) &=& \frac{2}{\b} \arctan{\left\{ \tan{(\frac{\d}{2})} \Big[\frac{\sinh{(2 k_1 x)}}{ \cosh{(2 k_1 x)} + \frac{\sin{(\a_1-\d/2)}+\cos{(\a_1-\d/2)}}{\cos{(\d/2)}}} \Big]\right\}} \label{phi1imp1} \\
\Phi_2(x) &=& - \frac{1}{\b} \log{\Big[|\rho| \frac{1+ 2 \cos{(\a_2)} e^{2 k_1 x} +e^{4 k_1 x} }{1+ 2 \cos{(\a_1)} e^{2 k_1 x} +e^{4 k_1 x} }\Big]} \label{phi2p2}
\er
It is clear that these functions satisfy
\br
\label{par12ii}
 \Phi_1(-x) &=& - \Phi_1(x),\\
\label{par12iii}
 \Phi_2(-x) &=& \Phi_2(x).
 \er
Moreover, from the spinor solutions (\ref{xiA})-(\ref{xiA}) and taking into account (\ref{parcent122}) one can check the following  symmetry
\br
\label{ptA1}
{\cal P}{\cal T} ( \xi ) & = & Z \, \xi ,\,\,\,\,\,\,  \xi = \(\begin{array}{c}
 \xi_{A} \\ 
\xi_{B}
\end{array}\) \\
Z  &  \equiv & \(\begin{array}{cc}
  e^{-2i \zeta_1} & 0 \\ 
0 &   e^{-2i \zeta_2}
\end{array}\)
  \label{ptB1}.
\er
Note that assuming $\zeta_2 = \zeta_1 - \pi/2$ one gets $Z = e^{-2i \zeta_1}  \g_5$; so, from (\ref{ptA1})-(\ref{ptB1}) and taking into account $\g_5^2 =1$, one has
\br
\label{ptA11}
\g_ 5 {\cal P}{\cal T} ( \xi ) & = & e^{-2i \zeta_1} \, \xi.
\er
Remarkably, one notices that $\xi$ is an eigenvector with a pure phase eigenvalue of the anti-linear operator  $\g_5 {\cal P}{\cal T}$. This operator commutes with $H$ as defined in (\ref{c11})  for $M \in \IR\, (\d_m =0)$ in the {\bf Case C} of symmetry classification, i.e.
\br
\label{symmg5pt}
[\g_ 5 {\cal P}{\cal T}\,, \, H] =0.
\er
So, due to the relationships (\ref{ptA11})-(\ref{symmg5pt}) one can regard the model as a $\g_ 5 {\cal P}{\cal T}$-symmetric one, such that  (\ref{Ah1})-(\ref{Axi1}) hold corresponding to the case i) with real eigenvalues (\ref{reald1}). In fact, the reality condition (\ref{EPs1}) is trivially satisfied in this case since one has $\d_k=\d_m =0$. So, the eigenvalue equation corresponds to (\ref{lom1A2}) in the {\bf Case A2} of eigenvalue classification.  In (\ref{lom1A2}) the energy will become real for $|\frac{k_1}{M_1}| <1$; so, $|\frac{k_1}{M_1}| > 1$ is the region of broken $\g_5 {\cal P}{\cal T}$ symmetry and $|\frac{k_1}{M_1}| < 1$  is the region of unbroken $\g_5 {\cal P}{\cal T}$ symmetry, as presented in  the Fig. 4.   

However, if one considers (\ref{ptA1})-(\ref{ptB1}) (with $\zeta_2 \neq \zeta_1 - \pi/2$) and (\ref{symmg5pt}), one can say that the system has undergone a spontaneously broken $\g_ 5 {\cal P}{\cal T}$-symmetry. In fat, this corresponds to the case ii) of spontaneously broken $ {\cal A}$-symmetric regime with complex conjugate pairs of eigenvalues (\ref{complex12}) when  (\ref{Ah1}) and (\ref{Axi12}) hold.

\subsection{Spectral topology and winding number}
\label{subsec:spectopo}

One can analyze the spectral properties of the model by  examining the behavior of the curves on the complex plane defined by the uniformization variable $z$ and the energy $\l$ and wave number $k$ expressions in (\ref{z2})-(\ref{polar}). Notice that for the bound states in the Hermitian case one has the real parameters $M=M_1$, $\l = \l_1$. So,  when $k=k_1 (k_1 \in \IR)$ one has the unit circle $|z| =1$ (see Fig.1). This implies that the bound states lie on this unit circle ($\l_1= \pm \sqrt{M_1^2- k_1^2}$). However, for purely imaginary $k = i k_2$  one notices that the scattering states are allowed for $z \in \IR$ ($\l_1= \pm \sqrt{M_1^2 + k_2^2}$). The so called half-bound states at the edges of the continuous spectrum are given at $\l_1 = \pm M_1 \rightarrow   z = \pm 1$. So, the thresholds of the energy gap of the free massive fermions are at $\l_1 = \pm M_1$. 
 
In the non-Hermitian case one has a complex eigenvalue spectrum,
which must be studied on the complex plane,  and so leads to the emergence of eigenvalue topology, or spectral topology. So, one has a continuum of bound states (CBM), such that the energy $\l$ is defined as an analytical function of the complex variable $z$.  This additional input of spectral topological concepts is a unique feature in non-Hermitian systems. In this context, the zero-energy points at $ z = \pm i$ in (\ref{z2}) can arise as branch points in the $k-$space eigenvalue manifold as can be written from (\ref{lmk1})
\br
\l = \pm \sqrt{M^2 - k^2}.
\er  
As we will see below, the branch points ($k = \pm M$) exhibit exotic topological phenomena associated with the winding of eigenvalues and eigenvectors. The complex plane defined by the uniformization variable $z$ turns out to be convenient for performing the relevant calculations in order to study the analytical properties of the energy $\l(z)$ and wave number $k(z)$ functions, i.e. in order to examine the zero modes and the points corresponding to the edges of the model, which we will discuss below. The analytic function $\l(z)$ in (\ref{z2}) has two zeros and one pole, located on  the imaginary axis at the points $z=\pm i  \rightarrow k = \pm M$, each one of first order, and the pole at the origin $z=0$. Next, let us define the winding number
\br
\label{wind1}
W &=& \frac{1}{2 \pi i}\oint_{\G} \, \frac{d\l}{\l}\\
     &=& Z - P\label{wind11}.
\er
In the second line above we have used the Cauchy's argument principle, such that $Z$ is the number of zeros and $P$ is the number of poles of the function $\l$ inside the contour $\G$ (see Fig. 1). The $Z$  and $P$ must be counted with their multiplicities and order, respectively.
It is interesting to compute the winding number of $\l(z)$ in (\ref{lvr}) as it traverses the contour $\G$ encircling the zero modes $\l = 0 $ located at the points $z = \pm i$ and the pole at $z=0$ in the complex plane $z$. So, for the function $\l(z)$ in (\ref{z2}) and using (\ref{wind1}) one has  
\br
\nonumber
W &=& \frac{1}{2 \pi i}\oint_{\G} \, \frac{z^2 - 1}{z (z^2+1)}\, dz\\
&=& 1. \label{wind111}
\er 
Then, one can see that the eq. (\ref{wind11}) is also verified, i.e. $W =1$, since $Z=2$ and $P=1$ are the number of zeros and poles of the function $\l(z)$ inside $\G$.  

\section{Continuum of bound states/solitons (CBM) and extended ELC modes}
\label{sec:class}

In this section we discuss  the general solution presented above and classify it into two types of solutions: The dual continuum of bound states/solitons and the extended modes embedded into the localized states (ELC). Moreover, we also compute analytically  the biorthogonality relationships between the relevant left and right eigenvectors.

\subsection{Continuum of bound state modes/solitons (CBM)}
\label{sec:class1}

In a Hermitian $sl(2)$ Toda system coupled to fermion, which has recently been presented in \cite{jhep22}, the in-gap and BIC bound states have quantized energies, whereas the free states form a continuum. We have shown above how this principle fails for non-Hermitian Toda system coupled to fermions, by analyzing the non-Hermitian Hamiltonian with a complex Toda field and the properties of pseudo-chirality and pseudo-Hermicity. In fact, the biorthogonal eigenstates, which we may call ``continuum bound state modes'' (CBM), have spatial envelopes and form a continuum filling the complex energy plane. We have shown that their strong coupling sector host a continuum complex Toda field solitons, such that  the fermion eigenstates are localized inside the topological solitons with topological invariant quantities (\ref{bilint1})-(\ref{bilint2}). In the strong coupling sector the family of complex solitons can be dubbed as continuum of topological solitons (CTS). These type of novel features of a non-Hermitian Hamiltonian, i.e. the spatially localized eigenstates at every value of  the energy in the complex energy plane $\l$, have recently been uncovered also in \cite{wang2}.

The spinor bound states requires the condition $k_1 \neq 0$ in the next relationships
\br
\label{kk1}
k_1 &=& \pm  \frac{\sqrt{2}}{2} \sqrt{\D + (\widetilde{M}- \widetilde{\l})}\\
k_2 &=& \pm \frac{\sqrt{2}}{2} \sqrt{\D - (\widetilde{M}- \widetilde{\l})}
\label{kk2}
\\
\D &\equiv& \sqrt{|M-\l|^2|M+\l|^2}\\
\widetilde{M} &\equiv&  M_1^2- M_2^2,\\         
\widetilde{\l}  &\equiv&   \l_1^2- \l_2^2,
\er 
Notice that the parameters defined above satisfy  $\D > |\widetilde{M}- \widetilde{\l}|$ without any further restriction on the complex plane $k= k_1 + i k_2$ for a fixed value of the complex mass  parameter $M= M_1 + i M_2$. So, for each complex  eigenvalue $\l \in \IC$ one has a  continuum of complex scalar solutions with wave numbers $k_1$ and $k_2$ given in (\ref{kk1})-(\ref{kk2}). As discussed above, a continuum of bound states emerge coupled to a continuous family of topological solitons (see Fig.2)  such that their topological charges depend on the energy eigenvalues as in (\ref{top1})-(\ref{top22}). In addition, the zero mode $\l=0$ implies $k=\pm M$ in (\ref{kk1})-(\ref{kk2}).
\begin{figure}
\centering
\includegraphics[width=8cm,scale=1, angle=0,height=5cm]{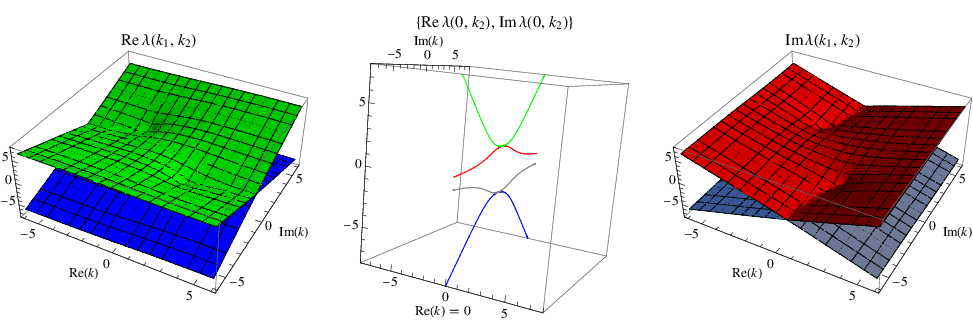}
\parbox{5in}{\caption{The left and right plots show the energy surfaces $\l(k_1, k_2)$ for $M_1 = 2, \,M_2= -2$. The middle figure shows the curves $\l(k_1=0, k_2)$ (line colors distiguish the left(right) figure surface each one belongs to) which correspond to the extended waves in the localized continuum (ELC) states (\ref{phi1e})-(\ref{xbe}). The points of the left and right manifolds (excluding the curves $\l(k_1=0, k_2)$ of the middle plot) represent the localized continuum states (\ref{phi1sol})-(\ref{phi2sol}) and (\ref{xiA})-(\ref{xiB}). Notice the presence of the points $k = \pm M$ (left and right plots) which correspond to EPs $(w=\frac{k}{M} = \pm 1)$ and zero modes $(\l =0)$.}}
\end{figure}

On the other hand, the complex energy components become
\br
\label{l11}
\l_1 &=& \pm \frac{\sqrt{2}}{2} \sqrt{\D + (\widetilde{M}- \widetilde{k})}\\
\label{l22}
\l_2 &=& \pm \frac{\sqrt{2}}{2} \sqrt{\D - (\widetilde{M}- \widetilde{k})}\\
\D &\equiv& \sqrt{|M-k|^2|M+k|^2}\\
&=&  \sqrt{[(k_1-M_1)^2+(k_2-M_2)^2][(k_1+M_1)^2+(k_2+M_2)^2]}\\
\widetilde{M} &\equiv&  M_1^2- M_2^2,\\         
\widetilde{k}  &\equiv&   k_1^2- k_2^2.
\label{l1122}
\er 
From the above equations (\ref{l11})-(\ref{l1122}) one has the following $M= \pm k  \Rightarrow \D = 0$ and $\widetilde{M} - \widetilde{k}= 0$, which imply $\l =0$. This is just the zero mode $\l=0$. In Fig. 8 the left and right plots show the energy surfaces $\l_1(k_1, k_2)$ and $\l_2(k_1, k_2)$ for $M_1 = 2, \,M_2= -2$ in the eqs. (\ref{l11})-(\ref{l22}). The middle figure shows the curves for extended wave states in the localized continuum (ELC) states (\ref{phi1e})-(\ref{xbe})(line colors distinguish the left(right) figure surface each one belongs to). The points of the left and right manifolds (excluding the curves $\l(k_1=0, k_2)$ of the middle plot) represent the localized continuum states (\ref{phi1sol})-(\ref{phi2sol}) and (\ref{xiA})-(\ref{xiB}). The points $k = \pm M$ (left and right plots) correspond to the zero modes $\l = 0$. Note that these points also define the EPs, since according to (\ref{lom1}) one has $w=\frac{k}{M} = \pm 1$.  
 
Moreover, from the complex dispersion relationship (\ref{lmk1}) one can get the next two equations, considering the real and imaginary parts, respectively 
\br
\label{dr1}
k_1^2 - k_2^2-M_1^2+ M_2^2 + \l_1^2-\l_2^2&=&0\\
k_1 k_2 - M_1 M_2 + \l_1 \l_2 &=&0.
\label{dr2}
\er 
These two relationships are equivalent to the next system of equations
\br
\label{dr11}
k_1^2(k_1^2 -M_1^2+ M_2^2 + \l_1^2-\l_2^2) - (M_1 M_2 - \l_1 \l_2)^2&=&0,\,\,\, k_1 \neq 0,\\
k_2^2(k_2^2 +M_1^2- M_2^2 - \l_1^2+\l_2^2) - (M_1 M_2 - \l_1 \l_2)^2&=&0,\,\,\, k_2 \neq 0.
\label{dr22}
\er
Once the values of $M_1$ and  $M_2$ are fixed one has that the equations (\ref{dr11})-(\ref{dr22}) define a couple of families of curves in the pane $( \l_1 , \l_2)$ parametrized by the variables $k_1$ and $k_2$, respectively. The Fig. 9 shows the spectral curves filling some regions of the complex plane corresponding to the family of curves (\ref{dr11})-(\ref{dr22}). Notice the overlap regions around the points $(M_1,M_2) = (\pm2 , \pm2)$ (left Fig.) and $(M_1,M_2) = (\pm 2, \mp 2)$ (right Fig.). In the Fig. 10 we plot the family of curves $(\l_1, \l_2)$, some of which intersect to each other at the origin $\l =0$. Note the coalescence of some states at the EPs $\l =0$. At this point the zero mode eigenvectors become biorthogonal as we will show below in (\ref{nor1EP111}) and  in section \ref{sec:Majo1} the biorthogonal Majorana zero mode-topological soliton duality will be examined.

\begin{figure}
\centering
\includegraphics[width=8cm,scale=1, angle=0,height=6cm]{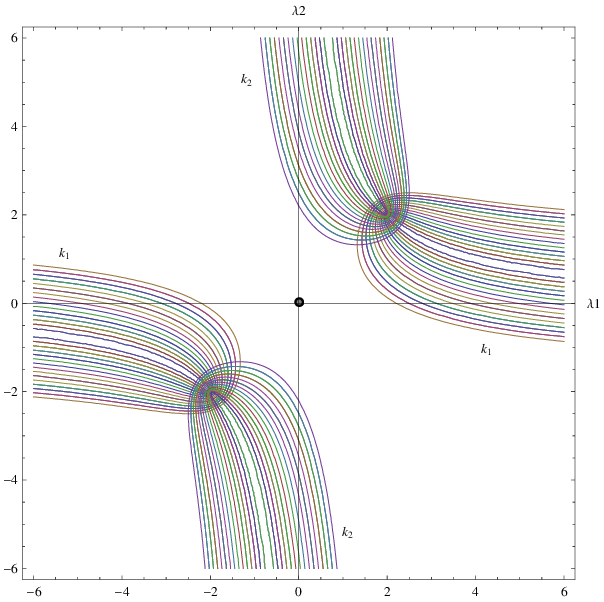}
\includegraphics[width=8cm,scale=1, angle=0,height=6cm]{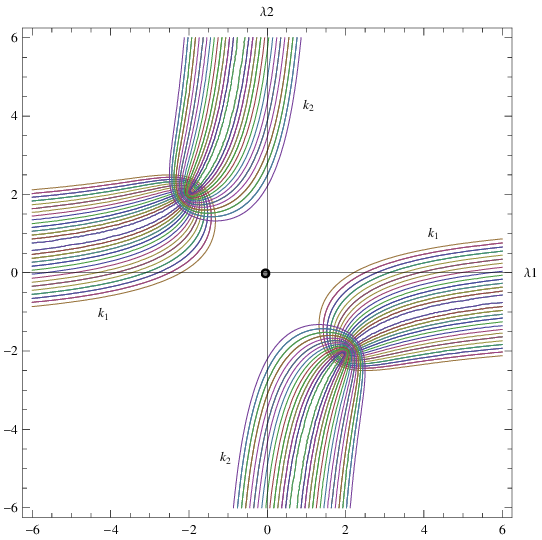}
\parbox{5in}{\caption{Spectral curves on the plane $\l=(\l_1,\l_2)$ for $(M_1,M_2) = (\pm2 , \pm2)$ (left) and $(M_1,M_2) = (\pm 2, \mp 2)$ (right). The $k_1$ and $k_2$ families correspond to (\ref{dr11}) and  (\ref{dr22}), respectively. The figures show the curves for the parameter ranges $ |k_a|<1.5\, (a=1,2)$. The zero modes are at $\l = 0\, (k_1 = M_1, k_2 =M_2)$. The points $\l =(M_1, M_2)$ correspond to the threshold points $k=0$.}}
\end{figure}
The parabolas $\l_1 \l_2 = M_1 M_2 = const.$ in the plane $(\l_1 , \l_2)$ can be obtained when one of the components  $k_1$ or $k_2$ vanishes, and correspond to the extended wave states in the localized continuum for $k_1=0$, or to a subset of the continuum bound states for $k_2 =0$. 

So, the confluence of non-Hermiticity, nonlinearity and spectral topology generate new phenomena such as the non-
Hermitian currents equivalence (\ref{equi1}), the complex valued topological charges (${\cal Q} \equiv {\cal Q}_1 +i  {\cal Q}_2$) and the presence of the winding numbers (\ref{wind111}) related to the spectral topology. The model exhibits the three phenomena unique to non-Hermitian systems: the non-vanishing fermion current, nonzero winding number of energy and the presence of the pair of topological solitons in the model. One can argue that the validity of any one of the three properties is the sufficient and necessary condition for the validity of the other two in this non-Hermitian system. 

In summary, we have shown that the non-Hermitian
Hamiltonian $H$ can allow an uncountably infinite set of bound
states, contrary to the results in \cite{jhep22} in which the in-gap and BIC bound states appear with quantized energies. The biorthogonal continuum bound states possess localized envelopes and
are related to a continuum of topological complex Toda solitons. Regarding the zero modes, one can also argue that the biorthogonal Majorana zero modes are embedded into a  
family of continuum of bound states, with properties different from those studied in their Hermitian counterparts \cite{jhep22}. 

\begin{figure}
\centering 
\includegraphics[width=8cm,scale=1, angle=0,height=6cm]{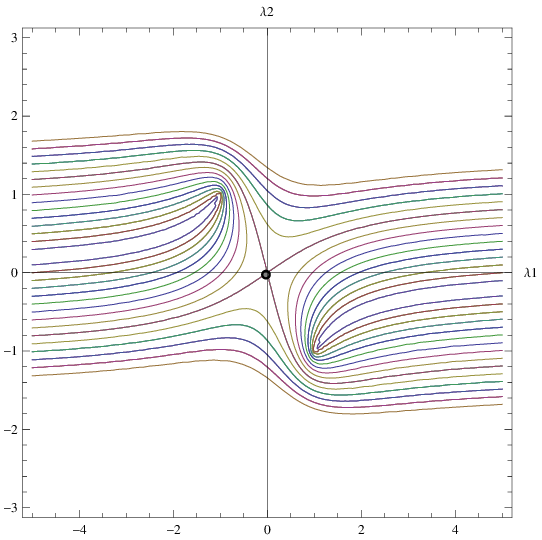}
\includegraphics[width=8cm,scale=1, angle=0,height=6cm]{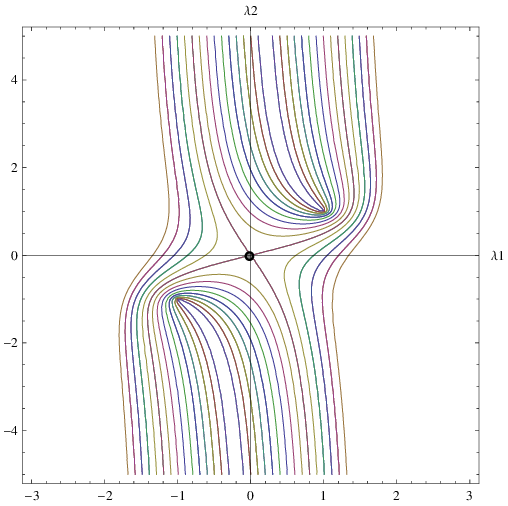}
\parbox{5in}{\caption{The family of curves $(\l_1, \l_2)$ for $M_1= - M_2 =1$ (left) and $M_1= M_2 =1$ (right) in the parameter ranges $|k_{1,\,2}|< 1.5$. Notice the coalescence of some states at the EPs $\l =0$.}}
\end{figure}

\subsection{Extended wave states embedded in the continuum of bound states} 
\label{sec:class2}

Next, we construct the extended wave states localized in the continuum of bound states (ELC). These states can be obtained by setting $k_1=0$ into the tau functions (\ref{tau01x})-(\ref{tauabxx}). So, from (\ref{tau01}) the scalar components become  
\br
\nonumber
\Phi_1 &=& - \frac{2}{\b} \arctan{\Big[\frac{|a| \sin{(2k_2 x + \a_1 - \frac{\d}{2})} - |b| \sin{(2k_2 x + \a_2 + \frac{\d}{2})} + |a||b| \sin{(\a_1-\a_2-\frac{\d}{2})}  - \sin{\frac{\d}{2}}}{|a| \cos{(2k_2 x + \a_1 - \frac{\d}{2})} + |b|  \cos{(2k_2 x + \a_2 + \frac{\d}{2})} + |a||b| \cos{(\a_1-\a_2-\frac{\d}{2})}  + \cos{\frac{\d}{2}} } \Big]},\\
\label{phi1e}
\\
\Phi_2 &=& - \frac{1}{\b} \log{\Big[|\rho| \frac{1+|b|^2 + 2 |b| \cos{(2 k_2 x + \a_2)}}{1+|a|^2 + 2 |a| \cos{(2 k_2 x + \a_1)}}  \Big]}. \label{phi2e}
\er
Similarly, inserting (\ref{tau01x})-(\ref{tauabxx}) into (\ref{tauab}) for $k_1=0\, (\d_k = \pi/2)$  we have the spinor components
\br
\nonumber
\xi_A &=& |c|\, \frac{|b| \cos{(k_2 x + \a_2 - \zeta_1)} + \cos{(k_2 x + \zeta_1)}}{1+ |b|^2 + 2 |b| \cos{(2 k_2 x + \a_2)}}-\\
&& i |c|\, \frac{|b| \sin{(k_2 x + \a_2 - \zeta_1)} - \sin{(k_2 x + \zeta_1)}}{1+ |b|^2 + 2 |b| \cos{(2 k_2 x + \a_2)}},
\label{xae}\\
\nonumber
\xi_B &=&|d|\, \frac{|a| \cos{(k_2 x + \a_1 - \zeta_2)} + \cos{(k_2 x + \zeta_2)}}{1+ |a|^2 + 2 |a| \cos{(2 k_2 x + \a_1)}}-\\
&& i |d|\, \frac{|a| \sin{(k_2 x + \a_1 - \zeta_2)} - \sin{(k_2 x + \zeta_2)}}{1+ |a|^2 + 2 |a| \cos{(2 k_2 x + \a_1)}}.
\label{xbe}
\er
Using the relationships (\ref{AB1}) from (\ref{xae})-(\ref{xbe}) one can get $\eta_A = \xi_B$ and $\eta_B= \xi_A$ with the $\xi_{A, B}$'s given above.

\begin{figure}
\centering
\includegraphics[width=6cm,scale=1, angle=0,height=3cm]{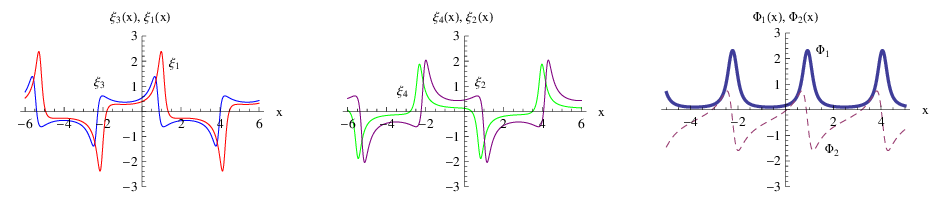}
\parbox{5in}{\caption{A representative of the extended wave states in the continuum of bound states. The components of $\Phi= \Phi_1 + i \Phi_2$ and $\xi^T =(\xi_3+i \xi_4,\, ,\xi_1+i \xi_2)$  for $|a|=1.5,|b|=1.5, |c|=1, |d|=1.3, \zeta_1 = 1.2, \zeta_2=1.5, \a_1=\pi/3,\a_2=\pi/2, k_1=0, k_2=1,\b=1,|\rho|=1.5, M_1 = -0.05,\,M_2=3.9$. Notice the periods $2\pi$ and $\pi$ for the spinors and scalars, respectively.}}
\end{figure}

The Fig. 11 shows a representative plot of this type of extended states. Notice that the spinor waves exhibit a period $\frac{2\pi}{k_2}$, whereas the scalar waves present a periodic profile with period $\frac{\pi}{k_2}$. The middle plot in Fig. 8 shows the curves $\l(k_1=0, k_2)$ which correspond to the ELC extended wave states (\ref{phi1e})-(\ref{xbe}). 

The family of spectral curves for $k_1=0$ in (\ref{dr11})-(\ref{dr22}) can be written as
\br
\label{elc1}
k_2^2 M_1^2 - (\l_1^2 -M_1^2)(\l_2^2  + M_2^2) =0.
\er
The Fig. 12 shows this family of curves parametrized by $k_2$ for $M_1^2=M_2^2 =2$. Notice that these curves lie inside the continuum of bound states in the Fig. 9. So, these states constitute the extended waves in the localized continuum (ELC), in contradistinction to the well known Hermitian bound states in the continuum (BIC), which appear embedded into the continuum of scattering states.
\begin{figure}
\centering 
\includegraphics[width=8cm,scale=1, angle=0,height=6cm]{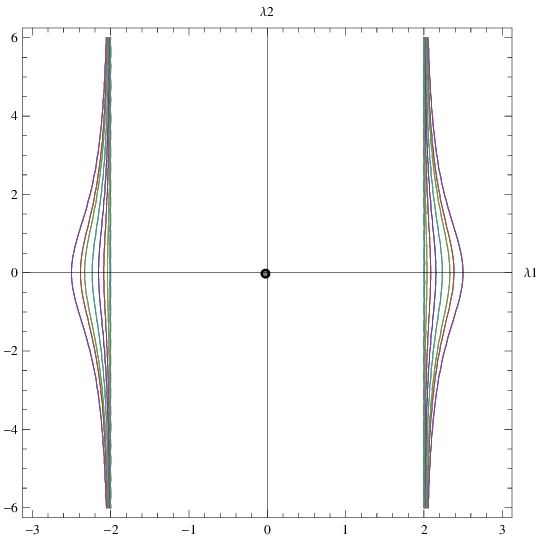}
\parbox{5in}{\caption{The family of curves $(\l_1, \l_2)$ for $k_1=0$ and $M_1^2=M_2^2 =2$ in the parameter range $|k_2|< 1.5$ which correspond to the extended waves in the localized continuum (ELC) states parametrized by $k_2$ (\ref{elc1}). Notice the ELC zero  modes $\l=0\, (k_2= \pm M_2)$. }}
\end{figure}

Next, we examine the ${\cal P}{\cal T} $ symmetric sector of the ELC solutions. Let us assume
\br
\a_1 = -\a_2 = \frac{\d}{2},\, \a_1 = \zeta_2,\, \a_2 = \zeta_1,\,\,   \zeta_{a} = n\pi\, (a=1,2),\,\,\,\,\,\, |a| |b| =1 (|a| \neq |b|).
\er
Then one can show the next definite parity symmetries of the scalar field components in (\ref{phi1e})-(\ref{phi2e})
\br
\label{elcsca1}
\Phi_1(-x) &=& - \Phi_1(x)\\
\Phi_2(-x) & = &  \Phi_2(x). \label{elcsca2}
\er
Then, due to (\ref{elcsca1})-(\ref{elcsca2}) the Hamiltonian can be regarded as ${\cal P}{\cal T} $ symmetric provided that $\d_m = \frac{\pi}{2}$ as defined in (\ref{ptsym}) in the {\bf Case E} symmetry classification,  i.e. one must have $x_0 =0,\, \d_m = \pi/2$ in (\ref{pt10})-(\ref{pt2}). 
Next, one has the ${\cal P}{\cal T} $ symmetry of the spinor solution  (\ref{xae})-(\ref{xbe}) 
\br
\label{symmsp1}
{\cal P}{\cal T}  (\xi) = \xi.
\er
Next, let us examine the reality condition of the eigenvalues. So, due to the ${\cal P}{\cal T} $ symmetry of $H$ and the eigenvalue equation (\ref{symmsp1}) one has that (\ref{Ah1})-(\ref{Axi1}) hold corresponding to the case i) with real eigenvalues (\ref{reald1}). In fact, one can verify that the parameters $\d_m=\d_k= \pi/2$ and $\phi = (2n-1)\pi$ satisfy the eigenvalue reality conditions (\ref{phi11})-(\ref{realpara1}). So, the eigenvalue equation corresponds to (\ref{lom1B2}) in the {\bf Case B2} eigenvalue classification.  In (\ref{lom1B2}) the energy will become real for $|\frac{k_2}{M_2}| >1$; so, $|\frac{k_2}{M_2}| < 1$ is the region of broken ${\cal P}{\cal T}$ symmetry and $|\frac{k_2}{M_2}| > 1$  is the region of unbroken ${\cal P}{\cal T}$ symmetry, as presented in  the Fig. 5. At the points $\frac{k_2}{M_2}=\pm 1$ the real eigenvalues merge and become complex.  

Note that the bound states in the continuum (BIC) are
localized states such that spatially and spectrally overlap with a
continuum of free states. However, those states  do not hybridize with any of the extended or radiating states. It has recently been established that those bound states
remain localized even though they coexist with the continuous spectrum of radiating waves
that can carry energy away, i.e. for energies above or below the threshold of Hermitian Toda coupled to fermion \cite{jhep22} . 
On the  other hand, the extended wave states on the localized continuum defined in (\ref{xae})-(\ref{xbe}) can be regarded as some kind of dual or inverted BIC states due to the non-Hermicity of the model (\ref{lag1}). Recently, these novel type of states in a NH system have been dubbed as extended states in a localized continuum (ELC) \cite{wwang, amir}. 

\subsection{Left and right eigenvectors and biorthogonality relationships}
\label{sec:biorth}

In the NH case there are some types of time-independent inner products \cite{sato, das}. Thus, for the model above one can write: $\eta^T\xi$, $\xi^T\Gamma \xi$, $\eta^T\Gamma\eta$, $\bar{\xi}^{T}\xi$, $\xi^T\cal{G}\xi$,  $\bar{\eta}^T\eta$, $\eta^T\cal{G}\eta$, $\bar{\xi}^T{\cal G}^{-1}\bar{\xi}$ , $\bar{\eta}^T{\cal G}^{-1}\bar{\eta}$, and the analog products associated to the $\chi,\phi,\hat{\chi}, \hat{\phi}$ eigenvectors. In the following, before verifying the biorthogonality property (\ref{bior}), let us compute the NH inner product (\ref{nhinn}) between two right spinors of types $\eta$ and $\xi$ at the point $\l =0$. So, let us consider the inner product 
\br
\label{prod1}
(\eta_{-\l}|\xi_{\l}) = \int^{+\infty}_{-\infty}[ \eta_{-\l,A} \xi_{\l,A} + \eta_{-\l,B} \xi_{\l,B}]\, dx. 
\er 
The soliton parameters in (\ref{para1})-(\ref{lvr}) can be mapped as $\l \rightarrow -\l,\,\, k \rightarrow k$ by making the complex variable transformation $z \rightarrow - \frac{1}{z}$; in addition, through this transformation one can also relate the parameters of  $\eta_{-\l}$ to that of $\xi_{\l}$. So, from (\ref{prod1}) one can write
\br
\label{prod11}
(\eta_{-\l}|\xi_{\l}) &=& \frac{1}{2k} \( \frac{\log{(a'/b)}}{a'-b} +  \frac{\log{(a/b')}}{a-b'}\),\\
&=&  \pm  \frac{M}{k \b^2} \frac{(1-z^2)^2}{\widetilde{c}c z^2 + \widetilde{c}'c'}  \frac{1+z^2}{z} \log{\(\frac{\widetilde{c}c}{\widetilde{c}'c'} z^2\)}. \label{prod111}
\er  
where the parameters $a', b', c', \widetilde{c}'$ and $a, b, c, \widetilde{c}$ correspond to the spinors $\eta_{-\l}$ and $\xi_{\l}$, respectively, according to the notation in (\ref{para1})-(\ref{lvr}). For the zero modes, i.e. $\l =0$ ($z = \pm i$), from (\ref{prod111}) one has
\br
(\eta_{0}|\xi_{0}) &=& 0.
\er 
In the next steps we will compute the biorthogonality relationship (\ref{bior}) for some special cases. So, let us find  the left eigenvectors satisfying the equations (\ref{lefteigen1})-(\ref{equiv22}) for the relevant eigenvalue $\bar{\l}$. Following similar steps  as above we consider the tau functions $\bar{\tau}_{0, 1}$, $\bar{\tau}_{A,  B},\bar{\tau}'_{A, B}$ such that 
\br
\label{tau01f}
e^{i \b \Phi} &=& \bar{\rho}\, (\frac{\bar{\tau}_0}{\bar{\tau}_1})^2,\\
\label{tauabf}
\bar{\xi}_{A} &=& \frac{\bar{\tau}_A}{\bar{\tau}_1},\,\,\,\bar{\xi}_{B} = \frac{\bar{\tau}_B}{\bar{\tau}_0},\,\,\,\bar{\eta}_{A} = \frac{\bar{\tau}'_A}{\bar{\tau}_0},\,\,\,
\bar{\eta}_{B} = \frac{\bar{\tau}'_B}{\bar{\tau}_1},
\er
with $\bar{\rho}$ being a constant complex parameter. 

So, let us assume the following tau functions
\br
\label{tau01xf}
\bar{\tau}_1 &=& 1 + \bar{a}\, e^{2\bar{k} x},\,\,\,\,\bar{\tau}_0 = 1 + \bar{b} \, e^{2 \bar{k} x},\\
\label{tauabxf}
\bar{\tau}_A &=& \bar{c} \,e^{ \bar{k} x},\,\,\,\bar{\tau}_B = \bar{d} \, e^{ \bar{k} x},\\
\label{tauabxxf}
\bar{\tau}'_A &=& \bar{c}' \, e^{ \bar{k} x},\,\,\,\bar{\tau}'_B = \bar{d}'\,  e^{ \bar{k} x},
\er  
with the complex parameters $ \bar{a},\bar{b},\bar{c},\bar{d}, \bar{c}', \bar{d}'$ and $\bar{k}$. Next, making use of these tau functions and substituting the expressions (\ref{tau01f})-(\ref{tauabf}) into  the eigenvalue equations for the left eigenvectors (\ref{lefteigen1}) with eigenvalue $\l$, one can show that the equations are satisfied provided that the following parameter relationships hold
\br
\label{para1f}
\bar{a} &=&  \frac{\bar{c}' \bar{c}}{2M} \frac{\b^2 w^3}{(w^2-1)^2},\,\,\,\,\,\,\,
\bar{d}'= \frac{\bar{c}'\, \bar{c}}{\bar{d}},\,\,\,\,\,\,\,
 \bar{\rho} = i \frac{\bar{c}}{\bar{d}}\, w,\\
\label{para111f}
 \bar{b}  &=&\frac{\bar{c}' \bar{c}}{2M} \frac{\b^2 \, w}{(w^2-1)^2},\,\,\,\,\,
w\equiv -\frac{\bar{\l}}{M} \pm \sqrt{\frac{\bar{\l}^2}{M^2} - 1} \\
\label{lvrf}
\bar{\l}& = & -\frac{M}{2} (w + w^{-1}),\,\,\,\,\,\,\, \bar{k}= \frac{i M}{2} (w-w^{-1}).
\er
Notice that the form of the relationship (\ref{lmk1}) also holds in this case, i.e. $\bar{\l}^2 + \bar{k} = M^2$, and the $\bar{\l}$ and $\bar{k}$ parameters depend on the uniformization parameter $w$  in the same form as in the right eigenvalue case (\ref{lvr}).
 
Next, let us compute $(\bar{\xi}_{\bar{\l}}|\xi_{\l})$ for $\bar{\l} = \pm \l\, $ by using the expressions (\ref{tauab}) and (\ref{tauabf}) and the relevant tau functions (\ref{tau01x})-(\ref{tauabxx}) and (\ref{tau01xf})-(\ref{tauabxxf}), respectively. First, let us consider $\bar{\l} = \l\, (\l \neq 0)$. The transformation $w = \frac{1}{z}$ maps $\bar{\l}(w) \rightarrow \l(z)$ and $\bar{k}\rightarrow  -k$. So, let us compute the next integral in $x$ for $\bar{k}=-k$
\br
\nonumber
(\bar{\xi}_{\l}|\xi_{\l})  &=&  \int_{-\infty}^{\infty} [\bar{\xi}_{\l,A}  \xi_{\l,A} + \bar{\xi}_{\l,B}  \xi_{\l,B}  ]    \, dx\\
 &=&  \int_{-\infty}^{\infty} [\eta_{-\l,A}  \xi_{\l,A} - \eta_{-\l,B}  \xi_{\l,B}  ]    \, dx \nonumber\\
&=& \frac{1}{2 k} \( \frac{\bar{c} c \log{\(\bar{a} b\)}}{\bar{a}b -1} + \frac{\bar{d} d \log{\( \bar{b} a\)}}{a \bar{b}-1}\)
\label{nor1EP}
\\
 &=& 1. \label{nor1}
\er
Then, in order to satisfy the biorthogonality relationship (\ref{bior}) the parameters must satisfy the normalization condition (\ref{nor1}) at the points $z \neq \pm i$. 

Next, let us compute the biorthogonality relationship (\ref{bior}) at the EPs, which in the $z$ plane correspond to the points $z= \pm i$.  So, let us write  (\ref{nor1EP}) as 
\br
(\bar{\xi}_{\l_{EP}}|\xi_{\l_{EP}})
 &=&\frac{1}{2 k} \Big[ (\frac{\bar{c} c}{A-1} +  \frac{\bar{d} d}{A z^4-1}) \log{(A)} + \frac{\bar{d} d}{A z^4-1}  \log{(z^4)}\Big],\,\,\,\,\,\,\,A\equiv \frac{\widetilde{c} c \bar{c}'\bar{c} \b^4}{4 M^2} z^2.\label{nor1EP1}\\
&=& \frac{1}{2 k} \Big[ (\frac{\bar{c} c + \bar{d} d}{A-1}) \log{(A)} \Big]\label{nor1EP11},\\
&=& 0.\label{nor1EP111}
\er  
In order to write (\ref{nor1EP11}) in the above derivations  we have evaluated the r.h.s. of the equation (\ref{nor1EP1}) at the EPs, i.e. $z= \pm i\, (z^4 =1)$, and assumed that $A \neq 1$. Below we compute the $0$ in the r.h.s. of (\ref{nor1EP111}). In order to show the vanishing of the r.h.s. of  (\ref{nor1EP11}) let us impose the pseudo-chirality relationship (\ref{ABl}) to the left and right eigenvectors, i.e. $\bar{\xi}_{-\l,\, A} =\xi_{\l,\, B},\,\, \bar{\xi}_{-\l,\, B} = - \xi_{\l,\, A}$. So, from  the expressions (\ref{tauab}) and (\ref{tauabf}) and the relevant tau functions (\ref{tau01x})-(\ref{tauabxx}) and (\ref{tau01xf})-(\ref{tauabxxf}), respectively, one has the relationships
\br
\label{relat12}
\bar{c} = d,\,\,\bar{a} = a,\,\,\bar{d} = -c,\,\,\bar{b} = b\,\,\,\,\Rightarrow\,\,\,\,\,\,  \bar{c}c + \bar{d}d =0. 
\er 
Then, the $0$ of the r.h.s. in the equation (\ref{nor1EP111}) is obtained by considering the relationship (\ref{relat12}) into the r.h.s. of (\ref{nor1EP11}). We clearly see that  $(\bar{\xi}_{\l_{EP}}|\xi_{\l_{EP}})=0$ at the exceptional points, so we cannot normalize the states at the EPs.

Next,  let us compute (\ref{bior}) for the special case $(\bar{\xi}_{-\l}|\xi_{\l}) =0$. Notice that the transformation $w = -\frac{1}{z}$ maps $\bar{\l}(w) \rightarrow -\l(z)$ and $\bar{k} \rightarrow k$. So, one has $\bar{\l} = -\l$. Let us compute the integral in $x$ for $\bar{k}=k$
\br
\label{ortho1}
(\bar{\xi}_{-\l}|\xi_{\l})  &=&  \int_{-\infty}^{\infty} [\bar{\xi}_{-\l,A}  \xi_{\l,A} + \bar{\xi}_{-\l,B}  \xi_{\l,B} ]    \, dx\\
&=& \frac{1}{2 k} \( \frac{\bar{c} c \log{\(\bar{a}/b\)}}{\bar{a}-b} + \frac{\bar{d} d \log{\(\bar{b}/a\)}}{\bar{b}-a}\)
\label{ortho11}
\er
Therefore, making use of the relationships (\ref{relat12}) into (\ref{ortho11}) one can show that 
\br
\label{xi00v}
(\bar{\xi}_{-\l}|\xi_{\l})  = 0. \er

The zero mode biorthogonality equation (\ref{xi00}) follows by setting $\l = 0$ in the last equation (\ref{xi00v}), i.e. $(\bar{\xi}_{0}|\xi_{0}) =0$ . Notice that for the general case of arbitrary states with $\l_i$ and $\l_j$ ($\l_i \neq \l_j \, \rightarrow \,k_i \neq k_j  $ ),  the $x-$integration $(\bar{\xi}_{\l_j}|\xi_{\l_j}) = \int^{+\infty}_{-\infty} \bar{\xi}_{\l_j}^T \xi_{\l_j}\,dx$, in general, will not provide an exact integral, and one could perform the $x-$integration using numerical methods. A similar procedure as above can be performed in the case of $\eta$ and $\bar{\eta}$ eigenvectors and show $(\bar{\eta}_{\l_i}|\eta_{\l_j}) = \d_{ij}$, except at  $\l=0$.

\section{Biorthogonal Majorana zero mode-topological soliton duality}
\label{sec:Majo1}

Firstly, we find the zero mode {\sl right} eigenvectors. We develop the zero mode construction for a general set of parameters in order to examine the associated symmetries and topological charges at the singular point $\l=0$. 

Let us set $\l =0$ into the system of equations (\ref{tau11})-(\ref{tau44}). So, the tau functions (\ref{tau01x})-(\ref{tauabxx}) satisfy the equations of motion provided that 
\br
\label{parazeromode1}
k &=& \pm M,\,\,\,\,\,
b = -a ,\,\,\,\,\, \rho = \pm \frac{\widetilde{c}}{\widetilde{d}}\,,\,\,\,\,\,
\widetilde{d} = \frac{c \widetilde{c}}{d}.
\er
and the charge densities equivalence equation (\ref{equi1}) for 
\br
\label{parazeromode2}
a &=& \pm i \frac{c d \b^2}{8 M},\\
&=&\frac{|c| |d| \b^2}{8 |M|} e^{i ( \frac{2n+1}{2}\pi - \d_m  + \zeta_1 + \zeta_2)},
\er
without any further restriction. Then, the one-soliton solution satisfy the condition (\ref{equi1}) for the equivalence between the spinor bilinear $\eta^{T}\xi$ and the $x-$derivative of the complex field $\Phi$. An analog relationship in the one-soliton sector has been defined as  the equivalence
between Noether and topological currents in the Hermitian version of the model \cite{npb1, aop1, jhep22}.

For the bove parameters the scalar fields can be written as
\br
\label{zmkink}
\Phi_1  &=& \frac{2}{\b} \arctan{ (\frac{A_o}{B_o})},\\
\nonumber
A_o &\equiv& e^{2 M_1 x} \Big[ 8 |c||d| |M| \b^2 (\sin{\Theta(x)} + \cos{\Theta(x)}) + |c|^2|d|^2 \b^4  \sin^2{\Theta(x)}  \, e^{2 M_1 x} \Big],\\ \nonumber
B_o &\equiv& - 64 |M|^2 - 8 |c||d| |M| \b^2 \sin{\Theta(x)}  e^{2 M_1 x} +  |c|^2|d|^2 \b^4  \sin{\Theta(x)} \cos{\Theta(x)} \, e^{4 M_1 x},\\ \nonumber
\Theta(x) &\equiv& \d_m -\zeta_1-\zeta_2 - 2 M_2 \, x,\,\,\,\,\,\,\, \d_m = \arctan{(\frac{M_2}{M_1})}.
\er
and
\br
\label{phi2mo}
\Phi_2 = \frac{1}{\b}  \log{\Big\{ 1 + \frac{32 |c||d|  |M| \b^2 \sin{\Theta(x)}\, e^{ 2 M_1 x} }{64 |M|^2 - 16 |c||d| |M| \b^2 \sin{\Theta(x)} \, e^{ 2 M_1 x} + |c|^2|d|^2 \b^4  \,e^{ 4 M_1 x} }\Big\} }.
\er
The functions above  represent a family of general zero-mode solutions, with some of them being singular solutions. Next, we examine some localized solutions exhibiting  parity symmetries. Considering the limit $M_2=0\, \rightarrow\, \d_m =0$  one has
\br
\label{zmkinkdm01}
\Phi_1^{h}  &=& -\frac{2}{\b} \arctan{ (\frac{A_h}{B_h})},\\
\nonumber
A_h &\equiv& e^{2 M_1 x} \Big[ 8 |c||d| |M| \b^2 (\cos{(\zeta_1+\zeta_2)} - \sin{(\zeta_1+\zeta_2)}) + |c|^2|d|^2 \b^4  \sin^2{(\zeta_1+\zeta_2)}  \, e^{2 M_1 x} \Big],\\ \nonumber
B_h &\equiv& 64 |M|^2 - 8 |c||d| |M| \b^2 \sin{(\zeta_1+\zeta_2)}  e^{2 M_1 x} +  |c|^2|d|^2 \b^4  \sin{(\zeta_1+\zeta_2)} \cos{(\zeta_1+\zeta_2)} \, e^{4 M_1 x},\\
\Phi_2^h &=& \frac{1}{\b}  \log{\Big\{ 1 - \frac{ 2 \sin{(\zeta_1+\zeta_2)}}{\sin{(\zeta_1+\zeta_2) + \, \frac{|c||d| \b^2}{16 M_1} e^{ 2 M_1 x} + \frac{4 M_1}{|c||d| \b^2}  \,e^{- 2 M_1 x} }}\Big\} }\label{zmkinkdm02}
\er
One can compute the topological charges associated to the solution (\ref{zmkinkdm01})-(\ref{zmkinkdm02}). So, one has  
\br
{\cal Q}_1 &\equiv & \frac{\b}{2\pi} [\Phi_1( +\infty) - \Phi_1(  -\infty)] \label{Q11c}\\
&=& - \mbox{sign}(M_1) \( \frac{\zeta_1+\zeta_2}{\pi}\),\,\,\,(\zeta_1+\zeta_2) \notin  \{ \frac{\pi}{2},\pi \}\\
{\cal Q}_2 &\equiv & \frac{1}{2\pi} [\Phi_2(  +\infty) - \Phi_2(  -\infty)].\label{Q22c}\\
&=& \frac{2}{\pi} \log{(1)} \\
&=& 0.
\er 
Let us examine the parity symmetry of the solution. Taking $|M| = \frac{|c||d| \b^2}{8}$ one has 
\br
\Phi_2^h &=& \frac{1}{\b}  \log{\Big\{ 1 - \frac{ 2 \sin{(\zeta_1+\zeta_2)}}{\sin{(\zeta_1+\zeta_2)} + \,  \cosh{( 2 M_1 x)}}\Big\} },\label{zmkinkdm02par}
\er
which defines an even function  $\Phi_2^h (-x) = \Phi_2^h(x) $. In addition, taking the values $\zeta_1+ \zeta_2 =\{ \frac{\pi}{2}, \pi\}$ one can show 
\br
\label{kink1zm}
\Phi_1^{h}  &=& \frac{2}{\b} \arctan{ [ e^{\pm \frac{1}{4} |c| |d| \b^2 x}]}.
\er
This solution represents a sine-Gordon type kink (antikink) soliton with asymptotic values $\b \Phi_1(+ \infty) = \pi$, $\b \Phi_1(- \infty) = 0$ [$\b \Phi_1(+ \infty) = 0, \b \Phi_1(- \infty) = \pi$] which define the topological charges ${\cal Q}_1  = \mbox{sign}(M_1) \,\frac{1}{2}$. These charges can be computed using the eq.  (\ref{Q1}).  Notice that for $\zeta_1+ \zeta_2 = \pi $ the field $\Phi_2^h$ in (\ref{zmkinkdm02par}) becomes a trivial solution.

So, for the set $\zeta_1+ \zeta_2 \in \{ \frac{\pi}{2}, \pi\}$, the field components  satisfy the parity symmetries
\br
\label{parity11}
\b \Phi_1^h (-x) &=& - \b \Phi_1^h(x)  + \pi,\\
\b \Phi_2^h (-x) &=&  \b \Phi_2^h(x). 
\label{parity21}
\er
In the case $\d_m =0$ the Hamiltonian becomes ${\cal P} {\cal T} $ symmetric, i.e.
\br
\label{pth1zm}
{\cal P} {\cal T} H  ({\cal P} {\cal T})^{-1} = H
\er 
since the matrix component $i M e^{i \b \Phi^h}$ of $H$ is ${\cal P} {\cal T} $ symmetric, i. e.
\br
\label{dels10}
{\cal P} {\cal T} \left\{ i |M| e^{i \b (\Phi_1^h(x) + i \Phi^h_2(x))}\right\}  &= & - i |M| e^{i \b (\Phi^h_1(x) + i \Phi^h_2(x) + \pi)},\,\,\,\,\,\,\,\,\, M = |M| e^{i \d_m},\,\,\,\,\d_m =0 \\
&=& i |M|   e^{i \b (\Phi_1^h(x) + i \Phi_2^h(x))}, \label{dels1}
\er
where the parity symmetries (\ref{parity11})-(\ref{parity21}) have been used.   
 
The above  soliton and bound state solutions of the model, provided the parameters (\ref{parazeromode1})-(\ref{parazeromode2}) relationships are satisfied, allow us to construct the following general mappings between the  spinor components  and the complex scalar field
\br
\label{map1}
\xi_{0A} &=& \pm \frac{c}{2\sqrt{a}} e^{-i\b \Phi/2} (\rho - e^{i\b \Phi} )^{1/2} \\
\label{map2}
\xi_{0B} &=&  \frac{d}{2\sqrt{a \rho}} (\rho - e^{i\b \Phi} )^{1/2} \\
\label{map3}
\eta_{0A}  &=&  \frac{d}{2\sqrt{a \rho}} (\rho - e^{i\b \Phi} )^{1/2} \\
\label{map4}
\eta_{0B}  &=& \pm \frac{c}{2\sqrt{a}} e^{-i\b \Phi/2}  (\rho - e^{i\b \Phi} )^{1/2}.
\er
Notice that these spinor field components  (\ref{map1})-(\ref{map4}) satisfy the relationships (\ref{AB1}). 
One can show the  ${\cal P} {\cal T} $ symmetry of the above zero mode  spinor assuming the parity symmetry (\ref{parity11})-(\ref{parity21}) of the scalar components. So, from   (\ref{map1})-(\ref{map2}) and  taking into account (\ref{parity11})-(\ref{parity21})  one can write 
\br
\label{ptspinor11}
{\cal P}{\cal T} \(\begin{array}{c}
 \xi_{0A} \\ 
\xi_{0B}
\end{array}\)  = \(\begin{array}{c}
 \xi_{0A} \\ 
\xi_{0B}
\end{array}\),  
\er
with the set of parameters 
\br
\label{param00}
 \rho = |\rho| e^{i \d_o}\, (\d_o = \pi/2),\,\,\, d = |d| e^{i \pi/2}  (\zeta_2 = \pi/2),\,\,\, a, c \in \IR  (\a_1 =0,\, \zeta_1=0). 
\er
From (\ref{param00}) one notices that the parameter condition $\zeta_1+\zeta_2= \pi/2$ holds in order to have the scalar parity symmetries (\ref{parity11})-(\ref{parity21}). Since the Hamiltonian is ${\cal P} {\cal T} $ symmetric according to (\ref{pth1zm}) and $\xi_{0}$ is a  ${\cal P} {\cal T} $ eigenstate, one can argue that the zero mode with the parameter set (\ref{param00})  belongs to the ${\cal P} {\cal T} $ symmetric regime, as it satisfies the conditions (\ref{Ah1})-(\ref{Axi1}). Different set of parameters from those in (\ref{param00}) will define a spontaneoulsly broken  ${\cal P} {\cal T} $ symmetric phase, even though the eigenvalues remain zero. 

In the Fig. 13 we show the plots of general $\xi_0$ Majorana zero mode components and the associated localized scalar fields in (\ref{zmkink})-(\ref{phi2mo}) for a particular set of parameters. The field $\Phi_1$ is a kink-type soliton with topological charge ${\cal Q}_1=1$, whereas the excitation $\Phi_2$ develops a trivial topological charge ${\cal Q}_2 = 0$.  This Fig. corresponds to $\d_m = \d_k \neq 0$ since $M_2=k_2 \neq 0$,\,and  $M_1=k_1 \neq 0$; so, the associated Hamiltonian $H$ can not be regarded as ${\cal PT}$-symmetric  according  to  (\ref{dels10})-(\ref{dels1}). 

The second spinor $\eta_0$ can be plotted using $\eta_{0A}=\xi_{0B}, \eta_{0B}=\xi_{0A}$. One can see that this non-Hermitian Majorana zero mode real components $\xi_{j},\, j=1,2,3,4$ are independent from each other as compared to its Hermitian counterpart reported in \cite{jhep22}, in which  the four real components reduce to two independent fields due to the particle-hole symmetry, or, a charge conjugation symmetry which holds in the relativistic Hermitian Toda model coupled to one Dirac fermion.   

\begin{figure}
\centering 
\includegraphics[width=3cm,scale=1, angle=0,height=3cm]{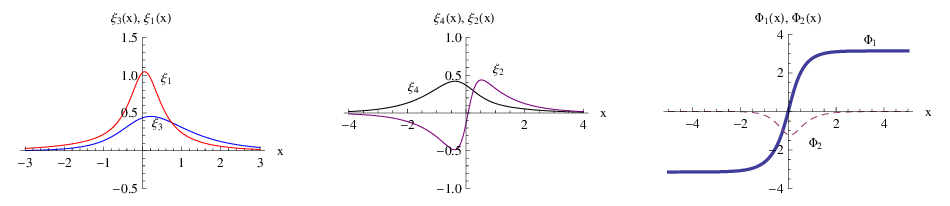}
\parbox{5in}{\caption{Majorana zero mode bound state $\xi_0$ and the associated scalar fields of eqs. (\ref{zmkink})-(\ref{phi2mo}) for $|a| =|b|=1, |c|=|d|=1, \zeta_1 =-\zeta_2 = 1.2, \a_1=\pi+1, \a_2 =1,k_1=1,k_2=-1,\b=1$. The field $\Phi_1$ is a soliton with ${\cal Q}_1=1$, whereas $\Phi_2$ possesses zero topological charge.}}
\end{figure}

Next, using the mapping (\ref{map1})-(\ref{map4}) into the equations of motion (\ref{sec1}) and defining $\rho \equiv e^{\rho_o} e^{i \d_o}$\ with $\rho_o\, \mbox{and} \, \d_o$\, being real parameters, one can write the next equation for  the complex scalar
\br
\label{NHSG}
-\pa^2_x  \Phi +   \frac{4 M^2}{\b} \sin{(\b \Phi - \d_{o} + i \rho_o )}=0.
\er
This is precisely the ${\cal P} {\cal T}-$symmetric sine-Gordon model introduced in \cite{bender22} by shifting the SG scalar by an imaginary constant, apart from the constant real number $\d_{o}$.  

Likewise, using the mapping (\ref{map1})-(\ref{map4}) into the system of equations of motion (\ref{H1})-(\ref{H11}) with $\l=0$ one can get  the following equations for the zero mode components
\br
\label{th1}
i \xi'_{0A} + i \frac{M}{\rho}\,  \xi_{0B}  + \frac{1}{2} \b^2 \, \xi_{0A}\, \xi_{0B}\,\eta_{0B} &=&0,\\
\label{th2}
-i \xi'_{0B} - i M \rho\,\, \xi_{0A}  + \frac{1}{2} \b^2 \,  \xi_{0A}\, \xi_{0B}\, \eta_{0A}  &=&0,\\  
\label{th3}
-i \eta'_{0A}  - i M \rho\,\, \eta_{0B} + \frac{1}{2} \b^2 \,  \eta_{0A}\, \eta_{0B} \, \xi_{0B} &=&0,\\  
\label{th4}
i \eta'_{0B} + i \frac{M}{\rho}\, \eta_{0A}  + \frac{1}{2} \b^2 \,  \eta_{0A}\, \eta_{0B}\, \xi_{0A}  &=&0.
\er
The model (\ref{NHSG}) can be considered as a NH sine-Gordon model describing the dynamics of the complex scalar field $\Phi$, whereas  the system (\ref{th1})-(\ref{th4}) as the NH Thirring model describing the biorthogonal Majorana fermions. Notice that the coupling parameter $\b^2$ appears in the three-fermion coupling (or four-fermion term in the Lagrangian description) of the system (\ref{th1})-(\ref{th4}), whereas in the Lagrangian description of the NHSG model (\ref{NHSG}) it would appear as $\frac{1}{\b^2}$ factor of  the $\cos{(\b \Phi - \d_{o} + i \rho_o )}$ term of the relevant potential. So, following the ideas of \cite{jhep22}, the mapping (\ref{map1})-(\ref{map4}) can be regarded as the strong $\rightleftarrows$ weak coupling mappings of the dual sectors of NH Toda model coupled to Majorana fermions. The Hermitian version of the model (\ref{th1})-(\ref{th4}) has recently been considered for two independent spinors and shown to be integrable in the bosonic and Grassmanian versions, respectively \cite{basu1,basu2}. A Hermitian (generalized) massive Thirring/sine-Gordon duality  
has also been discussed in \cite{jhep1, jhep2}. Non-Hermitian ${\cal PT}$-symmetric versions of the massive Thirring model and the sine-Gordon model have recently been considered, and some aspects of their duality relationships in the NH context were examined \cite{defenu, ashida}.  

So, the mapping (\ref{map1})-(\ref{map4}) turns out to be a new approach to generating solvable fermionic and bosonic  ${\cal PT}$-invariant models whenever there is a boson-fermion duality, since  it amounts to reproduce the ${\cal PT}$-symmetric versions of the massive Thirring and sine-Gordon models. The idea of generating a non-Hermitian but ${\cal PT}$-symmetric Hamiltonian from a Hermitian Hamiltonian by
shifting the boson field operator by an imaginary constant was suggested before, see e.g. \cite{bender22} and references therein.

Next, for completeness we provide the zero mode {\sl left} eigenvectors. So, one can set $\l =0$ into the system of equations (\ref{lefteigen1})-(\ref{equiv22}). Since we have already found above the left  eigenvectors for the general $\l$ eigenvalue following the tau functions formalism (\ref{tau01f})-(\ref{tauabf}) and (\ref{tau01xf})-(\ref{tauabxxf}), we simply set $w = \pm i\, (\l =0)$ for the zero modes into the relevant parameter relationships (\ref{para1f})-(\ref{lvrf}). So, one has 
\br
\bar{k} &=& \pm M,\,\,\,\,\,
\bar{b} = - \bar{a} ,\,\,\,\,\, \bar{\rho} = \pm \frac{\bar{c}'}{\bar{d}'} ,\,\,\,\,\,
\bar{d}' = \frac{\bar{c} \bar{c}'}{\bar{d}},\\
\bar{a}' &=& \pm i \frac{\bar{c}' \bar{c} \b^2}{8 M}.
\er 
Topological superconductors can host Majorana zero modes and are currently
intensively pursued as a promising candidate for topological quantum computation, but their experimental realization has so far been  inconclusive. It has recently been shown, using non-Hermitian exceptional points, that non-Hermiticity can strongly enhance topological superconductivity, implying more robust Majorana zero modes \cite{arouca}.

\section{Discussions and conclusions}
\label{sec:discussion}

We have studied a NH $sl(2)$ Toda model coupled to Dirac fields (\ref{lag1}) by examining the pseudo-chiral (\ref{spc1}) and (anti)-pseudo-Hermitian (\ref{psc2}) properties of the model and their relevant representations. The pseudo-chiral symmetry has recently been introduced as a manifestation of Noether's theorem in NH systems, such that a new inner product was defined different  from the one in standard quantum mechanics \cite{rivero}. We have discussed the properties of the right and left eigenvectors and the related eigenvalues, as well as the role played by the complex scalar field components.  Among the key characteristics of the model are the biorthogonality relationship between the left and right eigenvectos, and the fact that the complex eigenvalues come in pairs of type $\{\l, -\l\}$ and $\{\l, -\l^\star\}$. The Table 1 summarizes the main results related to these symmetries and their relevant representations, respectively, for the set of Hamiltonians $\{H, \widetilde{H}\}$ and  $\{H^\star, \widetilde{H}^\star\}$. This Table describes each type of symmetry and the corresponding eigenvectors and their mutual relationships, as well as the spectral parameter $\l$ properties allowed by the relevant symmetry. The biorthogonality relationships and the presence of the  Majorana zero modes at $\l=0$ and the coalescence of the eigenvectors are also indicated. 

Moreover, we present the interplay of the ${\cal P }$, ${\cal T }$, ${\cal PT}$ and $\g_5{\cal P} {\cal T }$symmetry transformations  with the pseudo-chiral and anti-pseudo-Hermiticity properties, which imply a symmetry of the Hamiltonian for each set of parities ($\nu_1,\,\nu_2$) which can take the scalar fields $\Phi_1$ and $\Phi_2$, respectively. In Table 2 we summarize the main results of the cases $A, B, C, D$ and $E$ discussed such that the last column of this Table indicates their relevant symmetries. Remarkably, in the cases $C$ and $E$ the both type of Hamiltonians $\{H, H^{\star}\}$ exhibit an anti-linear symmetry  ${\cal A}$\, (${\cal A} \in \{{\cal P} {\cal T },  \g_5{\cal P} {\cal T }\}$). These properties have been used in sections \ref{sec:real}, \ref{sec:class} and \ref{sec:Majo1}, respectively,  in order to examine the ${\cal A}$-symmetry unbroken  with real eigenvalues and the ${\cal A}$-symmetry broken phases of the model.  In the section \ref{sec:real} we have analyzed the Hamiltonian  with  scalar solitons  (\ref{phi1imp1})-(\ref{phi2p2}) and parity symmetry (\ref{par12ii})-(\ref{par12iii}), in section \ref{sec:class}  the scalar extended wave  (\ref{phi1e})-(\ref{phi2e}) with parity symmetry (\ref{elcsca1})-(\ref{elcsca2}), and in section \ref{sec:Majo1} the solitons (\ref{zmkinkdm02par})-(\ref{kink1zm}) with parities (\ref{parity11})-(\ref{parity21}). The relevant spinor eigenvectors of the operator ${\cal A}$ have also been constructed in each case. 

Some relationships between the complex scalar fields $\Phi$ and $\bar{\Phi}$ and the biorthogonal spinor sectors $\{\xi, \eta\}$ and $\{\bar{\xi}, \bar{\eta}\}$ of the model are furnished by the correspondences between the spinors and topological currents densities in (\ref{equi1}) and (\ref{equiv22}), respectively, together with the equations (\ref{lrcurr11}) and (\ref{PPhi}) provided by the  pseudo-chirality symmetry. The emergence of the fields  $\bar{\Phi}$ and $\{\bar{\xi}, \bar{\eta}\}$ is a feature which characterizes a  non-Hermitian Toda model coupled to Dirac spinors.  We have shown that the solutions (\ref{tau01}) and (\ref{tauab}) with the tau functions in (\ref{tau01x})-(\ref{tauabxx}) satisfy the system of equations (\ref{H1})-(\ref{H11}) which coupled the spinors and the scalar field and the currents equivalence equation (\ref{equi1}). In the soliton sector, the $x-$integral in the whole real line of the current density  (\ref{bilint1}) provides a complex number, which are proportional to the topological charges ${\cal Q}_1$ and ${\cal Q}_2$, respectively. In fact, the real part of (\ref{bilint1}) is proportional to ${\cal Q}_1$, whereas the imaginary part to ${\cal Q}_2$ as in (\ref{Q11})-(\ref{Q22}), respectively. So, the above NH feature is the analog of the equivalence between the $U(1)$ Noether and real topological currents in the Hermitian version of the model \cite{jhep22}.

The construction of the analytical biorthogonal spinor bound states and their associated complex scalar topological solitons were performed. Analytical  solutions of the static version of the system were obtained by using the tau function approach. 
The complex eigenvalue $\l$, in each topological sector, satisfies a second order algebraic equation as the dispersion relation (\ref{lmk1}),  $\l^2= M^2 - k^2$, plus its complex conjugate. So, a related uniformization parameter $z$ defined in (\ref{lvr}) has played a key role in understanding the asymptotic behavior of the scalar field. The solutions exhibit some topological charges associated to the asymptotic values of the complex scalar field  (\ref{topzz}), i.e. $\Phi(+ \infty)-\Phi(- \infty)= \pm \frac{4}{i \b} \log{z}$, with  $z \in \IC \backslash \{0\}$ being the uniformization parameter. The biorthogonal spinor bound states are found for each topological configurations of the background scalar fields. In these developments the solitons shapes depend crucially on the spinor bound state complex energy and wave number parameters, $\l $ and $k$, i.e. the soliton shape depends on the fermionic state to which it is coupled. So, it is worth mentioning that the back-reactions of the spinors on the solitons are described nonperturbatively and exactly through our analytical methods.  

The spectral topology concept is a unique feature in non-Hermitian systems. The eigenvalue $\l$ has been defined as an analytical function of the complex variable $z$ (\ref{lvr}). The dispersion relation $\l(k, M)$ can also be suitably parametrized by $\om$ (\ref{lom1}) in order to define the EPs and the zero modes at $w= \pm 1$; whereas, the $z-$parametrization is suitable to exhibit the relevant zero-modes and the topological charges of the scalar fields. The both parameters $z$ and $\om$ are related in (\ref{tr22}). We have used the $z-$parametrization  in the computations of the winding numbers of eigenvalues and the biorthogonality relationships of the left and right eigenvectors. In this scenario, the EPs points at $\om =\pm1$ (\ref{lom1}) ($z = \pm i$ in (\ref{z2})) arise as branch points in the $k-$space eigenvalue manifold (see the points at $k = \pm M$ in the left and right plots of the Fig. 8). The winding number of $\l(z)$ has been computed as it traverses the contour $\G$ encircling the zero modes $\l = 0 $ located at the points $z = \pm i$ and the pole at $z=0$ in the complex plane $z$ (see Fig. 1). The contour integral of the function $\l(z)$ using (\ref{wind1}) provides the winding number $W=1$ in (\ref{wind111}),  and so, it is in accordance with the Cauchy's argument principle (\ref{wind11}).
 
The general solution (\ref{tau01}) and (\ref{tauab}) with the tau functions in (\ref{tau01x})-(\ref{tauabxx}) can be classified into two types of solutions: 

1. The continuum of bound states/solitons (\ref{xiA})-(\ref{xiB}) and (\ref{phi1sol})-(\ref{phi2sol}), respectively. For the wave number with non-vanishing real component, i.e. $k_1 \neq 0$, we have defined the topological invariants for each component of the scalar field (\ref{top1})-(\ref{top2}). So, the spinor solutions  (\ref{xiA})-(\ref{xiB}) become spatially localized eigenstates at every value of  the energy in the complex plane $\l$ which satisfies  the dispersion relation (\ref{lmk1}) provided that $\Re{(k)} \neq 0$. So, this feature violates the general wisdom that bound states should be quantized. In section \ref{sec:class} we have discussed in more detail this important class of solutions of a NH $sl(2)$ Toda model coupled to fermions, and presented the Fig. 2 as a representative plot.  Recently, this phenomenon has also been reported in the context of NH lattice models \cite{wang2}. The charges $\Re{(Q_{topol})} =\pm 1,\,\Im{(Q_{topol})} =0$ are special in the sense that the relevant solitons host the Majorana zero-modes, as we will discuss below. 
 
2. Another class of solutions are the extended waves embedded into the localized states (ELC). These states hold provided that one sets  $k_1=0$ into the tau functions (\ref{tau01x})-(\ref{tauabxx}). The scalar fields take the form (\ref{phi1e})-(\ref{phi2e}) and the extended wave spinors  become (\ref{xae})-(\ref{xbe}). The Fig. 11 shows a representative plot of this type of state. The middle plot in Fig. 8 shows the curves $\l(k_1=0, k_2)$ corresponding to the ELC states. These states can be regarded as a generic NH phenomenon representing some kind of inverted BIC states due to the non-Hermicity of the model (\ref{lag1}). Recently, these novel type of ELC states in a NH system have been reported in the context of an active mechanical lattice \cite{wwang}. 

The above  type of solutions may further be classified into subsets according to the ${\cal P }$, ${\cal T }$, ${\cal PT}$ and $\g_5{\cal PT}$  symmetries of the Hamiltonians $H$ and $H^{\star}$ for each set of parities ($\nu_1,\,\nu_2$) of the scalar field components $\Phi_1$ and $\Phi_2$, respectively. In this context, we have constructed  the representations of the symmetries of type ${\cal A}\in \{{\cal PT},\g_5{\cal PT} \}$ presented in the last column of Table 2  of section \ref{PT1}. In the section \ref{sec:real} we have shown that the non-Hermitian model (\ref{lag1}) exhibits real energy spectra in a special parameter subspace. The regions of real and complex eigenvalues, and the EPs points of eigenvector degeneracies, have been identified. In general, the real eigenvalue condition in parameter space becomes (\ref{EPs1}). It has also been shown that the real eigenvalues come in pairs $\pm \l_1$, so preserving the pseudo-chiral symmetry. These issues have been examined in sections \ref{sec:real} , \ref{sec:class} and \ref{sec:Majo1} where we have shown that  the anti-pseudo-Hermitian symmetry breaking occurs at the EPs defined as $\om = \pm1$ in (\ref{lom1}). Then, one can conclude that  the physical regimes of the model are determined by the discrete antilinear symmetries ${\cal P }{\cal T }$ and $\g_5{\cal P }{\cal T }$, respectively. 
   
In order to understand the NH soliton-particle duality we have examined the zero mode bound states and corresponding scalar field solutions in section \ref{sec:Majo1}. In contradistinction to the Hermitian case in \cite{jhep22} the NH  Majorana zero mode components $\xi_{j},\, j=1,2,3,4$ are independent from each other, and they are hosted by the complex scalar field supporting a topological soliton in its real component and a wave with vanishing  topological charge in its imaginary component (see Fig. 13). Then, one can argue that these non-Hermitian zero modes are stable even in the presence of the imaginary component scalar wave with vanishing topological charge. We uncovered the mapping (\ref{map1})-(\ref{map4}) which can be regarded as a strong $\rightleftarrows$ weak duality mappings of the dual sectors of NH Toda model coupled to Majorana fermions. In this picture the strong coupling sector is described by  a  NH sine-Gordon model (\ref{NHSG}), whereas  the weak coupling sector by the NH Thirring model (\ref{th1})-(\ref{th4}) of biorthogonal Majorana fermions. In future works it would be interesting to consider the NH  (generalized) massive Thirring/sine-Gordon duality extension of the relevant Hermitian models \cite{jhep1, jhep2}.  

Some analytical and numerical extensions of this work deserve to be considered, such as the NH kink-antikink scattering in order to understand  the behavior of the NH fermions and hosted bound states, as well as the extensions for more than one species of chiral fermions and scalars in the scalar-fermion interaction term. In particular, a scalar potential term can be introduced into the Lagrangian in order to study the NH version of the model studied in \cite{jhep22}. So, our work presented the pseudo-chiral and (anti)-pseudo-Hermitian symmetries and their representations from the field theory perspective which were mainly reported before in  the context of NH lattice models. We hope our approach will be relevant to the understanding of the physical origins of non-Hermiticity, the finite lifetimes of resonances, back reaction of fermions on solitons and the interplay of nonlinearity, topology, and integrability in the context of NH integrable field theory models. 

Finally, it would be interesting  to study the dynamics of kinks and bound state solutions of the NH model (\ref{lag1})  in the context of quasi-integrability, and the relevant quasi-conserved quantities in the so-called (quasi-)integrable approach \cite{arxiv2}. So far, to our knowledge, a NH model possessing a strong-weak duality property had not been studied in the quasi-integrability context. Besides, our findings could be relevant in the context of one-dimensional topological superconductors in order to understand how the non-hermiticity influence the shape and the lifetime of the bound states, in particular the Majorana zero modes. It will be an interesting issue to study the robustness against decoherence and disorder of our non-Hermitian Majorana zero modes, as compared to their Hermitian counterparts. These type of non-Hermitian modes, being more stable in the presence of imperfections, make them appealing for quantum information processing and fault-tolerant quantum computing. These and other extensions will be reported elsewhere. 

\vspace{1cm}

\noindent {\bf Acknowledgements}

The author is grateful to J.J. Monsalve and J. R. V. Pereira for useful discussions, and Dr. A. M. Jazayeri for calling his attention to reference  \cite{amir}.  
 
\appendix
 
\section{Equations in components}
\label{app:eqs11}
Taking $\lambda=\lambda_1+i\lambda_2,\,\, \Phi = \Phi_1+i\Phi_2,\,\, M = M_1+ i M_2$ and
\br
\label{xi11}
\psi & = & e^{-i\lambda t} \xi,\,\,\,\,\,\,\,\,\, \xi = \(\begin{array}{c}
 \xi_3+i \xi_4 \\ 
\xi_1+i \xi_2
\end{array}\),\\
\label{eta11}
\chi &=&  e^{i \l t} \eta,\,\,\,\,\,\,\,\,\, \eta =  \(\begin{array}{c}
\eta_3+i\eta_4\\ 
\eta_1+i\eta_2
\end{array}\),
\er
one has the next equations for the spinor components 
\begin{equation}\label{eq1}
\xi'_1  + (\lambda_1 \xi_2+\lambda_2\xi_1 )+ e^{-\beta\Phi_2} [ (M_1 \xi_3 -M_2 \xi_4 )\cos{(\beta\Phi_1)}  - (M_2 \xi_3 + M_1 \xi_4 ) \sin{(\beta\Phi_1)}] =0, \\ 
\end{equation}
\begin{equation}\label{eq2}
\xi'_2  - (\lambda_1 \xi_1-\lambda_2\xi_2)+ e^{-\beta\Phi_2} [ (M_2 \xi_3 +M_1 \xi_4 )\cos{(\beta\Phi_1)}  + (M_1 \xi_3 - M_2 \xi_4 ) \sin{(\beta\Phi_1)}] =0,\\ 
\end{equation}
\begin{equation}\label{eq3}
\xi'_3 - (\lambda_1 \xi_4+\lambda_2\xi_3)+e^{\beta\Phi_2} [ (M_1 \xi_1 -M_2 \xi_2 )\cos{(\beta\Phi_1)}  + (M_2 \xi_1 + M_1 \xi_2 ) \sin{(\beta\Phi_1)}] =0, \\ 
\end{equation}
\begin{equation}\label{eq4}
\xi'_4 + (\lambda_1 \xi_3-\lambda_2\xi_4)+e^{\beta\Phi_2} [ (M_2 \xi_1 +M_1 \xi_2 )\cos{(\beta\Phi_1)}  + (M_2 \xi_2 - M_1 \xi_1 ) \sin{(\beta\Phi_1)}] =0,\\ 
\end{equation}
and the equations for the $\eta_a's\,(a=1,2,3,4) $ can be obtained from (\ref{eq1})-(\ref{eq4}) by replacing  
\br 
\xi_3 &\rightarrow & \eta_1,\,\,\,\,\xi_4 \rightarrow \eta_2,\\
\xi_1&\rightarrow & \eta_3,\,\,\,\,\xi_2 \rightarrow \eta_4,
\er
in accordance to the relationship $\xi = \G \eta$ in (\ref{xieta1}). So, let us write 
\begin{equation}\label{eq11c}
\eta'_3  + (\lambda_1 \eta_4 +\lambda_2\eta_3 )+ e^{-\beta\Phi_2} [ (M_1 \eta_1 -M_2 \eta_2 )\cos{(\beta\Phi_1)}  - (M_2 \eta_1 + M_1 \eta_2 ) \sin{(\beta\Phi_1)}] =0, \\ 
\end{equation}
\begin{equation}\label{eq22c}
\eta'_4  - (\lambda_1 \eta_3 - \lambda_2\eta_4)+ e^{-\beta\Phi_2} [ (M_2 \eta_1 +M_1 \eta_2 )\cos{(\beta\Phi_1)}  + (M_1 \eta_1 - M_2 \eta_2 ) \sin{(\beta\Phi_1)}] =0,\\ 
\end{equation}
\begin{equation}\label{eq33c}
\eta'_1 - (\lambda_1 \eta_2 + \lambda_2\eta_1)+e^{\beta\Phi_2} [ (M_1 \eta_3 -M_2 \eta_4 )\cos{(\beta\Phi_1)}  + (M_2 \eta_3 + M_1 \eta_4 ) \sin{(\beta\Phi_1)}] =0, \\ 
\end{equation}
\begin{equation}\label{eq44c}
\eta'_2 + (\lambda_1 \eta_1- \lambda_2\eta_2)+e^{\beta\Phi_2} [ (M_2 \eta_3 +M_1 \eta_4 )\cos{(\beta\Phi_1)}  + (M_2 \eta_4 - M_1 \eta_3 ) \sin{(\beta\Phi_1)}] =0,\\ 
\end{equation}

For the scalar field components one has
\br\nonumber
&&\partial^2\Phi_1 + \beta  \cos{(\beta\Phi_1)} \Big[-e^{\beta\Phi_2}(\eta_4(M_2 \xi_1+M_1\xi_2) - \eta_3 (M_1 \xi_1-M_2\xi_2)) + \\ \nonumber
&&e^{-\beta\Phi_2} (\eta_1(M_1 \xi_3-M_2\xi_4) - \eta_2 (M_1 \xi_4+M_2\xi_3))\Big] +\\
\nonumber
&& \nonumber \beta    \sin{(\beta\Phi_1)} \Big[ e^{\beta\Phi_2} (\eta_3(M_2 \xi_1+M_1\xi_2) + \eta_4 (M_1 \xi_1-M_2\xi_2))- \\
&& e^{-\beta\Phi_2} (\eta_1(M_2 \xi_3 + M_1\xi_4) + \eta_2 (M_1 \xi_3-M_2\xi_4))\Big] =0 \label{eq9}
\er
and
\br
&&\nonumber
\partial^2\Phi_2 + \beta  \cos{(\beta\Phi_1)} \Big[e^{\beta\Phi_2}(\eta_3(M_2 \xi_1+M_1\xi_2) + \eta_4 (M_1 \xi_1-M_2\xi_2)) + \\ 
&& \nonumber e^{-\beta\Phi_2} (\eta_1(M_2 \xi_3+M_1\xi_4) - \eta_2 (M_2 \xi_4-M_1\xi_3))\Big] +\\
\nonumber
&&  \beta    \sin{(\beta\Phi_1)} \Big[ e^{\beta\Phi_2} (\eta_4(M_2 \xi_1+M_1\xi_2) - \eta_3 (M_1 \xi_1-M_2\xi_2))+ \\
&& e^{-\beta\Phi_2} (\eta_1(M_1 \xi_3-M_2\xi_4) - \eta_2 (M_2 \xi_3+M_1\xi_4))\Big] =0 \label{eq10}
\er
Notice that the scalar equation for $\Phi_2$ (\ref{eq10}) becomes trivial provided that 
\br
\label{phi20}
\Phi_2 =0,\,\,\,\,\,\eta = \pm \xi^{\star}, \,\,\,\,\, M_2 =0.
\er
Taking into account the condition (\ref{phi20}) one has that the system of eqs. (\ref{eq1})-(\ref{eq4}), (\ref{eq11c})-(\ref{eq44c}) and (\ref{eq9})-(\ref{eq10}) reduce to the Hermitian model recently studied in \cite{jhep22}. Moreover, the relationship (\ref{equi1}) becomes
\br
\label{phi1top}
\Phi'_1 &=& -\frac{\b}{2} [\eta_1 \xi_1-\eta_2 \xi_2+\eta_3 \xi_3-\eta_4 \xi_4 ] ,\\ \label{phi1top1}
&=& - \frac{\b }{4}[\eta_A \xi_A + \eta_A^{\star} \xi_A^{\star}+ \eta_B \xi_B + \eta_B^{\star} \xi_B^{\star}],\\
\label{phi1top2}
&=&  - \frac{\b }{2} [ \xi_A  \xi_B + \xi_A^{\star} \xi_B^{\star}  ],\\
\Phi'_2 &=& -\frac{\b}{2} [\eta_2 \xi_1+\eta_1 \xi_2+\eta_4 \xi_3+\eta_3 \xi_4 ],\label{phi2top}\\
\label{phi2top1}
&=& \frac{\b i}{4}[\eta_A \xi_A - \eta_A^{\star} \xi_A^{\star}+ \eta_B \xi_B - \eta_B^{\star} \xi_B^{\star}],\\
\label{phi2top2}
&=&  \frac{\b i }{2} [ \xi_B  \xi_A - \xi_B^{\star} \xi_A^{\star}  ],
\er
The relationship (\ref{phi2top}) becomes trivial if the condition (\ref{phi20}) is assumed.
However, for $M_2 \neq 0$, $\Phi_2 =0$ and $\eta \neq \pm  \xi^{\star}$ one has that the eq. (\ref{eq10}) provides the constraint
\br
\label{const12}
\tan{(\b \Phi_1)} = - \frac{ \eta_3(M_2 \xi_1+M_1\xi_2) + \eta_4 (M_1 \xi_1-M_2\xi_2) + \eta_1(M_2 \xi_3+M_1\xi_4) - \eta_2 (M_2 \xi_4-M_1\xi_3)}{\eta_4(M_2 \xi_1+M_1\xi_2) - \eta_3 (M_1 \xi_1-M_2\xi_2)+ \eta_1(M_1 \xi_3- M_2\xi_4) - \eta_2 (M_2 \xi_3+M_1\xi_4)}.
\er

\end{document}